\documentclass[prodtf]{mscs}

\pubyear{2005}
\pagerange{37--92}
\volume{15(1)}
\doi{S0960129504004426}

\usepackage{t1enc} 
\usepackage{amsmath}
\usepackage{makeidx}
\usepackage{latexsym}
\usepackage{theorem}
\usepackage{amssymb}
\usepackage{url}

\newcommand{\I}{}
\newcommand{\at}{}

\newdimen\proofrulebreadth \proofrulebreadth=.05em
\newdimen\proofdotseparation \proofdotseparation=1.25ex
\newdimen\proofrulebaseline \proofrulebaseline=2ex
\newcount\proofdotnumber \proofdotnumber=3
\let\then\relax
\def\hfi{\hskip0pt plus.0001fil}
\mathchardef\squigto="3A3B
\newif\ifinsideprooftree\insideprooftreefalse
\newif\ifonleftofproofrule\onleftofproofrulefalse
\newif\ifproofdots\proofdotsfalse
\newif\ifdoubleproof\doubleprooffalse
\let\wereinproofbit\relax
\newdimen\shortenproofleft
\newdimen\shortenproofright
\newdimen\proofbelowshift
\newbox\proofabove
\newbox\proofbelow
\newbox\proofrulename
\def\shiftproofbelow{\let\next\relax\afterassignment\setshiftproofbelow\dimen0 }
\def\shiftproofbelowneg{\def\next{\multiply\dimen0 by-1 }%
\afterassignment\setshiftproofbelow\dimen0 }
\def\setshiftproofbelow{\next\proofbelowshift=\dimen0 }
\def\setproofrulebreadth{\proofrulebreadth}

\def\prooftree{
\ifnum  \lastpenalty=1
\then   \unpenalty
\else   \onleftofproofrulefalse
\fi
\ifonleftofproofrule
\else   \ifinsideprooftree
        \then   \hskip.5em plus1fil
        \fi
\fi
\bgroup
\setbox\proofbelow=\hbox{}\setbox\proofrulename=\hbox{}%
\let\justifies\proofover\let\leadsto\proofoverdots\let\Justifies\proofoverdbl
\let\using\proofusing\let\[\prooftree
\ifinsideprooftree\let\]\endprooftree\fi
\proofdotsfalse\doubleprooffalse
\let\thickness\setproofrulebreadth
\let\shiftright\shiftproofbelow \let\shift\shiftproofbelow
\let\shiftleft\shiftproofbelowneg
\let\ifwasinsideprooftree\ifinsideprooftree
\insideprooftreetrue
\setbox\proofabove=\hbox\bgroup$\displaystyle 
\let\wereinproofbit\prooftree
\shortenproofleft=0pt \shortenproofright=0pt \proofbelowshift=0pt
\onleftofproofruletrue\penalty1
}

\def\eproofbit{
\ifx    \wereinproofbit\prooftree
\then   \ifcase \lastpenalty
        \then   \shortenproofright=0pt  
        \or     \unpenalty\hfil         
        \or     \unpenalty\unskip       
        \else   \shortenproofright=0pt  
        \fi
\fi
\global\dimen0=\shortenproofleft
\global\dimen1=\shortenproofright
\global\dimen2=\proofrulebreadth
\global\dimen3=\proofbelowshift
\global\dimen4=\proofdotseparation
\global\count255=\proofdotnumber
$\egroup  
\shortenproofleft=\dimen0
\shortenproofright=\dimen1
\proofrulebreadth=\dimen2
\proofbelowshift=\dimen3
\proofdotseparation=\dimen4
\proofdotnumber=\count255
}

\def\proofover{
\eproofbit 
\setbox\proofbelow=\hbox\bgroup 
\let\wereinproofbit\proofover
$\displaystyle
}%
\def\proofoverdbl{
\eproofbit 
\doubleprooftrue
\setbox\proofbelow=\hbox\bgroup 
\let\wereinproofbit\proofoverdbl
$\displaystyle
}%
\def\proofoverdots{
\eproofbit 
\proofdotstrue
\setbox\proofbelow=\hbox\bgroup 
\let\wereinproofbit\proofoverdots
$\displaystyle
}%
\def\proofusing{
\eproofbit 
\setbox\proofrulename=\hbox\bgroup 
\let\wereinproofbit\proofusing
\kern0.3em$
}

\def\endprooftree{
\eproofbit 
  \dimen5 =0pt
\dimen0=\wd\proofabove \advance\dimen0-\shortenproofleft
\advance\dimen0-\shortenproofright
\dimen1=.5\dimen0 \advance\dimen1-.5\wd\proofbelow
\dimen4=\dimen1
\advance\dimen1\proofbelowshift \advance\dimen4-\proofbelowshift
\ifdim  \dimen1<0pt
\then   \advance\shortenproofleft\dimen1
        \advance\dimen0-\dimen1
        \dimen1=0pt
        \ifdim  \shortenproofleft<0pt
        \then   \setbox\proofabove=\hbox{%
                        \kern-\shortenproofleft\unhbox\proofabove}%
                \shortenproofleft=0pt
        \fi
\fi
\ifdim  \dimen4<0pt
\then   \advance\shortenproofright\dimen4
        \advance\dimen0-\dimen4
        \dimen4=0pt
\fi
\ifdim  \shortenproofright<\wd\proofrulename
\then   \shortenproofright=\wd\proofrulename
\fi
\dimen2=\shortenproofleft \advance\dimen2 by\dimen1
\dimen3=\shortenproofright\advance\dimen3 by\dimen4
\ifproofdots
\then
        \dimen6=\shortenproofleft \advance\dimen6 .5\dimen0
        \setbox1=\vbox to\proofdotseparation{\vss\hbox{$\cdot$}\vss}%
        \setbox0=\hbox{%
                \advance\dimen6-.5\wd1
                \kern\dimen6
                $\vcenter to\proofdotnumber\proofdotseparation
                        {\leaders\box1\vfill}$%
                \unhbox\proofrulename}%
\else   \dimen6=\fontdimen22\the\textfont2 
        \dimen7=\dimen6
        \advance\dimen6by.5\proofrulebreadth
        \advance\dimen7by-.5\proofrulebreadth
        \setbox0=\hbox{%
                \kern\shortenproofleft
                \ifdoubleproof
                \then   \hbox to\dimen0{%
                        $\mathsurround0pt\mathord=\mkern-6mu%
                        \cleaders\hbox{$\mkern-2mu=\mkern-2mu$}\hfill
                        \mkern-6mu\mathord=$}%
                \else   \vrule height\dimen6 depth-\dimen7 width\dimen0
                \fi
                \unhbox\proofrulename}%
        \ht0=\dimen6 \dp0=-\dimen7
\fi
\let\doll\relax
\ifwasinsideprooftree
\then   \let\VBOX\vbox
\else   \ifmmode\else$\let\doll=$\fi
        \let\VBOX\vcenter
\fi
\VBOX   {\baselineskip\proofrulebaseline \lineskip.2ex
        \expandafter\lineskiplimit\ifproofdots0ex\else-0.6ex\fi
        \hbox   spread\dimen5   {\hfi\unhbox\proofabove\hfi}%
        \hbox{\box0}%
        \hbox   {\kern\dimen2 \box\proofbelow}}\doll%
\global\dimen2=\dimen2
\global\dimen3=\dimen3
\egroup 
\ifonleftofproofrule
\then   \shortenproofleft=\dimen2
\fi
\shortenproofright=\dimen3
\onleftofproofrulefalse
\ifinsideprooftree
\then   \hskip.5em plus 1fil \penalty2
\fi
}

\newcommand{\comment}[1]{}
\newcommand{\bu}{$\bullet$}

\newcommand{\vsp}[1][1mm]{\vspace*{#1}}
\newcommand{\moins}{\setminus}
\newcommand{\vide}{\emptyset}

\newcommand{\lfp}{\mr{lfp}}

\newcommand{\dom}{\mr{dom}}

\newcommand{\FV}{\mr{FV}}
\newcommand{\pos}{\mr{Pos}}

\renewcommand{\a}{\rightarrow}
\newcommand{\A}{\Rightarrow}
\renewcommand{\aa}{\leftrightarrow}

\newcommand{\ad}{\downarrow}

\renewcommand{\to}{\mapsto}

\newcommand{\ab}{\a_\b}
\newcommand{\ai}{\a_\io}
\renewcommand{\ae}{\a_\eta}
\newcommand{\abe}{\a_{\b\eta}}

\newcommand{\ar}{\a_\cR}
\newcommand{\abr}{\a_{\b\cR}}

\newcommand{\al}[1][]{{}_{#1}\!\!\leftarrow}
\newcommand{\als}[1][]{{}_{#1}^*\!\!\leftarrow}

\renewcommand{\I}[1]{[\![#1]\!]}

\newcommand{\ex}{\exists}
\newcommand{\all}{\forall}
\newcommand{\ou}{\vee}

\newcommand{\biget}{\bigwedge}
\newcommand{\et}{\wedge}
\newcommand{\non}{\neg}
\newcommand{\st}{\star}
\newcommand{\B}{\Box} 
\renewcommand{\th}{\vdash}

\newcommand{\sle}{\subseteq}

\newcommand{\tle}{\unlhd}
\newcommand{\tlt}{\lhd}
\newcommand{\tgt}{\rhd}

\newcommand{\cge}{\succeq}

\newcommand{\cgt}{\succ}

\newcommand{\qge}{\sqsupseteq}

\newcommand{\qgt}{\sqsupset}

\newcommand{\lex}{_\mr{lex}}
\newcommand{\mul}{_\mr{mul}}

\renewcommand{\u}[1]{{\underline{#1}}}

\renewcommand{\b}{\beta}
\newcommand{\g}{\gamma}
\newcommand{\G}{\Gamma}
\renewcommand{\d}{\delta}
\newcommand{\D}{\Delta}

\newcommand{\vep}{\varepsilon}
\newcommand{\z}{\zeta}
\renewcommand{\t}{\theta}

\newcommand{\io}{\iota}
\newcommand{\ka}{\kappa}
\newcommand{\la}{\lambda}
\renewcommand{\L}{\Lambda}
\renewcommand{\r}{\rho}
\newcommand{\s}{\sigma}
\renewcommand{\S}{\Sigma}

\newcommand{\vphi}{\varphi}
\newcommand{\w}{\omega}

\newcommand{\mc}{\mathcal}

\newcommand{\mr}{\mathrm}
\newcommand{\mb}{\mathbb}

\newcommand{\cB}{\mc{B}}
\newcommand{\cC}{\mc{C}}
\newcommand{\cD}{\mc{D}}
\newcommand{\cE}{\mc{E}}
\newcommand{\cF}{\mc{F}}
\newcommand{\cG}{\mc{G}}

\newcommand{\cI}{\mc{I}}
\newcommand{\cJ}{\mc{J}}
\newcommand{\cK}{\mc{K}}

\newcommand{\cN}{\mc{N}}

\newcommand{\cP}{\mc{P}}

\newcommand{\cR}{\mc{R}}
\newcommand{\cS}{\mc{S}}
\newcommand{\cT}{\mc{T}}

\newcommand{\cV}{\mc{V}}
\newcommand{\cW}{\mc{W}}
\newcommand{\cX}{\mc{X}}

\newcommand{\fa}{\mathfrak{a}}
\newcommand{\fb}{\mathfrak{b}}

\newcommand{\va}{{\vec{a}}}

\newcommand{\vl}{{\vec{l}}}
\newcommand{\vm}{{\vec{m}}}

\newcommand{\vt}{{\vec{t}}}
\newcommand{\vu}{{\vec{u}}}
\newcommand{\vv}{{\vec{v}}}
\newcommand{\vw}{{\vec{w}}}
\newcommand{\vx}{{\vec{x}}}
\newcommand{\vy}{{\vec{y}}}
\newcommand{\vz}{{\vec{z}}}

\newcommand{\vR}{{\vec{R}}}
\newcommand{\vS}{{\vec{S}}}
\newcommand{\vT}{{\vec{T}}}
\newcommand{\vU}{{\vec{U}}}
\newcommand{\vV}{{\vec{V}}}

\newenvironment{rul}%
  {$\begin{array}{rcl}}%
  {\end{array}$}
  {\begin{center}\begin{rul}}%
  {\end{rul}\end{center}}

\newenvironment{rew}[1][~~\a~~]%
  {$\begin{array}{r@{#1}l}}%
  {\end{array}$}
\newenvironment{rewc}[1][~~\a~~]%
  {\begin{center}\begin{rew}[#1]}%
  {\end{rew}\end{center}}

\newenvironment{typc}[1][\,:\,]{\begin{rewc}[#1]}{\end{rewc}}

\newcounter{counter}

{\theorembodyfont{\rmfamily} 
  \newtheorem{dfn}[counter]{Definition}
  \newtheorem{lem}[counter]{Lemma}
  \newtheorem{thm}[counter]{Theorem}

}

\newcommand{\cqfd}{\hfill$\blacksquare$} 
\newenvironment{prf}{{\bf Proof.}}{}

\newenvironment{lstgeneric}[2]
  {\begin{list}{#1}{\topsep=.5mm\itemsep=.5mm\parsep=0mm%
    \itemindent=-3ex\labelsep=1ex\labelwidth=0ex #2}}
  {\end{list}}

\newenvironment{lst}[1]
  {\begin{lstgeneric}{#1}{\itemindent=-1ex}}
  {\end{lstgeneric}}

\newenvironment{enumi}[1]
  {\begin{lstgeneric}{}{\usecounter{enumi}\leftmargin=7mm%
    }}
  {\end{lstgeneric}}

\newenvironment{bfenumi}[1]
  {\begin{lstgeneric}{}{\usecounter{enumi}\leftmargin=7mm%
    }}
  {\end{lstgeneric}}

\newenvironment{bfenumii}[1]
  {\begin{lstgeneric}{}{\usecounter{enumii}\leftmargin=7mm%
    }}
  {\end{lstgeneric}}

\newenvironment{enumalphai}
  {\begin{lstgeneric}{}{\usecounter{enumi}\leftmargin=7mm%
    }}
  {\end{lstgeneric}}

\newenvironment{bfenumalphai}
  {\begin{lstgeneric}{}{\usecounter{enumi}\leftmargin=7mm%
    }}
  {\end{lstgeneric}}

\renewenvironment{prf}{\begin{proof}}{\end{proof}}
\renewcommand{\cqfd}{}

\begin{document}




\title[Definitions by rewriting in the Calculus of Constructions]
{Definitions by rewriting\\ in the Calculus of Constructions}

\author[Fr\'ed\'eric Blanqui]{Fr\'ed\'eric Blanqui$^1$$^2$\\
$^1$ Laboratoire d'Informatique de l'\'Ecole Polytechnique (LIX)\addressbreak
91128 Palaiseau Cedex, France\addressbreak
(until 30 September 2003)\addressbreak
$^2$ Institut National de Recherche en Informatique et Automatique
(INRIA)\addressbreak
Laboratoire lorrain de Recherche en Informatique et ses Applications
(LORIA)\addressbreak
615 rue du Jardin Botanique, BP 101, 54602 Villers-lès-Nancy,
France\addressbreak
{\tt blanqui@loria.fr} (from 1st October 2003)}

\date{16 September 2002. Revised 12 November 2003.}

\maketitle

\noindent{\bf Abstract:} {\it This paper presents general syntactic
conditions ensuring the strong normalization and the logical
consistency of the Calculus of Algebraic Constructions, an extension
of the Calculus of Constructions with functions and predicates defined
by higher-order rewrite rules. On the one hand, the Calculus of
Constructions is a powerful type system in which one can formalize the
propositions and natural deduction proofs of higher-order logic. On
the other hand, rewriting is a simple and powerful computation
paradigm. The combination of both allows, among other things, to
develop formal proofs with a reduced size and more automation compared
with more traditional proof assistants. The main novelty is to
consider a general form of rewriting at the predicate-level which
generalizes the strong elimination of the Calculus of Inductive
Constructions.}




\section{Introduction}
\label{sec-intro}

\newcommand{\xu}{\{x\to u\}}

This work aims at defining an expressive language allowing to specify
and prove mathematical properties easily. The quest for such a
language started with Girard' system F \cite{girard72thesis} on the
one hand and De Bruijn's Automath project \cite{debruijn68} on the
other hand. Later, Coquand and Huet combined both calculi into the
Calculus of Constructions (CC) \cite{coquand85thesis}. As in system F,
in CC, data types are defined through impredicative encodings that are
difficult to use in practice. So, following Martin-L\"of's theory of
types \cite{martinlof84book}, Coquand and Paulin-Mohring defined an
extension of CC with inductive types and their associated induction
principles as first-class objects, the Calculus of Inductive
Constructions (CIC) \cite{coquand88colog}, which is the basis of the
proof-assistant Coq \cite{coq02}.

However, defining functions or predicates by induction is not always
convenient. Moreover, with such definitions, equational reasoning is
uneasy and leads to very large proof terms. Yet, for decidable
theories, equational proofs need not to be kept in proof terms. This
idea that proving is not only reasoning (undecidable) but also
computing (decidable) has been recently formalized in a general way by
Dowek, Hardin and Kirchner with the Natural Deduction Modulo (NDM) for
first-order logic \cite{dowek98trtpm}.

A more convenient and powerful way of defining functions and
predicates is by using rewrite rules \cite{dershowitz90book}. This
notion is very old but its study really began in the 70's with Knuth
and Bendix \cite{bendix70book} for knowing whether, in a given
equational theory, an equation is valid or not. Then, rewriting was
quickly used as a programming paradigm (see \cite{dershowitz90book})
since any computable function can be defined by rewrite rules.

In the following sub-sections, we present in more details our
motivations for extending CIC with rewriting, the previous works on
the combination of $\la$-calculus and rewriting, and our own
contributions.




\subsection{Advantages of rewriting}
\label{subsec-motiv}

In CIC, functions and predicates can be defined by induction on
inductively defined types. The case of the type $nat$ of natural
numbers, defined from $0:nat$ (zero) and $s:nat\A nat$ (successor
function), yields G\"odel' system T: a function $f:nat\A\tau$ is
defined by giving a pair of terms $(u,v)$, written $(rec~u~v)$, where
$u:\tau$ is the value of $f(0)$ and $v:nat\A\tau\A\tau$ is a function
which computes the value of $f(n+1)$ from $n$ and $f(n)$. Computations
proceeds by applying the following (higher-order) rewrite rules,
called {\em $\io$-reduction}:

\begin{rewc}[~~\ai~~]
rec~u~v~0 & u\\
rec~u~v~(s~n) & v~n~(rec~u~v~n)\\
\end{rewc}

\newcommand{\xt}{\{x\to t\}}
\newcommand{\bi}{{\b\io}}
\newcommand{\br}{{\b\cR}}

For instance, addition can be defined by the term $\la xy.(rec~u~v~x)$
with $u=y$ and $v=\la nr.s(r)$ (definition by induction on $x$). Then,
one can check that:\footnote{$\ab^*$ is the transitive closure of the
$\b$-reduction relation: $(\la x.t~u)\ab u\xt$.}

\begin{center}
$2+2\ab^* rec~2~v~2\ai v~1~(rec~2~v~1)\ab^* s(rec~2~v~1)$\\
$\ai s(v~0~(rec~2~v~0))\ab^* s(s(rec~2~v~0))\ai s(s(2))=4$
\end{center}

Proofs by induction are formalized in the same way: if $P$ is a
predicate on natural numbers, $u$ a proof of $P0$ and $v$ a proof of
$(n:nat)Pn\A P(sn)$,\footnote{As often in type systems, we denote
universal quantification over a type $T$ by $(x:T)$.} then $rec~P~u~v$
is a proof of $(n:nat)Pn$, and $\io$-reduction corresponds to the
elimination of induction cuts. In fact, $(rec~u~v)$ is nothing but a
particular case of $(rec~P~u~v)$ with the non-dependent predicate
$P=\la n.\tau$.

In addition, deduction steps are made modulo
$\bi$-equivalence\footnote{Reflexive, symmetric and transitive closure
of the $\bi$-reduction relation (which is the union of the $\b$ and
$\io$ reduction relations).}, that is, if $\pi$ is a proof of $P$ and
$P =_\bi Q$, then $\pi$ is also a proof of $Q$. For instance, if $\pi$
is a proof of $P(2+2)$, then it is also a proof of $P(4)$, as one
would naturally expect. The verification that a term $\pi$ is indeed a
proof of a proposition $P$, called type-checking, is decidable since
$\bi$ is a confluent (the order of computations does not matter) and
strongly normalizing (there is no infinite computation) relation
\cite{werner94thesis}.

Although the introduction of inductive types and their induction
principles as first-class objects is a big step towards a greater
usability of proof assistants, we are going to see that the
restriction of function definitions to definitions by induction, and
the restriction of type conversion to $\bi$-equivalence, have several
important drawbacks. The use of rewriting, that is, the ability of
defining functions by giving a set of rewrite rules $\cR$, and the
possibility of doing deductions modulo $\b\cR$-equivalence, can remedy
these problems. It appears that $\io$-reduction itself is nothing but
a particular case of higher-order rewriting
\cite{klop93tcs,nipkow91lics} where, as opposed to first-order
rewriting, the constructions of the $\la$-calculus (application,
abstraction and product) can be used in the right hand-sides of
rules.\footnote{We will not consider higher-order pattern-matching
here although it should be possible as we show it for the simply-typed
$\la$-calculus in \cite{blanqui00rta}.} A common example of a
higher-order definition is the function $map$ which applies a function
$f$ to each element of a list:

\begin{rewc}
map~f~nil & nil\\
map~f~(cons~x~\ell) & cons~(f~x)~(map~f~\ell)\\
\end{rewc}

\noindent
where $nil$ stands for the empty list and $cons$ for the function
adding an element at the head of a list.\\


\noindent{\bf Easier definitions.} First of all, with rewriting,
definitions are easier. For instance, addition can be defined by
simply giving the rules:

\begin{rewc}
0+y & y\\
(s~x)+y & s~(x+y)\\
\end{rewc}

\noindent
Then, we have $2+2\a s(2+1)\a s(s(2+0))\a s(s(2))=4$. Of course, one
can make the definitions by induction look like this one, as it is the
case in Coq \cite{coq02}, but this is not always possible. For
instance, the definition by induction of the comparison function $\le$
on natural numbers requires the use of two recursors:

\begin{center}
$\la x.rec~(\la y.true)~(\la nry.rec~false~(\la n'r'.rn')~y)~x$
\end{center}

\noindent
while the definition by rewriting is simply:

\begin{rewc}
0 \le y & true\\
s~x \le 0 & false\\
s~x \le s~y & x \le y\\
\end{rewc}


\noindent{\bf More efficient computations.} From a computational point
of view, definitions by rewriting can be more efficient, although the
process of selecting an applicable rule may have a higher cost
\cite{augustsson85fcpa}. For example, since $+$ is defined by
induction on its first argument, the computation of $n+0$ requires
$n+1$ reduction steps. By adding the rule $x+0\a x$, this takes only
one step.\\


\noindent{\bf Quotient types.} Rewriting allows us to formalize some
quotient types in a simple way, without requiring any additional
extension \cite{barthe95csl,courtieu01csl}, by simply considering
rewrite rules on constructors, which is forbidden in CIC since
constructors must be free in this system. For instance, integers can
be formalized by taking $0$ for zero, $p$ for predecessor and $s$ for
successor, together with the rules:

\begin{rewc}
s~(p~x) & x\\
p~(s~x) & x\\
\end{rewc}

This technique applies to any type whose constructors satisfy a set of
equations that can be turned into a confluent and strongly normalizing
rewrite system \cite{jouannaud86lics}.\\


\noindent{\bf More automation.} We previously saw that, in CIC, if $P$
is a predicate on natural numbers, then $P(2+2)$ is $\bi$-equivalent
to $P(4)$ and, hence, that a proof of $P(2+2)$ is also a proof of
$P(4)$. This means that proving $P(4)$ from $P(2+2)$ does not require
any argument: this is automatically done by the system. But, because
functions must be defined by induction, this does not work anymore for
computations on open terms: since $+$ is defined by induction on its
first argument, $P(x+2)$ is not $\bi$-equivalent to
$P(s(s(x)))$. Proving $P(s(s(x)))$ from $P(x+2)$ requires a user
interaction for proving that $x+2$ is equal to $s(s(x))$, which
requires induction.

We may even go further and turn some lemmas into simplification
rules. Let us for instance consider the multiplication on natural
numbers:

\begin{rewc}
0\times y & 0\\
(s~x)\times y & y+(x\times y)\\
\end{rewc}

Then, the distributivity of the addition over the multiplication can
be turned into the rewrite rule:

\begin{rewc}
(x+y)\times z & (x\times z)+(y\times z)
\end{rewc}

\noindent
hence allowing the system to prove more equalities and more lemmas
automatically by simply checking the $\br$-equivalence with already
proved statements. In the case of an equality $u=v$, it suffices to
check whether it is $\br$-equivalent to the instance $u=u$ of the
identity axiom, which is the same as checking whether $u$ and $v$ have
the same $\br$-normal form.\\


\noindent{\bf Smaller proofs.} Another important consequence of
considering a richer equivalence relation on types is that it reduces
the size of proofs, which is currently an important limitation in
proof assistants like Coq. For instance, while the proof of
$P(s(s(x)))$ requires the application of some substitution lemma in
CIC, it is equal to the proof of $P(x+2)$ when rewriting is
allowed. The benefit becomes very important with equality proofs,
since they require the use of many lemmas in CIC (substitution,
associativity, commutativity, etc.), while they reduce to reflexivity
with rewriting (if one considers rewriting modulo associativity and
commutativity \cite{peterson81jacm}).\\


\noindent{\bf More typable terms.} The fact that some terms are not
$\bi$-equivalent as one would expect has another unfortunate
consequence: some apparently well-formed propositions are rejected by
the system. Take for instance the type $list:(n:nat)\st$ of lists of
length $n$ with the constructors $nil:list0$ and $cons:nat\A
(n:nat)listn\A list(sn)$. Let $app:(n:nat)listn\A (n':nat)listn'\A
list(n+n')$ be the concatenation function on $list$. If, as usual,
$app$ is defined by induction on its first argument then,
surprisingly, the following propositions are not typable in CIC:

\begin{rewc}[~=~]
app~n~\ell~0~\ell' & \ell\\
app~(n+n')~(app~n~\ell~n'~\ell')~n''~\ell'' &
app~n~\ell~(n'+n'')~(app~n'~\ell'~n''~\ell'')\\
\end{rewc}

In the first equation, the left hand-side is of type $list(n+0)$ and
the right hand-side is of type $listn$. Although one can prove that
$n+0=n$ holds for any $n$ in $nat$, the equality is not well-typed
since $n+0$ is not $\bi$-convertible to $n$ (only terms of equivalent
types can be equal).

In the second equation, the left hand-side is of type
$list((n+n')+n'')$ and the right hand-side is of type
$list(n+(n'+n''))$. Again, although one can prove that $(n+n')+n''=
n+(n'+n'')$ holds for any $n$, $n'$ and $n''$ in $nat$, the two terms
are not $\bi$-convertible. Therefore, the proposition is not
well-formed.

On the other hand, by adding the rules $x+0\a x$ and $(x+y)+z\a
x+(y+z)$, the previous propositions become well-typed as expected.\\


\noindent{\bf Integration of decision procedures.} One can also define
predicates by rewrite rules or having simplification rules on
propositions, hence generalizing the definitions by {\em strong
elimination} in CIC. For example, one can consider the set of rules of
Figure \ref{fig-hsiang} \cite{hsiang82thesis} where $\oplus$
(exclusive ``or'') and $\et$ are commutative and associative symbols,
$\bot$ represents the proposition always false and $\top$ the
proposition always true.

\begin{figure}[ht]
\caption{Decision procedure for classical propositional
tautologies\label{fig-hsiang}}
\vsp
\begin{rewc}
P \oplus \bot & P\\
P \oplus P & \bot\\[2mm]

P \et \top & P\\
P \et \bot & \bot\\
P \et P & P\\
P \et (Q \oplus R) & (P \et Q) \oplus (P \et R)\\
\end{rewc}
\end{figure}

Hsiang \cite{hsiang82thesis} showed that this system is confluent and
strongly normalizing, and that a proposition $P$ is a tautology ({\em
i.e.} is always true) iff $P$ reduces to $\top$. So, assuming
type-checking in CC extended with this rewrite system remains
decidable, then, to know whether a proposition $P$ is a tautology, it
is sufficient to submit an arbitrary proof of $\top$ to the
verification program. We would not only gain in automation but also in
the size of proofs (any tautology would have a proof of constant
size).

We can also imagine simplification rules on equalities like the ones
of Figure \ref{fig-int-ring} where $+$ and $\times$ are associative
and commutative, and $=$ commutative.

\begin{figure}[ht]
\caption{Simplification rules on equality\label{fig-int-ring}}
\vsp
\begin{rewc}
x = x & \top\\
s~x = s~y & x=y\\
s~x = 0 & \bot\\
x+y = 0 & x=0 \et y=0\\
x\times y=0 & x=0 \ou y=0\\
\end{rewc}
\end{figure}




\subsection{Problems}

We saw that rewriting has numerous advantages over induction but it is
not clear to which extent rewriting can be added to powerful type
systems like the Calculus of Constructions (CC) without compromising
the decidability of type-checking and the logical
consistency. Furthermore, since rewrite rules are user-defined, it is
not clear also whether $\b\cR$-equivalence/normalization can be made
as efficient as a fixed system with $\b\io$-reduction only
\cite{gregoire02icfp}, although some works on rewriting seem very
promising \cite{eker96wrla,kirchnerhelene01jfp}.

Since we want to consider deductions modulo $\br$-equivalence, we at
least need this equivalence to be decidable. The usual way of proving
the decidability of such an equivalence relation is by proving
confluence and strong normalization of the corresponding reduction
relation. Since these properties are not decidable in general, we will
look for decidable sufficient conditions as general as possible.

As for the logical consistency, we cannot deduce it from normalization
anymore as it is the case in CC \cite{barendregt92book}, since adding
function symbols and rewrite rules is like adding hypothesis and
equality/equivalence axioms. Therefore, for logical consistency also,
we will look for sufficient conditions as general as possible.

In the following sub-section, we present a short history of the
different results obtained so far on the combination of $\b$-reduction
and rewriting. Then, we will present our own contributions.




\subsection{Previous works}

The first work on the combination of typed $\la$-calculus and
(first-order) rewriting is due to Breazu-Tannen in 1988
\cite{breazu88lics}. He showed that the combination of simply-typed
$\la$-calculus and first-order rewriting is confluent if rewriting is
confluent. In 1989, Breazu-Tannen and Gallier \cite{breazu89icalp},
and Okada \cite{okada89issac} independently, showed that the strong
normalization also is preserved. These results were extended by
Dougherty \cite{dougherty91rta} to any ``stable'' set of pure
$\la$-terms. The combination of first-order rewriting and Pure Type
Systems (PTS) \cite{geuvers91jfp,barendregt92book} was also studied by
several authors
\cite{barbanera90ctrs,barthe96csl,barthe97alp,barthe98icalp}.

In 1991, Jouannaud and Okada \cite{jouannaud91lics} extended the
result of Breazu-Tannen and Gallier to the higher-order rewrite
systems satisfying the {\em General Schema}, an extension of primitive
recursion to the simply-typed $\la$-calculus. With higher-order
rewriting, strong normalization becomes more difficult to prove since
there is a strong interaction between rewriting and $\b$-reduction,
which is not the case with first-order rewriting.

In 1993, Barbanera, Fern\'andez and Geuvers
\cite{barbanera94lics,fernandez93thesis} extended the proof of
Jouannaud and Okada to the Calculus of Constructions (CC) with
object-level rewriting and simply-typed function symbols. The methods
used so far for non-dependent type systems
\cite{breazu89icalp,dougherty91rta} cannot be applied to dependent
type systems like CC since, in this case, rewriting is included in the
type conversion rule and, thus, allows more terms to be typable. This
was extended to PTS's in \cite{barthe95hoa}.

Other methods for proving strong normalization appeared. In 1993, Van
de Pol \cite{vandepol93hoa,vandepol95tlca,vandepol96thesis} extended
to the simply-typed $\la$-calculus the use of monotonic
interpretations. In 1999, Jouannaud and Rubio \cite{jouannaud99lics}
extended the Recursive Path Ordering (RPO) to the simply-typed
$\la$-calculus.

In all these works, even the ones on CC, function symbols are always
simply typed. It was Coquand \cite{coquand92types} in 1992 who
initiated the study of rewriting with dependent and polymorphic
symbols. He studied the completeness of definitions with dependent
types. He proposed a schema more general than the schema of Jouannaud
and Okada since it allows inductive definitions on strictly-positive
types, but it does not necessarily imply strong normalization. In
1996, Gim\'enez \cite{gimenez96thesis,gimenez98icalp} defined a
restriction of this schema for which he proved strong
normalization. In 1999, Jouannaud, Okada and the author
\cite{blanqui02tcs,blanqui99rta} extended the General Schema in order
to deal with strictly-positive types while still keeping simply-typed
symbols. Finally, in 2000, Walukiewicz
\cite{walukiewicz00lfm,walukiewicz02jfp} extended Jouannaud and
Rubio's HORPO to CC with dependent and polymorphic symbols.\\

All these works share a strong restriction: rewriting is restricted to
the object level.

In 1998, Dowek, Hardin and Kirchner \cite{dowek98trtpm} proposed a new
approach to deduction for first-order logic: Natural Deduction Modulo
(NDM) a congruence $\equiv$ on propositions representing the
intermediate computations between two deduction steps. This deduction
system consists in replacing the usual rules of Natural Deduction by
equivalent rules modulo $\equiv$. For instance, the elimination rule
for $\A$ ({\em modus ponens}) becomes:

\begin{center}
$\cfrac{\G\th R \quad \G\th P}{\G\th Q} \quad (R \equiv (P\A Q))$
\end{center}

They proved that the simple theory of types \cite{dowek01mscs} and
skolemized set theory can be seen as first-order theories modulo
congruences using {\em explicit substitutions} \cite{abadi91jfp}. In
\cite{dowek98types,dowek00note}, Dowek and Werner gave several
conditions ensuring strong normalization of cut elimination in NDM.




\subsection{Contributions}
\label{subsec-contrib}

Our main contribution is to establish general conditions ensuring the
strong normalization of the Calculus of Constructions (CC) extended
with predicate-level rewriting \cite{blanqui01lics}. In
\cite{blanqui01thesis}, we show that these conditions are satisfied by
most of the Calculus of Inductive Constructions (CIC) and by Natural
Deduction Modulo (NDM) a large class of equational theories.

Our work can be seen as an extension of both NDM and CC, where the
congruence not only includes first-order rewriting but also
higher-order rewriting since, in CC, functions and predicates can be
applied to functions and predicates.

It can therefore serve as a basis for a powerful extension of proof
assistants like Coq \cite{coq02} or LEGO \cite{lego92} which allow
definitions by induction only. For its implementation, it may be
convenient to use specialized rewriting-based applications like CiME
\cite{cime}, ELAN \cite{elan00} or Maude \cite{maude99}. Furthermore,
for program extraction \cite{paulin89popl}, one can imagine using
rewriting-based languages and hence get more efficient extracted
programs.\\

Considering predicate-level rewriting is not completely new. A
particular case is the ``strong elimination'' of CIC, that is, the
ability of defining predicates by induction on some inductively
defined data type. The main novelty here is to consider arbitrary
user-defined predicate-level rewrite rules.

Therefore, for proving the strong normalization property, we cannot
completely follow the methods of Werner \cite{werner94thesis} and
Altenkirch \cite{altenkirch93thesis} since they use in an essential
way the fact that function definitions are made by induction. And the
methods used in case of non-dependent first-order rewriting
\cite{breazu89icalp,barbanera90ctrs,dougherty91rta} cannot be applied
because higher-order rewriting has a strong interaction with
$\b$-reduction and because, in dependent type systems, rewriting
allows more terms to be typable. Our method is based on the notion of
reducibility candidates of Tait and Girard \cite{girard88book} and
extend Geuvers' method \cite{geuvers94types} for dealing with
rewriting.\\

Let us mention two other important contributions.

For allowing some quotient types (rules on constructors) and matching
on function symbols, which is not possible in CIC, we use a notion of
constructor more general than the usual one (see Section
\ref{subsec-ind-typ}).

For ensuring the subject reduction property, that is, the preservation
of typing under reduction, we introduce conditions more general than
the ones used so far. In particular, these conditions allow us to get
rid of non-linearities due to typing, which makes rewriting more
efficient and confluence easier to prove (see Section \ref{sec-sr}).




\section{The Calculus of Algebraic Constructions}
\label{sec-cac}

The Calculus of Algebraic Constructions (CAC) is an extension of the
Calculus of Constructions (CC) \cite{coquand88ic} with function and
predicate symbols defined by rewrite rules.


\subsection{Terms}

\renewcommand{\at}{\alpha}
\newcommand{\FB}{\cF^\B}
\newcommand{\Fs}{\cF^\st}
\newcommand{\XB}{\cX^\B}
\newcommand{\Xs}{\cX^\st}

CC is a particular Pure Type System (PTS) \cite{barendregt92book}
defined from a set $\cS=\{\st,\B\}$ of {\em sorts}. The sort $\st$ is
intended to be the type of data types and propositions, while the sort
$\B$ is intended to be the type of predicate types (also called {\em
kinds}). For instance, the type $nat$ of natural numbers is of type
$\st$, $\st$ is of type $\B$, the predicate $\le$ on natural numbers
is of type $nat\A nat\A\st$, and $nat\A nat\A\st$ is of type $\B$.

The terms of CC are usually defined by the following grammar rule:

\begin{center}
$t ::= s ~|~ x ~|~ [x:t]t ~|~ (x:t)t ~|~ tt$
\end{center}

\noindent
where $s$ is a sort, $x$ a variable, $[x:t]t$ an abstraction, $(x:t)t$
a (dependent) product, and $tt$ an application. We assume that the set
$\cX$ of variables is an infinite denumerable set disjoint from $\cS$.

We simply extend CC by considering a denumerable set $\cF$ of {\em
symbols}, disjoint from $\cS$ and $\cX$, and by adding the following
new construction:

\begin{center}
$t ::= \ldots ~|~ f\in\cF$
\end{center}

\noindent
We denote by $\cT(\cF,\cX)$ the set of terms built from $\cF$ and
$\cX$. Note that, in contrast with \cite{blanqui01thesis}, function
symbols are curried. No notion of arity is required.


\subsection{Notations}

\newcommand{\FVB}{\FV^\B}

\newcommand{\xup}{\{x\to u'\}}
\newcommand{\xv}{\{x\to v\}}
\newcommand{\yv}{\{y\to v\}}
\newcommand{\XU}{\{X\to U\}}
\newcommand{\XS}{\{X\to S\}}
\newcommand{\xS}{\{x\to S\}}

\newcommand{\vyu}{\{\vy\to\vu\}}
\newcommand{\vyt}{\{\vy\to\vt\}}
\newcommand{\vxt}{\{\vx\to\vt\}}
\newcommand{\vxtp}{\{\vx\to\vt'\}}
\newcommand{\vxl}{\{\vx\to\vl\}}
\newcommand{\vxv}{\{\vx\to\vv\}}
\newcommand{\vxS}{\{\vx\to\vS\}}
\newcommand{\vxu}{\{\vx\to\vu\}}
\newcommand{\vxup}{\{\vx\to\vu'\}}
\newcommand{\vzv}{\{\vz\to\vv\}}


\noindent{\bf Free and bound variables.} A variable $x$ in the scope
of an abstraction $[x:T]$ or a product $(x:T)$ is {\em bound}. As
usual, it may be replaced by any other variable. This is {\em
$\alpha$-equivalence}. A variable which is not bound is {\em free}. We
denote by $\FV(t)$ the set of free variables of a term $t$. A term
without free variable is {\em closed}. We often denote by $U\A V$ a
product $(x:U)V$ with $x\notin\FV(V)$ (non-dependent product). See
\cite{barendregt92book} for more details on these notions.\\


\noindent{\bf Vectors.} We often use vectors ($\vt,\vu,\ldots$) for
sequences of terms (or anything else). The size of a vector $\vt$ is
denoted by $|\vt|$. For instance, $[\vx:\vT] u$ denotes the term
$[x_1:T_1] \ldots [x_n:T_n]u$ where $n=|\vx|$.\\


\noindent{\bf Positions.} To designate a subterm of a term, we use a
system of {\em positions} {\em \`a la} Dewey (words over the alphabet
of positive integers). Formally, the set $\pos(t)$ of the positions in
a term $t$ is inductively defined as follows:

\begin{lst}{--}
\item $\pos(f)= \pos(s)= \pos(x)= \{\vep\}$,
\item $\pos((x:t)u)= \pos([x:t]u)= \pos(tu)= 1.\pos(t)\cup 2.\pos(u)$,
\end{lst}

\noindent
where $\vep$ denotes the empty word and '.' the concatenation. We
denote by $t|_p$ the subterm of $t$ at the position $p$, and by
$t[u]_p$ the term obtained by replacing $t|_p$ by $u$ in $t$. The
relation ``is a subterm of'' is denoted by $\tle$, and its strict part
by $\tlt$.

We denote by $\pos(f,t)$ the set of positions $p$ in $t$ such that
$t|_p=f$, and by $\pos(x,t)$ the set of positions $p$ in $t$ such that
$t|_p$ is a free occurrence of $x$ in $t$.\\


\noindent{\bf Substitutions.} A {\em substitution} $\t$ is an
application from $\cX$ to $\cT$ whose {\em domain} $\dom(\t)=
\{x\in\cX ~|~x\t\neq x\}$ is finite. Its set of free variables is $\FV(\t)=
\bigcup \{\FV(x\t)~|~ x\in\dom(\t)\}$. Applying a substitution $\t$ to
a term $t$ consists of replacing every variable $x$ free in $t$ by its
image $x\t$ (to avoid variable captures, bound variables must be
distinct from free variables). The result is denoted by $t\t$. We
denote by $\vxt$ the substitution which associates $t_i$ to $x_i$, and
by $\t\cup\xt$ the substitution which associates $t$ to $x$ and $y\t$
to $y\neq x$.\\


\noindent{\bf Relations.} Let $\a$ be a relation on terms. We denote
by:

\begin{lst}{--}
\item $\a\!\!(t)$ the set of terms $t'$ such that $t\a t'$,
\item $\al\,$ the inverse of $\a$,
\item $\a^+$ the smallest transitive relation containing $\a$,
\item $\a^*$ the smallest reflexive and transitive relation
containing $\a$,
\item $\aa^*$ the smallest reflexive, transitive and symmetric
relation containing $\a$,
\item $\ad\,$ the relation $\a^*\als$ ($t\ad u$ if there exists
$v$ such that $t\a^* v$ and $u \a^* v$).
\end{lst}

If $t\a t'$ then we say that $t$ {\em rewrites} to $t'$. If $t\a^* t'$
then we say that $t$ {\em reduces} to $t'$. A relation $\a$ is {\em
stable by context} if $u\a u'$ implies $t[u]_p \a t[u']_p$ for all
term $t$ and position $p\in\pos(t)$. The relation $\a$ is {\em stable
by substitution} if $t\a t'$ implies $t\t\a t'\t$ for all substitution
$\t$.

The {\em $\b$-reduction} (resp. {\em $\eta$-reduction}) relation is
the smallest relation stable by context and substitution containing
$[x:U]v~u\ab v\xu$ (resp. $[x:U]tx\ae t$ if $x\notin\FV(t)$).  A term
of the form $[x:U]v~u$ (resp. $[x:U]tx$ with $x\notin\FV(t)$) is a
{\em $\b$-redex} (resp. {\em $\eta$-redex}).

A relation $\a$ is {\em weakly normalizing} if, for all term $t$,
there exists an irreducible term $t'$ to which $t$ reduces. We say
that $t'$ is a {\em normal form} of $t$. A relation $\a$ is {\em
strongly normalizing} (well-founded, n\oe{}therian) if, for all term
$t$, any reduction sequence issued from $t$ is finite.

The relation $\a$ is {\em locally confluent} if, whenever a term $t$
rewrites to two distinct terms $u$ and $v$, then $u \ad v$. The
relation $\a$ is {\em confluent} if, whenever a term $t$ reduces to
two distinct terms $u$ and $v$, then $u \ad v$.

If $\a$ is locally confluent and strongly normalizing then $\a$ is
confluent (Newman's lemma). If $\a$ is confluent and weakly
normalizing then every term $t$ has a unique normal form denoted by
$t\ad$.\\


\noindent{\bf Orderings.} A {\em precedence} is a quasi-ordering on
$\cF$ whose strict part is well-founded. Let $>_1,\ldots,>_n$ be
orderings on the sets $E_1,\ldots,E_n$ respectively. We denote by
$(>_1,\ldots,>_n)\lex$ the {\em lexicographic} ordering on
$E_1\times\ldots\times E_n$. Now, let $>$ be an ordering on a set
$E$. We denote by $>\mul$ the ordering on finite multisets on $E$. An
important property of these extensions is that they preserve
well-foundedness. See \cite{baader98book} for more details on these
notions.


\subsection{Rewriting}

In first-order frameworks, that is, in a first-order term algebra, a
rewrite rule is generally defined as a pair $l\a r$ of terms such that
$l$ is not a variable and the variables occurring in $r$ also occur in
$l$ (otherwise, rewriting does not terminate). Then, one says that a
term $t$ rewrites to a term $t'$ at position $p$, written $t\a^p t'$,
if there exists a substitution $\s$ such that $t|_p=l\s$ and
$t'=t[r\s]_p$. See \cite{dershowitz90book} for more details on
(first-order) rewriting.

Here, we consider a very similar rewriting mechanism by restricting
left-hand sides of rules to be algebraic. On the other hand,
right-hand sides can be arbitrary. This is a particular case of {\em
Combinatory Reduction System} (CRS) \cite{klop93tcs} for which it is
not necessary to use {\em higher-order pattern-matching}. However, we
proved in \cite{blanqui00rta} that, in case of simply-typed
$\la$-calculus, our termination criteria can be adapted to rewriting
with higher-order pattern-matching.

\newcommand{\CF}{\cC\cF}
\newcommand{\DF}{{\cD\cF}}
\newcommand{\CFB}{\cC\FB}
\newcommand{\DFB}{{\cD\FB}}

\begin{dfn}[Rewriting]
  Terms only built from variables and applications of the form $f\vt$
  with $f\in\cF$ are said {\em algebraic}. A {\em rewrite rule} is a
  pair of terms $l\a r$ such that $l$ is algebraic, distinct from a
  variable and $\FV(r)\sle\FV(l)$. A rule $l\a r$ is {\em left-linear}
  if no variable occurs more than once in $l$. A rule $l\a r$ is {\em
  non-duplicating} if no variable has more occurences in $r$ than in
  $l$. A rule $f\vl\a r$ is {\em compatible with a precedence $\ge$}
  if, for all symbol $g$ occuring in $r$, $f\ge g$.

  Let $\cR$ be a denumerable set of rewrite rules. The {\em
  $\cR$-reduction relation} $\ar$ is the smallest relation containing
  $\cR$ and stable by substitution and context. A term of the form
  $l\s$ with $l\a r\in \cR$ is an {\em $\cR$-redex}. We assume that
  $\ar$ is finitely branching.
  
  Given a set $\cG\sle\cF$, we denote by $\cR_\cG$ the set of rules
  that {\em define} a symbol in $\cG$, that is, whose left-hand side
  is headed by a symbol in $\cG$. A symbol $f$ is {\em constant} if
  $\cR_{\{f\}} = \vide$, otherwise it is (partially) {\em defined}. We
  denote by $\CF$ the set of constant symbols and by $\DF$ the set of
  defined symbols.
\end{dfn}


\subsection{Typing}

We now define the set of {\em well-typed} terms. An {\em environment}
$\G$ is a list of pairs $x:T$ made of a variable $x$ and a term
$T$. We denote by $\vide$ the empty environment and by $\cE(\cF,\cX)$
the set of environments built from $\cF$ and $\cX$. The {\em domain}
of an environment $\G$, $\dom(\G)$, is the set of variables $x$ such
that a pair $x:T$ belongs to $\G$. If $x\in\dom(\G)$ then we denote by
$x\G$ the first term $T$ such that $x:T$ belongs to $\G$. The set of
{\em free variables} in an environment $\G$ is $\FV(\G)= \bigcup
\{\FV(x\G)~|~ x\in\dom(\G)\}$. Given two environments $\G$ and $\G'$,
$\G$ is {\em included} in $\G'$, written $\G\sle\G'$, if all the
elements of $\G$ occur in $\G'$ in the same order.


\newcommand{\tf}{{\tau_f}}
\newcommand{\tg}{{\tau_g}}
\newcommand{\tF}{{\tau_F}}
\newcommand{\tG}{{\tau_G}}
\newcommand{\tc}{{\tau_c}}
\newcommand{\td}{{\tau_d}}
\newcommand{\tC}{{\tau_C}}
\newcommand{\tD}{{\tau_D}}

\newcommand{\ti}{{\tau^i}}

\newcommand{\cv}[1][]{~\cC_{#1}^*~}
\newcommand{\cvG}{\cv[\G]}
\newcommand{\C}{\mb{C}}
\newcommand{\CV}[1][]{~\C_{#1}^*~}
\newcommand{\CVG}{\CV[\G]}
\newcommand{\CVGp}{\CV[\G']}
\newcommand{\CVD}{\CV[\D]}


\begin{dfn}[Typing]
\label{def-typing}

We assume that every variable $x$ is equipped with a sort $s_x$, that
the set $\cX^s$ of variables of sort $s$ is infinite, and that
$\alpha$-equivalence preserves sorts. Let $\FV^s(t)=\FV(t)\cap\cX^s$
and $\dom^s(\G)=\dom(\G)\cap\cX^s$. We also assume that every symbol
$f$ is equipped with a sort $s_f$ and a closed type $\tau_f=
(\vx:\vT)U$ such that, for all rule $f\vl\a r$, $|\vl|\le|\vx|$. We
often write $f:T$ for saying that $\tf=T$.

The typing relation of a CAC is the smallest ternary relation
$\th\,\sle \cE\times \cT\times \cT$ defined by the inference rules of
Figure \ref{fig-th} where $s,s'\in\cS$. A term $t$ is {\em typable} if
there exists an environment $\G$ and a term $T$ such that $\G\th t:T$
($T$ is a {\em type} of $t$ in $\G$). In the following, we always
assume that $\th\tf:s_f$ for all $f\in\cF$.

An environment is {\em valid} if a term is typable in it. A
substitution $\t$ is {\em well-typed from $\G$ to $\D$},
$\t:\G\leadsto\D$, if, for all $x\in\dom(\G)$, $\D\th x\t:x\G\t$. We
denote by $T ~\cC_\G~ T'$ the fact that $T\ad T'$ and $\G\th T':s'$,
and by $T ~\C_\G~ T'$ the fact that $T ~\cC_\G~ T'$ and $\G\th T:s$.
\end{dfn}


\begin{figure}
\centering
\caption{Typing rules\label{fig-th}}
\begin{tabular}{rcc}
(ax) & $\cfrac{}{\th\st:\B}$\\

\\ (symb) & $\cfrac{\th\tf:s_f}{\th f:\tf}$\\

\\ (var) & $\cfrac{\G\th T:s_x}{\G,x:T\th x:T}$
& $(x\notin\dom(\G))$\\

\\ (weak) & $\cfrac{\G\th t:T \quad \G\th U:s_x}{\G,x:U\th t:T}$
& $(x\notin\dom(\G))$\\

\\ (prod) & $\cfrac{\G\th U:s \quad \G,x:U \th V:s'}
{\G\th (x:U)V:s'}$\\

\\ (abs) & $\cfrac{\G,x:U \th v:V \quad \G\th (x:U)V:s}
{\G\th [x:U]v:(x:U)V}$\\

\\ (app) & $\cfrac{\G\th t:(x:U)V \quad \G\th u:U}
{\G\th tu:V\xu}$\\

\\ (conv) & $\cfrac{\G\th t:T \quad \G\th T':s'}{\G\th t:T'}$
& $(T \ad_{\b\cR} T')$\\
\end{tabular}
\end{figure}


Compared with CC, we have a new rule, (symb), for typing symbols and,
in the type conversion rule (conv), we have $\ad_{\b\cR}$ (that we
simply denote by $\ad$ in the rest of the paper) instead of the
$\b$-conversion $\aa^*_\b=\ad_\b$ (since $\b$ is confluent).

Well-typed substitutions enjoy the following important substitution
property: if $\G\th t:T$ and $\t:\G\leadsto\D$ then $\D\th t\t:T\t$.

The relations $\cC_\G$ (not symmetric) and $\C_\G$ (symmetric) are
useful when inverting typing judgements. For instance, a derivation of
$\G\th uv:W'$ necessarily terminates by an application of the (app)
rule, possibly followed by applications of the rules (weak) and
(conv). Therefore, there exists $V$ and $W$ such that $\G\th
u:(x:V)W$, $\G\th v:V$ and $W\xv \cvG W'$. Since, in the (conv) rule,
$T$ is not required to be typable by some sort $s$ (as it is the case
for $T'$), it is not {\em a priori} the case that $W\xv$ is typable
and therefore that, in fact, $W\xv \CVG W'$.\\


Many of the well-known basic properties of Pure Type Systems (PTS's)
\cite{barendregt92book} also hold for CAC's. In
\cite{blanqui01thesis}, we study these properties in an abstract way
by considering a PTS equipped with an unspecified type conversion rule
(instead of $\ad_\b$ or $\ad_{\b\cR}$ for instance), hence factorizing
several previous proofs for different PTS extensions. The properties
we use in this paper are:

\begin{lst}{}
\item [\bf (type correctness)] If $\G\th t:T$ then either $T=\B$ or
$\G\th T:s$.
\item [\bf (conversion correctness)] If $\G\th T:s$ and $T \CVG T'$
then $\G\th T':s$.
\item [\bf (convertibility of types)] If $\G\th t:T$ and $\G\th t:T'$
then $T \CVG T'$.
\end{lst}

\noindent
Only convertibility of types requires confluence (conversion
correctness is proved in Section \ref{subsec-sr-beta} without using
confluence).\\

Among well-typed terms, we distinguish:

\begin{lst}{--}
\item The set $\mb{K}$ of {\em predicate types} or {\em kinds} made of
the terms $K$ such that $\G\th K:\B$. It is easy to check that every
predicate type is of the form $(\vx:\vT)\st$.
\item The set $\mb{P}$ of {\em predicates} made of the terms $T$ such
that $\G\th T:K$ and $\G\th K:\B$.
\item The set $\mb{O}$ of {\em objects} made of the terms $t$ such
that $\G\th t:T$ and $\G\th T:\st$.
\end{lst}




\section{Subject reduction}
\label{sec-sr}

Before studying the strong normalization or the logical consistency of
our system, we must make sure that the reduction relation $\abr$ is
indeed correct w.r.t. typing, that is, if $\G\th t:T$ and $t\abr t'$
then $\G\th t':T$. This property is usually called {\em subject
reduction}. Once it holds, it can be easily extended to types,
environments and substitutions:

\begin{lst}{--}
\item If $\G\th t:T$ and $T\a T'$ then $\G\th t:T'$.
\item If $\G\th t:T$ and $\G\a\G'$ then $\G'\th t:T$.
\item If $\t:\G\leadsto\D$ and $\t\a\t'$ then $\t':\G\leadsto\D$.
\end{lst}

In presence of dependent types and rewriting, the subject reduction
for $\b$ appears to be a difficult problem. Indeed, in the case of a
head-reduction $[x:U']v~u\ab v\xu$ with $\G\th [x:U']v:(x:U)V$ and
$\G\th u:U$, we must prove that $\G\th v\xu:V\xu$. By inversion, we
have $\G,x:U'\th v:V'$ with $(x:U')V' \CVG (x:U)V$. We can conclude
that $\G\th v\xu:V\xu$ only if:

\begin{center}
$(x:U')V' \CVG (x:U)V$ implies $U' \CVG U$ and $V' \CV[\G,x:U] V$,
\end{center}

\noindent
a property that we call {\em product compatibility}.

This is immediate as soon as $\abr$ is confluent. Unfortunately, there
are very few results on the confluence of higher-order rewriting and
$\b$-reduction together (see the discussion after Definition
\ref{def-cond}). Fortunately, confluence is not the only way to prove
the product compatibility. In \cite{geuvers93thesis}, Geuvers proves
the product compatibility for the Calculus of Constructions (CC) with
$\aa^*_{\b\eta}$ as type conversion relation, although $\abe$ is not
confluent on untyped terms: $[x:T]x ~_\b\al{} [x:T]([y:U]y~x) \ae
[y:U]y =_\alpha [x:U]x$ \cite{nederpelt73thesis}. And, in
\cite{barbanera97jfp}, Barbanera, Geuvers and Fern\'andez prove the
product compatibility for CC with $\ad_\b\cup\ad_\cR$ as type
conversion relation, where $\cR$ is a set of simply-typed object-level
rewrite rules.

In Section \ref{subsec-sr-beta}, we prove the product compatibility,
hence the subject reduction of $\b$, for a large class of rewrite
systems, including predicate-level rewriting, without using
confluence, by generalizing the proof of Barbanera, Fern\'andez and
Geuvers \cite{barbanera97jfp}. Before that, we study the subject
reduction for rewriting.




\subsection{Subject reduction for rewriting}

In first-order sorted algebras, for rewriting to preserve sorts, it
suffices that both sides of a rule have the same sort. Carried over to
type systems, this condition gives: there exists an environment $\G$
and a type $T$ such that $\G\th l:T$ and $\G\th r:T$. This condition
is the one which has been taken in all previous work combining typed
$\la$-calculus and rewriting. However, it has an important
drawback. With polymorphic or dependent types, it leads to strongly
non left-linear rules, which has two important consequences. First,
rewriting is strongly slowed down because of the necessary equality
tests. Second, it is more difficult to prove confluence.

Let us take the example of the concatenation of two polymorphic lists
(type $list:\st\A\st$ with the constructors $nil:(A:\st)listA$ and
$cons:(A:\st)A\A listA\A listA$):

\begin{rewc}
app~A~(nil~A)~\ell' & \ell'\\
app~A~(cons~A~x~\ell)~\ell' & cons~A~x~(app~A~\ell~\ell')\\
\end{rewc}

This definition satisfies the usual condition by taking $\G= A:\st,
x:A, \ell:listA, \ell':listA$ and $T= listA$. But one may wonder
whether it is really necessary to do an equality test between the
first argument of $app$ and the first argument of $cons$ when one
wants to apply the second rule. Indeed, if
$app~A~(cons~A'~x~\ell)~\ell'$ is well-typed then, by inversion,
$cons~A'~x~\ell$ is of type $listA$ and, by inversion again, $listA'$
is convertible to $listA$. Thus, $A$ is convertible to $A'$.

In fact, what is important is not that the left-hand side of a rule be
typable, but that, if an instance of the left-hand side of a rule is
typable, then the corresponding instance of the right-hand side has
the same type. We express this by requiring that there exists an
environment $\G$ in which the right-hand side is typable, and a
substitution $\r$ which replaces the variables of the left-hand side
not belonging to $\G$ by terms typable in $\G$. Hence, one can
consider the following rules instead:

\begin{rewc}
app~A~(nil~A')~\ell' & \ell'\\
app~A~(cons~A'~x~\ell)~\ell' & cons~A~x~(app~A~\ell~\ell')\\
\end{rewc}

\noindent
by taking $\G= A:\st, x:A, \ell:listA, \ell':listA$ and $\r=\{A'\to
A\}$.


\begin{dfn}[Well-typed rule]
\label{def-wt-rule}

A rule $l\a r$ with $l=f\vl$, $f:(\vx:\vT)U$ and $\g\!=\!\vxl$ is {\em
well-typed} if there exists an environment $\G$ and a substitution
$\r$ such that:\footnote{The conditions {\bf(S1)} $\dom(\r)\cap
\dom(\G)= \vide$ and {\bf(S2)} $\G\th l\r:U\g\r$ given in
\cite{blanqui01lics} are not necessary for proving the subject
reduction property, but they are necessary for proving the strong
normalization property of the higher-order rewrite rules (see
Definition \ref{def-wf-rule}).}

\begin{bfenumi}{S}
\setcounter{enumi}{2}
\item $\G\th r:U\g\r$,
\item $\all \D,\s,T$, if $\D\th l\s:T$ then $\s:\G\leadsto\D$,
\item $\all \D,\s,T$, if $\D\th l\s:T$ then $\s\ad\r\s$.
\end{bfenumi}

\noindent
In the following, we write $(l\a r,\G,\r)\in\cR$ when the previous
conditions are satisfied.
\end{dfn}


An example with dependent types is given by the concatenation of two
lists of fixed length (type $list:nat\A\st$ with the constructors
$nil:list\,0$ and $cons:nat\A(n:nat)$ $list\,n\A list\,(s\,n)$) and
the function $map$ which applies a function $f$ to every element of a
list:

\begin{typc}
app & (n:nat)list\,n\A (n':nat)list\,n'\A list\,(n+n')\\
map & (nat\A nat)\A (n:nat)list\,n\A list\,n\\
\end{typc}

\noindent
where $app$ and $map$ are defined by:

\begin{rewc}
app~0~\ell~n'~\ell' & \ell'\\
app~p~(cons~x~n~\ell)~n'~\ell' & cons~x~(n+n')~(app~n~\ell~n'~\ell')\\
\end{rewc}

\begin{rewc}
map~f~0~\ell & \ell\\
map~f~p~(cons~x~n~\ell) & cons~(f~x)~n~(map~f~n~\ell)\\
map~f~p~(app~n~\ell~n'~\ell') & app~n~(map~f~n~\ell)~n'~(map~f~n'~\ell')\\
\end{rewc}

For the second rule of $app$, we take $\G= x:nat, n:nat, \ell:list\,n,
n':nat, \ell':list\,n'$ and $\r= \{p\to sn\}$. This avoids checking
that $p$ is convertible to $sn$. For the third rule of $map$, we take
$\G= f:nat\A nat, n:nat, \ell:list\,n, n':nat, \ell':list\,n'$ and
$\r= \{p\to n+n'\}$. This avoids checking that $p$ is convertible to
$n+n'$. The reader will find more examples at the end of Section
\ref{sec-conditions}.


\begin{lem}
\label{lem-app}
If $\b\cR$ is product compatible, $f:(\vx:\vT)U$, $\t=\vxt$ and $\G\th
f\vt:T$ then $\t:\G_f\leadsto\G$ and $U\t\CVG T$.
\end{lem}

\begin{prf}
By inversion, there is a sequence of products $(x_i:T_i')U_i$ ($1\le
i\le n=|\vx|$) such that $\G\th ft_1\ldots t_{n-1}:(x_n:T_n')U_n$,
$\G\th t_n:T_n'$, $U_n\t\CVG T$, \ldots, $\G\th f:(x_1:T_1')U_1$,
$\G\th t_1:T_1'$, $U_1\t\CVG (x_2:T_2')U_2$ and $(\vx:\vT)U\CVG
(x_1:T_1')U_1$. Let $V_i={(x_{i+1}:T_{i+1})}\ldots{(x_n:T_n)}U$. By
product compatibility, $T_1\t=T_1\CVG T_1'$ and $V_1\CV[\G,x_1:T_1]
U_1$. Hence, $V_1\t= {(x_2:T_2\t)}V_2\t\CVG U_1\t\CVG
(x_2:T_2')U_2$. Therefore, by induction, $T_2\t\CVG T_2'$, \ldots,
$T_n\t\CVG T_n'$ and $U\t\CVG U_n\t\CVG T$. Hence, by conversion,
$\G\th t_i:T_i\t$, that is, $\t:\G_f\leadsto\G$.
\end{prf}


\begin{thm}[Subject reduction for $\cR$]
\label{lem-cor-ar}

If $\b\cR$ is product compatible and $\cR$ is a set of well-typed
rules then $\cR$ preserves typing.
\end{thm}

\begin{prf}
  As usual, we prove by induction on $\D\th t:T$ that, if $t\ar t'$
  then $\D\th t':T$, and if $\D\ar\D'$ then $\D'\th t:T$. We only
  detail the (app) case. Assume that $\D\th l\s:T$, $(l\a
  r,\G,\r)\in\cR$, $l=f\vl$, $f:(\vx:\vT)U$ and $\g=\vxl$. Let
  $\t=\g\s$. After Lemma \ref{lem-app}, $\t:\G_f\leadsto\D$ and
  $U\t\CVD T$. By {\bf(S4)}, $\s:\G\leadsto\D$. By {\bf(S3)}, $\G\th
  r:U\g\r$. Therefore, by substitution, $\D\th r\s:U\g\r\s$. By
  {\bf(S5)}, $\r\s\ad\s$. Therefore, by conversion, $\D\th r\s:U\t$
  and $\D\th r\s:T$.\cqfd
\end{prf}


How to check the conditions (S3), (S4) and (S5) ? In all their
generality, they are certainly undecidable. On the one hand, we do not
know whether $\th$ and $\ad$ are decidable and, on the other hand, in
(S4) and (S5), we arbitrarily quantify over $\D$, $\s$ and $T$. It is
therefore necessary to put additional restrictions. In the following,
we successively consider the three conditions.


Let us look at (S3). In practice, the symbols and their defining rules
are often added one after another (or by groups but the following
argument can be generalized). Let $(\cF,\cR)$ be a system in which
$\th$ is decidable, $f\notin\cF$ and $\cR_f$ a set of rules defining
$f$ and whose symbols belong to $\cF'=\cF\cup\{f\}$. Then, in
$(\cF',\cR)$, $\th$ is still decidable. One can therefore try to check
(S3) in this system. This does not seem an important restriction: it
would be surprising if the typing of a rule requires the use of the
rule itself !


We now consider (S4).


\begin{dfn}[Canonical and derived types]
\label{def-der-typ}
\label{def-can-typ}

Let $t$ be a term of the form $l\s$ with $l=f\vl$ algebraic,
$f:(\vx:\vT)U$, $n=|\vx|=|\vl|$ and $\g=\vxl$. The term $U\g\s$ will
be called the {\em canonical type} of $t$. Let $p\in\pos(l)$ of the
form $(1^*2)^+$. We inductively define the {\em type of $t|_p$ derived
from $t$}, $\tau(t,p)$, as follows:

\begin{lst}{--}
\item if $p=1^{n-i}2$ then $\tau(t,p)= T_i\g\s$,
\item if $p=1^{n-i}2q$ and $q\neq\vep$ then $\tau(t,p)= \tau(t_i,q)$.
\end{lst}
\end{dfn}

The type of $t|_p$ derived from $t$ only depends on the term above
$t|_p$.


\begin{lem}[S4]
\label{lem-S4}

  If, for all $x\in\dom(\G)$, there is $p\in\pos(x,l)$ such that
  $x\G=\tau(l,p)$, then (S4) is satisfied.
\end{lem}

\begin{prf}
  We prove (S4) by induction on the size of $l$. Assume that $\D\th
  l\s:T$. We must prove that, for all $x\in\dom(\G)$, $\D\th
  x\s:x\G\s$. By assumption, there is $p\in\pos(x,l)$ such that
  $x\G=\tau(l,p)$. Since $l=f\vl$, $p=jq$. Assume that
  $f:(\vx:\vT)U$. Let $\g=\vxl$ and $\t=\g\s$. If $q=\vep$ then
  $x=l_j$ and $x\G=T_j\g$. Now, after Lemma \ref{lem-app},
  $\t:\G_f\leadsto\D$. So, $\D\th x_j\t:T_j\t$, that is, $\D\th
  x\s:x\G\s$. Assume now that $q\neq\vep$. Since $\D\th l_j\s:T_j\t$,
  $l_j$ is of the form $g\vm$ and $x\G=\tau(l_j,q)$, by induction
  hypothesis, $\D\th x\s:x\G\s$.\cqfd
\end{prf}


For (S5), we have no general result. By inversion, (S5) can be seen as
a unification problem modulo $\ad^*$. The confluence of $\a$ (which
implies that $\ad^*=\ad$) can therefore be very useful. Unfortunately,
there are very few results on the confluence of the combination of
higher-order rewriting and $\b$-reduction (see the discussion after
Definition \ref{def-cond}). On the other hand, one can easily prove
that local confluence is preserved.


\begin{thm}[Local confluence]
\label{lem-loc-confl}

If $\cR$ is locally confluent on algebraic terms then $\b\cR$ is
locally confluent on any term.
\end{thm}

\begin{prf}
  Assume that $t\a^p t_1$ and $t\a^q t_2$. We prove by induction on
  $t$ that there exists $t'$ such that $t_1\a^* t'$ and $t_2\a^*
  t'$. There are three cases:

\begin{lst}{\bu}
\item $p ~\sharp~ q$ ($p$ and $q$ have no common prefix). The reductions
at $p$ and $q$ can be done in parallel: $t_1\a^q t_1'$, $t_2\a^p t_2'$
and $t_1'=t_2'$.

\item $p=ip'$ and $q=iq'$. We can conclude by induction hypothesis on
$t|_i$.
  
\item $p=\vep$ or $q=\vep$. By exchanging the roles of $p$ and $q$, we
can assume that $p=\vep$. There are two cases:

\begin{lst}{--}
\item $t=[x:V]u~v$ and $t_1=u\xv$. We distinguish three sub-cases:

\begin{lst}{$\circ$}
\item $q=11q'$ and $V\a^{q'} V'$. Then, $t'= t_1$ works.
\item $q=12q'$ and $u\a^{q'} u'$. Then, $t'= u'\xv$ works.
\item $q=2q'$ and $v\a^{q'} v'$. Then, $t'= u\{x\to v'\}$ works.
\end{lst}

\item $t=l\s$, $l\a r\in\cR$ and $t_1=r\s$. There exists an algebraic
term $u$ of maximal size and a substitution $\t$ such that $t=u\t$ and
$x\t=y\t$ implies $x=y$ ($u$ and $\t$ are unique up to the choice of
variables and $u$ has the same non-linearities than $t$). As the
left-hand sides of rules are algebraic, $u=l\s'$ and $\s=\s'\t$. Now,
we distinguish two sub-cases:

\begin{lst}{$\circ$}
\item $q\in\pos(u)$. As the left-hand sides of rules are algebraic, we
have $u\ar r\s'$ and $u\ar v$. By local confluence of $\ar$ on
algebraic terms, there exists $u'$ such that $r\s' \a^* u'$ and $v
\a^* u'$. Then, $t'=u'\t$ works.

\item $q=q_1q'$ and $u|_{q_1}= x$. Let $q_2$, \ldots, $q_n$ be the
positions of the other occurrences of $x$ in $u$. If one reduces $t_2$
at each position $q_iq'$, one obtains a term of the form $l\s'\t'$
where $\t'$ is the substitution such that $x\t'$ is the reduct of
$x\t$, and $y\t'= y\t$ if $y\neq x$. Then, $t'= r\s'\t'$ works.\cqfd
\end{lst}
\end{lst}
\end{lst}
\end{prf}




\subsection{Subject reduction for $\b$}
\label{subsec-sr-beta}

In this section, we prove the product compatibility, hence the subject
reduction of $\b$, for a large class of rewrite systems, including
predicate-level rewrite rules, without using confluence, by
generalizing the proof of Barbanera, Fern\'andez and Geuvers
\cite{barbanera97jfp}. It is worth noting that no result of this
section assumes the subject reduction property for rewriting. They
only rely on simple syntactic properties of $\b$-reduction and
rewriting with respect to predicates and kinds (Lemma \ref{lem-kind}).

The idea is to $\b$-weak-head normalize all the intermediate terms
between $(x:U')V'$ and $(x:U)V$ so that we obtain a sequence of
conversions between product terms only. We first show that the subject
reduction property can indeed be studied in a system whose conversion
relation is like the one used in \cite{barbanera97jfp}.

\newcommand{\nh}{{\not\!h}}
\newcommand{\ah}{\a_h}
\newcommand{\anh}{\a_{\not h}}
\newcommand{\arnh}{\a_{\cR\not h}}


\begin{lem}
Let $\L$ be a CAC with conversion relation $\ad_{\b\cR}$ and $\L'$ be
the same CAC but with conversion relation $\ad_\b\cup\ad_\cR$. If
$\abr$ has the subject reduction property in $\L'$ then $\L=\L'$ (and
$\abr$ has the subject reduction property in $\L$).
\end{lem}

\begin{prf}
Let $\th$ (resp. $\th'$) be the typing relation of $\L$
(resp. $\L'$). Since $\ad_\b\cup\ad_\cR \,\sle\, \ad_{\b\cR}$, we
clearly have $\th'\,\sle\,\th$. We prove by induction on $\th$ that
$\th\,\sle\,\th'$. The only difficult case is of course (conv). By
induction hypothesis, we have $\G\th' t:T$ and $\G\th'
T':s'$. Furthermore, we have $T\a^*_{r_1}\a^*_{r_2}\ldots
{}_{s_2}^*\al {}_{s_1}^*\al T'$ with $r_k,s_k\in\{\b,\cR\}$. By type
correctness, either $T=\B$ or there is a sort $s$ such that $\G\th'
T:s$. If $T=\B$ then $T'\a^* \B$. But, since $\a$ has the subject
reduction property in $\L'$, we get that $\G\th' \B:s'$, which is not
possible. Therefore, $T$ and $T'$ are typable in $\L'$ and, since $\a$
has the subject reduction property in $\L'$, all the terms between $T$
and $T'$ are also typable in $\L'$. Therefore, we can replace the
conversion in $\L$ by a sequence of conversions in $\L'$.\cqfd
\end{prf}


We now prove a series of useful results about kinds and predicates
which will allow us to prove the subject reduction property on types
for the $\b$-weak-head reduction relation $h$: $t\ah t'$ if
$t=[\vx:\vT]([x:U]vu\vt)$ and $t'=[\vx:\vT](v\xu\vt)$. The
$\b$-internal reduction relation will be denoted by $\nh$. To this
end, we introduce several sets of terms.

\begin{lst}{--}
\item $\cK$: terms of the form $(\vx:\vT)\st$, usually called {\em
kinds}.
\item $\cP$: smallest set of terms, called {\em predicates}, such that
$\cX^\B\cup \cF^\B\sle \cP$ and, if $pt\in\cP$ or $[x:t]p\in\cP$ or
$(x:t)p\in\cP$, then $p\in\cP$.
\item $\cW$: terms having a subterm of the form $[y:W]K$ or $wK$,
called a {\em bad kind}.
\item $\cB$: terms containing $\B$.
\end{lst}

\begin{lem}
\begin{lst}{}
\item [($\alpha$)] No term in $\cB$ is typable.
\item [($\b$)] If $\G\th t:\B$ then $t\in\cK$.
\item [($\g$)] If $t\t\in\cB$ then $t\in\cB$ or $x\t\in\cB$ for some $x$.
\item [($\d$)] If $t\t\in\cK$ then $t\in\cK$ or $x\t\in\cK$ for some $x$.
\end{lst}
\end{lem}

\begin{prf}
\begin{lst}{}
\item [($\alpha$)] $\B$ is not typable and every subterm of a typable
term is typable.

\item [($\b$)] By induction on the size of $t$ (no conversion can take
place since $\B$ is not typable).

\item [($\g$)] Trivial.

\item [($\d$)] If $t\t\in\cK$ and $t\notin\cK$ then $t=(\vx:\vT)x$ with
$x\t\in\cK$.\cqfd
\end{lst}
\end{prf}

\begin{lem}
\label{lem-kind}
If, for every rule $l\a r\in\cR$, $r\notin\cB\cup\cK\cup\cW$, then:

\begin{enumalphai}{}
\item If $t\a t'$ and $t'\in\cB$ then $t\in\cB$.
\item If $\B\cvG T$ then $T=\B$.
\item If $K\in\cK$ and $\G\th K:L$ then $L=\B$.
\item No term in $\cW$ is typable.
\item If $t\a K\in\cK$ then $t\in\cK\cup\cW$.
\item If $t\a t'\in\cW$ then $t\in\cW$.
\item If $\G\th T:s$ and $T\a^* K\in\cK$ then $T\in\cK$ and $s=\B$.
\item If $T \CVG K$ and $\G\th K:\B$ then $\G\th T:\B$ and $T\in\cK$.
\item If $(\vx:\vT)\st \CVG (\vy:\vU)\st$ then $|\vx|=|\vy|$ and,
for all $i$, $T_i \CV[\G_i] U_i\{\vy\to\vx\}$ with $\G_i= \G,
x_1:T_1, \ldots, x_i:T_i$.
\item If $T \CVG T'$ and $\G\th T:\st$ then $\G\th T':\st$.
\item If $\G\th t:T$ and $t\in\cP$ then $T\in\cK$.
\item If $\G\th t:K$ and $\G\th K:\B$ then $t\in\cP$.
\end{enumalphai}
\end{lem}

\begin{prf}
\begin{enumalphai}{}
\item Assume that $t\a^p t'$ and $t'|_q=\B$. If $p~\sharp~q$ then
$t|_q=\B$ and $t\in\cB$. Otherwise, $p\le q$. If $t|_p=[x:U]v~u$ and
$t'|_p=v\xu$ then, by ($\g$), $v\in\cB$ or $u\in\cB$. Thus,
$t\in\cB$. Now, if $t|_p=l\s$, $t'|_p=r\s$ and $l\a r\in\cR$ then, by
($\g$), $r\in\cB$ or $x\s\in\cB$ for some $x$. Since $r\notin\cB$,
$x\s\in\cB$ and $t\in\cB$.

\item Assume that $\B\ad T'\CVG T$. Then, $T'\a^* \B$ and $\G\th
T':s$. By (a), $T'\in\cB$ and $T'$ cannot be typable. Thus, $T=\B$.

\item By induction on the size of $K$. If $K=\st$ then, by inversion,
$\B\cvG L$ and, by (b), $L=\B$. If $K=(x:T)K'$ then, by inversion,
$\G,x:T\th K':s$ and $s\cvG L$. By induction hypothesis, $s=\B$ and,
by (b), $L=\B$.

\item Assume that $\G\th [y:W]K:T$. By inversion, $\G,y:W\th K:L$ and
$\G\th (y:W)L:s$. By (c), $L=\B$ and $(y:W)L$ cannot be
typable. Assume now that $\G\th wK:T$. By inversion, $\G\th w:(x:L)V$,
$\G\th K:L$ and $\G\th (x:L)V:s$. By (c), $L=\B$ and $(x:L)V$ cannot
be typable.

\item Assume that $t\a K\in\cK$ and $t\notin\cK$. We prove that $t\in\cW$
by induction on the size of $t$. The only possible cases are
$t=(x:T)u$, $t=[x:U]v~u$ if $t\ab K$, and $t=l\s$ with $l\a r\in\cR$
if $t\ar K$. If $t=(x:T)u$ then $K=(x:T)L$ and $u\a L$. By induction
hypothesis, $u\in\cW$. If $t=[x:U]v~u$ then $K=v\xu$. By ($\d$), either
$v\in\cK$ or $u\in\cK$. In both cases, $t\in\cW$. Assume now that
$t=l\s$ with $l\a r\in\cR$. Then, $K=r\s$. By ($\d$), either $r\in\cK$ or
$x\s\in\cK$ for some $x$. Since $r\notin\cK$, $x\s\in\cK$ and
$t=l\s\in\cW$ since $x$ is the argument of some symbol ($l$ is
algebraic).

\item Assume that $t\a^p t'\in\cW$, $t'|_q=wK$ and $K\in\cK$ (the case
$t'|_q=[x:w]K$ is dealt with in the same way). There are several cases:

\begin{lst}{--}
\item $q~\sharp~ p$. Then, $t|_q=wK$ and $t\in\cW$.

\item $q<p$.
\begin{lst}{$\circ$}
\item $p=q1m$. Then, $t|_q=w'K$ with $w'\a w$ and $t\in\cW$.
\item $p=q2m$. Then, $t|_q=wu$ with $u\a K\in\cK$. By (e), $u\in\cK\cup\cW$.
Thus, $t\in\cW$.
\end{lst}

\item $q\ge p$. Then, $q=pm$. Assume that $t|_p=l\s$, $t'|_p=r\s$ and
$l\a r\in\cR$ (the case $t\ab t'$ is dealt with in the same way). Let
$\{p_1,\ldots,p_n\}= \{p\in\pos(x,r)~|~ x\in\FV(r)\}$. There are
saveral cases:
\begin{lst}{$\circ$}
\item $m~\sharp~ p_i$ for all $i$, or $m<p_i$ for some $i$. Then,
$r|_m\s=wK$, $r=uv$ and $v\s=K$. By ($\d$), $v\in\cK$ or $x\s\in\cK$ for
some $x$. If $v\in\cK$ then $r\in\cW$, which is not possible. Thus,
$x\s\in\cK$ and $l\s\in\cW$.
\item $m\ge p_i$ for some $i$. Then, there is $x\in\FV(l)$ such that
$x\s\in\cW$. Thus, $t\in\cW$.
\end{lst}
\end{lst}

\item By (e) and (f), if $T\a^* K\in\cK$ then $T\in\cK\cup\cW$. Since
$\G\th T:s$, $T\notin\cW$. Thus, $T\in\cK$ and $s=\B$.

\item By induction on the number of conversions between $T$ and $K$.
Assume that $\G\th T:s$, $T\ad K$ and $\G\th K:\B$. Then, there is
$K'\in\cK$ such that $K\a^* K'$ and $T\a^* K'$. By (g), $T\in\cK$ and
$s=\B$.

\item By (h), all the intermediate well-typed terms between
$K=(\vx:\vT)\st$ and $L=(\vy:\vU)\st$ are kinds and, if $K\ad L$
then, clearly, $|\vx|=|\vy|$ and $T_i\ad U_i\{\vx\to\vy\}$ for all $i$.

\item Immediate consequence of (i).

\item By induction on $\G\th t:T$.

\item By induction on $\G\th t:K$.\cqfd
\end{enumalphai}
\end{prf}


\begin{lem}
Given a rule $l\a r$ with $l=f\vl$, $f:(\vx:\vT)U$ and $\g=\vxl$,
$r\notin\cB\cup\cK\cup\cW$ if there is an environment $\G$ and a
substitution $\r$ such that $\G\th l\r:U\g\r$ and $\G\th r:U\g\r$.
\end{lem}

\begin{prf}
Since $r$ is typable, $r\notin\cB\cup\cW$. We now prove that
$r\notin\cK$. Since $\G\th l\r:U\g\r$, by inversion, we get that
$\g\r:\G_f\leadsto\G$. Since $\th\tf:s_f$, by inversion, we get that
$\G_f\th U:s_f$. So, by substitution, $\G\th U\g\r:s_f$. Now, if
$r\in\cK$ then, by (c), $U\g\r=\B$ but $\B$ is not typable. Therefore,
$r\notin\cK$.
\end{prf}


\newcommand{\bp}{{\b^{P\w}}}
\newcommand{\abp}{\a_\bp}

\begin{thm}[Subject reduction for $h$]
{\bf\cite{barbanera97jfp}} Assume that no right hand-side is in
$\cB\cup\cK\cup\cW$. Then, the restriction $\bp$ of $\b$ to the
redexes $[x:T]U~t\in\cP$ preserves typing. Therefore, $h$ preserves
typing on terms of type $\st$.
\end{thm}

\begin{prf}
The proof is as usual by induction on $\G\th t:T$ and by proving at
the same time that, if $\G\abp\G'$, then $\G'\th t:T$. The only
difficult case is the case of a head-reduction $[x:U']v~u\abp v\xu$
with $\G\th [x:U']v:(x:U)V$ and $\G\th u:U$. We must prove that $\G\th
v\xu:V\xu$. By inversion, we have $\G, x:U'\th v:V'$ with $(x:U')V'
\CVG (x:U)V$. Since $v\in\cP$, by (k), $V'\in\cK$. Therefore, by
Lemma \ref{lem-kind} (h) and (i), $(x:U)V\in\cK$, $U' \CVG U$ and $V'
\CV[\G,x:U] V$. Hence, by environment conversion and type conversion,
$\G,x:U\th v:V$ and, by substitution, $\G\th v\xu:V\xu$.

Now, if $\G\th t:\st$ then, by (l), $t=[x:U]vu\vt\in\cP$ and
$v\in\cP$. So, if $t\ah t'$ then $t\abp t'$ and $\G\th t':\st$.\cqfd
\end{prf}


\begin{lem}[Commutation]
If $t\ah^* u$ and $t\ar^* v$ then there exists $w$ such that $u\ar^*
w$ and $v\ah^* w$.
\end{lem}

\begin{prf}
By induction on the number of $h$-steps, it suffices to prove that, if
$[x:U]v~u\ah v\xu$ and $[x:U]v~u\ar^* t$, then there exists $w$ such
that $v\xu\ar^* w$ and $t\ah w$. Since left hand-sides of rules are
algebraic, $t$ is of the form $[x:U']v'~u'$ with $U\ar^* U'$, $v\ar^*
v'$ and $u\ar^* u'$. So, it suffices to take $w= v'\{x\to u'\}$.\cqfd
\end{prf}


\begin{lem}[Postponement]
Assume that no right hand-side is in $\cB\cup\cK\cup\cW$ and that the
right hand-side of every type-level rule is either a product or a
predicate symbol application. If $\G\th t:\st$ and $t\ar^* u\ah^* v$
then there exists $w$ such that $t\ah^* w\ar^* v$.
\end{lem}

\begin{prf}
By induction on the number of $h$-steps. Assume that $t\ar^* u\ah^*
u'\ah v$. By induction hypothesis, there exists $w'$ such that $t\ah^*
w'\ar^* u'$. By subject reduction on types, $\G\th w':\st$. So, by
(l), $w'$ is either of the form $(x:U)V$, $x\vt$, $f\vt$ with
$f\in\FB$, or $[x:B]ab\vt$. Since $w'\ar^* u'\ah v$, $w'$ cannot be of
the form $(x:U)V$ or $x\vt$. Since right hand-sides of type-level
rules are either a product or a predicate symbol application, $w'$
cannot be of the form $f\vt$. Therefore, $w'=[x:B]ab\vt$,
$u'=[x:B']a'b'\vt'$ with $B,a,b,\vt\ar^* B',a',b',\vt'$, and
$v=a'\{x\to b'\}\vt'$. Hence, by taking $w=a\{x\to b\}\vt$, we have
$t\ah w'\ah w\ar^* v$.\cqfd
\end{prf}


\begin{thm}[Subject reduction for $\b$]
If no right hand-side is in $\cB\cup\cK\cup\cW$ and the right
hand-side of every type-level rule is a symbol application then $\b$
preserves typing.
\end{thm}

\begin{prf}
The proof is as usual by induction on $\G\th t:T$ and by proving that,
if $\G\ab\G'$, then $\G'\th t:T$. The only difficult case is the case
of a head-reduction $[x:U']v~u\ab v\xu$ with $\G\th [x:U']v:(x:U)V$
and $\G\th u:U$. We must prove that $\G\th v\xu:V\xu$. We already know
that it is true when $v$ is a predicate. We must now prove it when $v$
is an object, that is, when $\G\th (x:U)V:\st$. By inversion, we have
$\G\th [x:U']v:(x:U')V'$ with $(x:U')V' \CVG (x:U)V$. By Lemma
\ref{lem-kind} (j), we have all the intermediate well-typed terms
between $(x:U')V'$ and $(x:U)V$ of type $\st$. Without loss of
generality, we can assume that $T_0= (x:U')V' \ad_\b T_1 \ad_\cR T_2
\ad_\b \ldots T_n= (x:U)V$. Let $T_i'$ be the common reduct between
$T_i$ and $T_{i+1}$. We now prove by induction on the number of
conversions that there is a sequence of well-typed product terms
$\pi_1, \ldots, \pi_n$ such that $\pi_0= T_0 \ad_\b \pi_1 \ad_\cR
\pi_2 \ad_\b \ldots \pi_n= T_n$.

Since $T_0$ is a product, $\pi_0'= T_0'$ is also a product. Since
$T_1\ab^* \pi_0'$, by standardization, there is a product term $\pi_1$
such that $T_1\ah^* \pi_1\anh^* \pi_0'$. Since $h$ has the subject
reduction property on types, $\pi_1$ is well-typed. Now, since
$T_1\ar^* T_1'$, by commutation, there is a product term $\pi_1'$ such
that $\pi_1\ar^* \pi_1'$ and $T_1'\ah^* \pi_1'$. Furthermore, since
$T_2\ar^* T_1'$, by postponement, there is a term $t$ such that
$T_2\ah^* t\ar^* \pi_1'$. Since $h$ has the subject reduction property
on types, $t$ is a well-typed term of type $\st$. We now proceed by
case on $t$.

\begin{lst}{--}
\item If $t$ is an abstraction $[x:T]w$ then, by inversion, there is
$W$ such that $(y:T)W\CVG \st$. By Lemma \ref{lem-kind} (h) and (i),
this is not possible.
\item If $t$ is an application but not a symbol application then,
since left hand-sides of rules are algebraic, $\pi_1'$ is an
application, which is not possible either.
\item If $t$ is a symbol application then, since right hand-sides of
type-level rules are symbol applications, $\pi_1'$ is a symbol
application too, which is not possible either.
\item Therefore, $t$ is a well-typed product term $\pi_2$.
\end{lst}

Now, since $T_2\ab^* T_2'$ and $\b$ is confluent, there is a product
term $\pi_2'$ such that $\pi_2\anh^* \pi_2'$ and $T_2'\ab^*
\pi_2'$, and we can now conclude by induction.\cqfd
\end{prf}





\section{Logical consistency}
\label{sec-consistency}

In the case of the pure Calculus of Constructions without symbols and
rewrite rules, logical consistency easily follows from normalization
by proving that there can be no normal proof of $\bot=
(\alpha:\st)\alpha$ in the empty environment
\cite{barendregt92book}. But, having symbols and rewrite rules is like
having hypothesis and axioms. Thus, in this case, logical consistency
does not directly follow from normalization. We can however give
general conditions ensuring logical consistency:

\begin{thm}[Logical consistency]
Assume that $\a$ is confluent and that every object symbol $f$
satisfies one of the following conditions:

\begin{enumi}{}
\item $f:(\vx:\vT)C\vv$ with $C\in\CFB$,
\item $f:(\vx:\vT)T_i$,
\item $f:(x_1:T_1)\ldots (x_n:T_n)U$ with $x_n\notin\FVB(U)$ and, for
all normal substitution $\g:(\vx:\vT)\leadsto(\alpha:\st)$, $f\vx\g$
is reducible.
\end{enumi}

\noindent
Then, there is no normal proof of $\bot=(\alpha:\st)\alpha$ in the
empty environment. Therefore, if $\a$ is also normalizing, then there
is no proof of $\bot$ in the empty environment.
\end{thm}

\begin{prf}
Assume that $\th t:\bot$, $t$ is normal and of minimal size, that is,
there is no term $u$ smaller than $t$ such that $\th u:\bot$. For
typing reasons, $t$ cannot be a sort or a product. Assume that $t$ is
an application. Since $t$ is typable in the empty environment, it
cannot have free variables and, since $t$ is normal, it must be of the
form $f\vt$. Assume that $|\vt|=k$ and that $f$ is of type
$(\vx:\vT)U$ with $|\vx|=n$. Let $\g_i=\{x_1\to t_1,\ldots,x_i\to
t_i\}$ ($i\le n$).
 
\begin{enumi}{}
\item In this case, $k\le n$ since $f$ cannot be applied to more than
$n$ arguments. Indeed, if $f$ is applied to $n+1$ arguments then, by
inversion, $\th ft_1\ldots t_n:(x_{n+1}:T_{n+1})V$. But, since $\th
ft_1\ldots t_n:C\vv\g_n$, by convertibility of types and confluence,
we must have $(x_{n+1}:T_{n+1})V\ad C\vv\g_n$, which is not
possible. Thus, $k\le n$ and $(x_{k+1}:T_{k+1}\g_k) \ldots
(x_n:T_n\g_k)C\vv\g_k\ad\bot$, which is not possible either.

\item There are 2 cases:

\begin{lst}{\bu}
\item $k<n$. Since $\th f\vt:(x_{k+1}:T_{k+1}\g_k) \ldots
(x_n:T_n\g_k)T_i\g_k$, we must have $n=k+1$ and, by taking
$x_n=\alpha$, $T_n\g_k\ad\st$ and $T_i\g_k\ad\alpha$. Hence
$T_i\g_k\a^*\alpha$ but $T_i\g_k$ is closed since
$\FV(T_i)\sle\{x_1,\ldots,x_{i-1}\}$, $\g_k$ is closed and $i-1\le
k$. So, $T_i\g_k\a^*\alpha$ is not possible.

\item $k\ge n$. We have $\vt=\vu\vv$ with $|\vu|=n$. Let $p=k-n$. By
inversion, there is a sequence of products $(y_1:V_1)W_1$, \ldots,
$(y_p:V_p)W_p$ such that $T_i\g_n=U\g_n\ad (y_1:V_1)W_1$, for all
$i<p$, $W_i\{y_i\to v_i\}\ad (y_{i+1}:V_{i+1})W_{i+1}$, and
$W_p\{y_p\to v_p\}\ad\bot$. Then, $\th u_i\vv:\bot$ and $u_i\vv$ is
smaller than $t$.
\end{lst}

\item If $k\ge n$ then $t$ is reducible, which is not possible. If
$k<n$ then $n=k+1$, $x_n=\alpha$ and $U\g_k\a^*\alpha$. But
$\FV(U)\sle\{x_1,\ldots,x_k,\alpha\}$ and $\g_k$ is closed. So,
$x_n\in\FVB(U)$, which is excluded.
\end{enumi}

Assume now that $t=[\alpha:T]v$. Then, by inversion, we must have
$\alpha:T\th v:V$ and $(\alpha:T)V \ad (\alpha:\st)\alpha$. Therefore,
$T=\st$, $V=\alpha$ and $\alpha:\st\th v:\alpha$. For typing reasons,
$v$ cannot be a sort, a product or an abstraction. Since it is normal,
it must be of the form $x\vu$ with $x$ a variable, or of the form
$f\vt$. Since $\alpha$ is the only variable that may freely occur in
$v$, $x=\alpha$. Since $\alpha$ can be applied to no argument,
$v=\alpha$. Then, we get $\alpha:\st\th \alpha:\alpha$, which is not
possible. Therefore, $v$ is of the form $f\vt$.

\begin{enumi}{}
\item In this case, $k\le n$ since $f$ cannot be applied to more than
$n$ arguments. Thus, $(x_{k+1}:T_{k+1}\g_k) \ldots
(x_n:T_n\g_k)C\vv\g_k\ad\alpha$, which is not possible.

\item If $k<n$ then $(x_{k+1}:T_{k+1}\g_k) \ldots
(x_n:T_n\g_k)T_i\g_k\ad\alpha$, which is not possible. Thus,
$\vt=\vu\vv$ with $|\vu|=n$. Let $p=k-n$. By inversion, there is a
sequence of products $(y_1:V_1)W_1$, \ldots, $(y_p:V_p)W_p$ such that
$T_i\g_n=U\g_n\ad (y_1:V_1)W_1$, for all $i<p$, $W_i\{y_i\to v_i\}\ad
(y_{i+1}:V_{i+1})W_{i+1}$, and $W_p\{y_p\to v_p\}\ad\alpha$. Then,
$\th [\alpha:\st]u_i\vv:\bot$ and $[\alpha:\st]u_i\vv$ is smaller than
$t$.

\item In this case too, $k\ge n$. Thus, $t$ is reducible, which is not
possible.\cqfd
\end{enumi}
\end{prf}


Note that, as opposed to the third condition, the first two conditions
do not care about the rewrite rules defining $f$.

To see the interest of the third condition, consider the following
example. Assume that the only symbols of the calculus are $nat:\st$,
$0:nat$, $s:nat\a nat$ and $rec: (P:nat\a\st)$ $P0\a ((n:nat)Pn\a
P(sn))\a (n:nat)Pn$ defined by the usual rules for recursors:

\begin{rewc}
rec~P~u~v~0 & u\\
rec~P~u~v~(s~n) & v~n~(rec~P~u~v~n)\\
\end{rewc}

This calculus is confluent since the combination of an orthogonal
system (the recursor rules) with the $\b$-reduction preserves
confluence. In this calculus, it is possible to express any function
whose existence is provable in intuitionistic higher-order
arithmetic.

Now, let us look at the normal terms of type $nat$ in the environment
$\alpha:\st$. Let $\cN$ be the set of these terms. A term in $\cN$
cannot be a sort, a product, an abstraction, nor a variable. It can
only be of the form $0$, $(s~t)$ with $t$ itself in $\cN$, or of the
form $(rec~P~u~v~t~\vu)$ with $t\in\cN$ also. But the last case is not
possible since, at some point, the argument $t$ of $(rec~P~u~v~t)$
must be of the form $0$ or $(s~t')$, and hence $(rec~P~u~v~t)$ must be
reducible. Therefore, all the normal terms of type $nat$ typable in
$\alpha:\st$ must be of the form $0$ or $(s~t)$, and if $t$ is such a
term then $(rec~P~u~v~t)$ is reducible. We also say that functions
defined by induction are {\em completely defined}
\cite{guttag78acta,thiel84popl,kounalis85eurocal,coquand92types}. Therefore,
after the previous theorem, this calculus is consistent.

This may certainly be extended to the Calculus of Inductive
Constructions and even to the Calculus of Inductive Constructions
extended with functions defined by rewrite rules whenever all the
symbols are completely defined.




\section{Conditions of Strong Normalization}
\label{sec-conditions}

We now present the conditions of strong normalization.




\subsection{Inductive types and constructors}
\label{subsec-ind-typ}

Until now we made few hypothesis on symbols and rewrite
rules. However, Mendler \cite{mendler87thesis} showed that the
extension of the simply-typed $\la$-calculus with recursion on
inductive types is strongly normalizing if and only if the inductive
types satisfy some positivity condition.

A base type $T$ occurs positively in a type $U$ if all the occurrences
of $T$ in $U$ are on the left of an even number of $\A$. A type $T$ is
positive if $T$ occurs positively in the type of the arguments of its
constructors. Usual inductive types like natural numbers and lists of
natural numbers are positive.

Now, let us see an example of a non-positive type $T$. Let $U$ be a
base type. Assume that $T$ has a constructor $c$ of type $(T\A U)\A
T$. $T$ is not positive because $T$ occurs at a negative position in
$T\A U$. Consider now the function $p$ of type $T\A (T\A U)$ defined
by the rule $p(cx)\a x$. Let $\w= \la x.(px)x$ of type $T\A U$. Then
the term $\w(c\w)$ of type $U$ is not normalizable:

\begin{center}
$\w(c\w) \,\ab\, p(c\w)(c\w) \,\ar\, \w(c\w) \,\ab\, \ldots$
\end{center}

In the case where $U=\st$, we can interpret this as Cantor's theorem:
there is no surjection from a set $T$ to the set of its subsets
$T\A\st$. In this interpretation, $p$ is the natural injection between
$T$ and $T\A\st$. Saying that $p$ is surjective is equivalent to
saying (with the Axiom of Choice) that there exists $c$ such that $p
\circ c$ is the identity, that is, such that $p(cx) \a x$. In
\cite{dowek99hab}, Dowek shows that such an hypothesis is
incoherent. Here, we show that this is related to the
non-normalization of non-positive inductive types.

Mendler also gives a condition, strong positivity, in the case of
dependent and polymorphic types. A similar but more restrictive
notion, called strict positivity, is used by Coquand and Paulin in the
Calculus of Inductive Constructions \cite{coquand88colog}.

Hereafter we introduce the more general notion of {\em admissible
inductive structure}. In particular, we do not consider that a
constructor must be constant: it is possible to have rewrite rules on
constructors. This allows us to formalize quotient types like the type
$int$ of integers by taking $0:int$ for zero, $s:int\A int$ for
successor, and $p:int\A int$ for predecessor, together with the rules:

\begin{rewc}
s~(p~x) & x\\
p~(s~x) & x\\
\end{rewc}


\newcommand{\mon}{\mr{Mon}}
\newcommand{\acc}{\mr{Acc}}

\begin{dfn}[Inductive structure]
\label{def-ind-str}

  An {\em inductive structure} is given by:
\begin{lst}{\bu}
\item a precedence $\ge_\cC$ on $\CFB$,
\item for every $C:(\vx:\vT)\st$ in $\CFB$, a set $\mon(C)\sle
\{i\le|\vx|~|~ x_i\in\XB\}$ for the {\em monotonic arguments} of $C$,
\item for every $f:(\vy:\vU)C\vv$ with $C\in\CFB$, a set $\acc(f)\sle
\{1,\ldots,|\vy|\}$ for the {\em accessible positions} of $f$.
\end{lst}

\noindent
For convenience, we assume that $\mon(f)=\vide$ if $f\notin\CFB$, and
$\acc(f)=\vide$ if $f$ is not of type $(\vy:\vU)C\vv$ with $C\in\CFB$.
\end{dfn}

The accessible positions of $f$ denote the arguments of $f$ that one
can use in the right hand-sides of rules. The monotonic arguments of
$C$ denote the parameters in which $C$ is monotonic.


\newcommand{\pp}{\pos^+}
\renewcommand{\pm}{\pos^-}
\newcommand{\pz}{\pos^0}
\newcommand{\pnz}{\pos^{\neq 0}}
\newcommand{\pd}{\pos^\d}
\newcommand{\pmd}{\pos^{-\d}}

\begin{dfn}[Positive and negative positions]
\label{def-neg-pos}

The set of {\em positive positions} in $t$, $\pp(t)$, and the set of
{\em negative positions} in $t$, $\pm(t)$, are simultaneously defined
by induction on the structure of $t$:

\begin{lst}{--}
\item $\pd(s)= \pd(x)= \{\vep~|~\d=+\}$,
\item $\pd((x:U)V)= 1.\pmd(U)\cup 2.\pd(V)$,
\item $\pd([x:U]v)= 2.\pd(v)$,
\item $\pd(tu)= 1.\pd(t)$ if $t\neq f\vt$,
\item $\pd(f\vt)= \{1^{|\vt|}~|~\d=+\}\cup
  \,\bigcup\{1^{|\vt|-i}2.\pd(t_i)~|~ i\in\mon(f)\}$,
\end{lst}

\noindent
where $\d\in\{-,+\}$, $-+=-$ and $--=+$ (usual rule of signs).
\end{dfn}


\begin{dfn}[Admissible inductive structures]
\label{def-adm-ind-str}

An inductive structure is {\em admissible} if, for all $C\in\CFB$, for
all $f:(\vy:\vU)C\vv$, and for all $j\in\acc(f)$:\footnote{In
\cite{blanqui01lics}, we give 6 conditions, (I1) to (I6), for defining
what is an admissible inductive structure. But we found that (I1) can
be eliminated if we modify (I2) a little bit. This is why, in the
following definition, there is no (I1) and (I2) is placed after (I6).}

\begin{bfenumi}{I}
\setcounter{enumi}{2}
\item $\all D\in\CFB, D=_\cC C \A \pos(D,U_j)\sle\pp(U_j)$\\
(symbols equivalent to $C$ must be at positive positions),

\item $\all D\in\CFB, D>_\cC C \A \pos(D,U_j)=\vide$\\
(no symbol greater than $C$ can occur in $U_j$),

\item $\all F\in\DFB, \pos(F,U_j)=\vide$\\
(no defined symbol can occur in $U_j$),

\item $\all Y\in\FVB(U_j), \ex\io_Y, v_{\io_Y}= Y$\\
(predicate variables must be parameters of $C$),

\setcounter{enumi}{1}
\item $\all Y\in\FVB(U_j), \io_Y\in\mon(C)\A \pos(Y,U_j)\sle\pp(U_j)$\\
(monotonic arguments must be at positive positions).
\end{bfenumi}
\end{dfn}

For instance, with $list:\st\A\st$, $nil:(A:\st)listA$ and
$cons:(A:\st)A\A listA\A listA$, $\mon(list)= \{1\}$, $\acc(nil)=
\{1\}$ and $\acc(cons)= \{1,2,3\}$ is an admissible inductive
structure. If we add $tree:\st$ and $node: list~tree\A tree$ with
$\mon(list)= \{1\}$, $\mon(tree)=\vide$ and $\acc(node)=\{1\}$, we
still have an admissible structure.

The condition (I6) means that the predicate arguments of a constructor
must be parameters of their type. A similar condition appears in the
works of Stefanova \cite{stefanova98thesis} (``safeness'') and
Walukiewicz \cite{walukiewicz02jfp} (``$\st$-dependency''). On the
other hand, in the Calculus of Inductive Constructions (CIC)
\cite{werner94thesis}, there is no such restriction.


We distinguish several kinds of inductive types.

\begin{dfn}[Primitive, basic and strictly-positive predicates]
\label{def-pred}

\hfill\\ A constant predicate symbol $C$ is:

\begin{lst}{--}
\item {\em primitive} if for all $D =_\cC C$, for all
$f:(\vy:\vU)D\vw$ and for all $j\in \acc(f)$, $U_j=E\vt$ with $E <_\cC
D$ and $E$ primitive, or $U_j=E\vt$ with $E =_\cC D$.
  
\item {\em basic} if for all $D =_\cC C$, for all $f:(\vy:\vU)D\vw$
and for all $j\in \acc(f)$, if $E =_\cC D$ occurs in $U_j$ then $U_j$
is of the form $E\vt$.
  
\item {\em strictly positive} if for all $D =_\cC C$, for all
$f:(\vy:\vU)D\vw$ and for all $j\in \acc(f)$, if $E =_\cC D$ occurs in
$U_j$ then $U_j=(\vz:\vV)E\vt$ and no $D' =_\cC D$ occurs in $\vV$.
\end{lst}
\end{dfn}

Primitive predicates are basic and basic predicates are strictly
positive. Note that primitive predicates not only include the usual
first-order data types. They also include some dependent type like the
type of lists of fixed length. On the other hand, the type of
polymorphic lists is basic but not primitive.

\newcommand{\form}{\mbox{\it form}}

The strictly positive predicates are the predicates allowed in the
Calculus of Inductive Constructions (CIC). For example, this includes
the type $ord$ of Brouwer's ordinals whose constructors are $0:ord$,
$s:ord\A ord$ and $lim:(nat\A ord)\A ord$, the process algebra
$\mu$CRL which can be formalized as a type $proc$ with a choice
operator $\S:(data\A proc)\A proc$ \cite{sellink93ssl}, or the type
$\form$ of the formulas of first-order predicate calculus whose
constructors are $\neg:\form\A\form$, $\ou:\form\A\form\A\form$ and
$\all:(term\A\form)\A\form$.

For the moment, we do not forbid non-strictly positive predicates but
the conditions we describe in the next section do not allow the
definition of functions by recursion on such predicates. Yet, these
predicates can be very useful as shown in \cite{matthes00} or in
\cite{abel01merlin}. In \cite{matthes00}, a type $cont$ with the
constructors $D:cont$ and $C:((cont\A list)\A list)\A cont$,
representing continuations, is used to define a breadth-first label
listing function on labelled binary trees. In particular, it uses a
function $ex:cont\A list$ defined by the rules:

\begin{rewc}
ex~D & nil\\
ex~(C~f) & f~ex\\
\end{rewc}

\noindent
It is not clear how to define a syntactic condition ensuring the
strong normalization of such a definition: in the right hand-side of
the second rule, $ex$ is explicitly applied to no argument smaller
than $f$. And, although $ex$ can only be applied to subterms of
reducts of $f$, not every subterm of a ``computable'' term (notion
used for proving strong normalization) is {\em a priori} computable
(see Section \ref{subsec-schema}).




\subsection{General Schema}




\subsubsection{Higher-order rewriting}

Which conditions on rewrite rules would ensure the strong
normalization of $\a= \ar\cup\ab$ ? Since the works of Breazu-Tannen
and Gallier \cite{breazu89icalp} and Okada \cite{okada89issac} on the
simply-typed $\la$-calculus or the polymorphic $\la$-calculus, and
later the works of Barbanera \cite{barbanera90ctrs} on the Calculus of
Constructions and of Dougherty \cite{dougherty91rta} on the untyped
$\la$-calculus, it is well known that adding first-order rewriting to
typed $\la$-calculi preserves strong normalization. This comes from
the fact that first-order rewriting cannot create $\b$-redexes. We
will prove that this result can be extended to predicate-level
rewriting if some conditions are fulfilled.

However, there are many useful functions whose definition do not enter
the first-order framework, either because some arguments are not
primitive (the concatenation function $app$ on polymorphic lists), or
because their definition uses higher-order features like the function
$map:(A:\st)(B:\st)(A\A B)\A listA\A listB$ which applies a
function to every element of a list:

\begin{rewc}
map~A~B~f~(nil~A') & nil~B\\
map~A~B~f~(cons~A'~x~\ell) & cons~B~(f~x)~(map~A~B~f~\ell)\\
map~A~B~f~(app~A'~\ell~\ell') & app~B~(map~A~B~f~\ell)~(map~A~B~f~\ell')\\
\end{rewc}

This is also the case of recursors like the recursor on natural
numbers $natrec:(A:\st)$ $A\A (nat\A A\A A)\A nat\A A$:

\begin{rewc}
natrec~A~x~f~0 & x\\
natrec~A~x~f~(s~n) & f~n~(natrec~A~x~f~n)\\
\end{rewc}

\noindent
and of induction principles (recursors are just non-dependent versions
of the corresponding induction principles), like the induction
principle on natural numbers $natind:(P:nat\A\st)P0\A((n:nat)Pn\A
P(sn))\A(n:nat)Pn$:

\begin{rewc}
natind~P~h_0~h_s~0 & h_0\\
natrec~P~h_0~h_s~(s~n) & h_s~n~(natind~P~h_0~h_s~n)\\
\end{rewc}

The methods used by Breazu-Tannen and Gallier \cite{breazu89icalp} or
Dougherty \cite{dougherty91rta} cannot be applied to our calculus
since, on the one hand, higher-order rewriting can create $\b$-redexes
and, on the other hand, rewriting is included in the type conversion
rule (conv), hence more terms are typable. But there exists other
methods, available in the simply-typed $\la$-calculus only or in
richer type systems, for proving the termination of this kind of
definitions:

\begin{lst}{\bu}
\item The {\em General Schema}, initially introduced by Jouannaud and
Okada \cite{jouannaud91lics} for the polymorphic $\la$-calculus and
extended to the Calculus of Constructions by Barbanera, Fern\'andez
and Geuvers \cite{barbanera94lics}, is an extension of the primitive
recursion schema: in the right hand-side of a rule $f\vl\a r$, the
recursive calls to $f$ must be done on strict subterms of $\vl$. It
can treat object-level rewriting with simply-typed symbols defined on
primitive types. It has been reformulated and extended to
strictly-positive types by Jouannaud, Okada and the author for the
simply-typed $\la$-calculus \cite{blanqui02tcs} and the Calculus of
Constructions \cite{blanqui99rta}.

\item The {\em Higher-Order Recursive Path Ordering} (HORPO)
\cite{jouannaud99lics} is an extension of RPO
\cite{plaisted78tr,dershowitz82tcs} to the simply-typed
$\la$-calculus. It has been recently extended by Walukiewicz
\cite{walukiewicz00lfm} to the Calculus of Constructions with strictly
positive types \cite{walukiewicz02jfp}. It can treat object-level
rewriting with polymorphic and dependent symbols defined on strictly
positive types. The General Schema can be seen as a non-recursive
version of HORPO.

\item It is also possible to look for an interpretation of the symbols
such that the interpretation of a term strictly decreases when a rule
is applied. This method, introduced by Van de Pol for the simply-typed
$\la$-calculus \cite{vandepol96thesis}, extends to the higher-order
framework the method of interpretations known for the first-order
framework \cite{zantema94jsc}. This is a very powerful method but
difficult to use in practice because the interpretations are
themselves higher-order and also because it is not modular: adding new
rules or new symbols may require finding new interpretations.
\end{lst}

For dealing with higher-order rewriting at the predicate-level
together with polymorphic and dependent symbols and strictly-positive
predicates, we have chosen to extend the method of the General
Schema. For first-order symbols, we use other conditions like in
\cite{jouannaud97tcs}.




\subsubsection{Definition of the schema}
\label{subsec-schema}

This method is based on Tait and Girard's method of reducibility
candidates \cite{tait67jsl,girard88book} for proving the strong
normalization of simply-typed or polymorphic $\la$-calculi. This
method consists of interpretating each type as a subset of the
strongly normalizable terms, the {\em computable} terms, and proving
that each well-typed term is computable. Indeed, a direct proof of
strong normalization by induction on the structure of terms does not
go through because of the application case: if $u$ and $v$ are
strongly normalizable then it is not clear how to prove that $uv$ also
is strongly normalizable.

The idea of the General Schema is then, from a left hand-side $f\vl$
of rule, to define a set of terms, called the {\em computability
closure} of $f\vl$, whose elements are computable whenever the $l_i$'s
so are. Then, to prove the strong normalization, it suffices to check
that, for each rule, the right hand-side belongs to the computability
closure of the left hand-side.

To build the computability closure, we first define a subset of the
subterms of $\vl$, called the {\em accessible} subterms of $\vl$, that
are computable whenever the $l_i$'s so are (not all the subterms of a
computable term are {\em a priori} computable). Then, we build the
computability closure by closing the set of accessible variables of
the left hand-side with computability-preserving operations.

In order to have interesting functions, we must be able to accept
recursive calls and, to preserve strong normalization, recursive calls
must decrease in a well-founded ordering. The strict subterm relation
$\tgt$ (in fact, restricted to accessible subterms for preserving
computability) is sufficient for dealing with definition on basic
predicates. In the definition of $map$ for instance, $\ell$ and
$\ell'$ are accessible subterms of $app~A'~\ell~\ell'$. But, for
non-basic predicates, it is not sufficient as examplified by the
following addition on Brouwer's ordinals:

\begin{rewc}
x+0 & x\\
x+(s~y) & s~(x+y)\\
x+(lim~f) & lim~([n:nat]x+fn)\\
\end{rewc}

Another example is given by the following simplification rule in
$\mu$CRL \cite{sellink93ssl}:

\begin{rewc}
(\S~f)\cdot p & \S~([d:data]fd\cdot p)\\
\end{rewc}

This is why, in our conditions, we use two distinct orderings. The
first one, $>_1$, is used for the arguments of basic type and the
second one, $>_2$, is used for the arguments of strictly-positive
type.

Finally, to have a finer control of the comparison of the arguments,
to each symbol, we associate a {\em status} describing how to compare
the arguments by using a simple combination of lexicographic and
multiset comparisons \cite{jouannaud97tcs}.


\begin{dfn}[Accessibility]
\label{def-acc}

We say that $u:U$ is {\em accessible modulo $\r$} in $t:T$, written
$t:T ~\tgt_\r~ u:U$, if $t=f\vu$, $f:(\vy:\vU)C\vv$, $C\in\CFB$,
$u=u_j$, $j\in\acc(f)$, $T\r= C\vv\g\r$, $U\r= U_j\g\r$, $\g=\vyu$ and
no $D=_\cC C$ occurs in $\vu\r$.
\end{dfn}

For technical reasons, we take into account not only the terms
themselves but also their types. This comes from the fact that we are
able to prove that two convertible types have the same interpretation
only if the two types are computable. This may imply some restrictions
on the types of the symbols.

Indeed, accessibility requires the equality (modulo the application of
$\r$) between canonical types and derived types (see Definition
\ref{def-can-typ}). More precisely, for having $t:T \tgt_\r u:U$, $T$
must be equal (modulo $\r$) to the canonical type of $t$ and $U$ must
be equal (modulo $\r$) to the type of $u$ derived from $t$. In
addition, if $u:U \tgt_\r v:V$, then $U$ must also be equal (modulo
$\r$) to the canonical type of $u$.


\renewcommand{\sf}{_{stat_f}}
\newcommand{\sF}{_{stat_F}}
\newcommand{\sg}{_{stat_g}}

\begin{dfn}
\label{def-stat}

Let $(x_i)_{i \ge 1}$ be an indexed family of variables.

\begin{lst}{}
\item [\bf Status.] A {\em status} is a term of the form
$(lex~m_1\ldots m_k)$ with $k\ge 1$ and each $m_i$ of the form
$(mul~x_{k_1}\ldots x_{k_p})$ with $p\ge 1$. The {\em arity} of a
status $stat$ is the greatest index $i$ such that $x_i$ occurs in
$stat$.
  
\item [\bf Status assignment.] A {\em status assignment} is an
application $stat$ which associates a status $stat_f$ to every
$f\in\cF$.

\item [\bf Predicate arguments.] Let $C:(\vz:\vV)\st$ and $\vu$ with
$|\vu|=|\vz|$. By $\vu|_C$, we denote the sub-sequence $u_{j_1}\ldots
u_{j_n}$ such that $j_1<\ldots<j_n$ and $\{j_1,\ldots,j_n\}=
\{j\le|\vz|~|~ z_j\in\XB\}$.

\item [\bf Strictly positive positions.] Let $f:(\vx:\vT)U$ with
$stat_f= lex~\vm$. The set of {\em strictly positive positions} of
$f$, $SP(f)$, is defined as follows. Assume that $m_i=
mul~x_{k_1}\ldots x_{k_p}$. Then, $i\in SP(f)$ iff there exist
$T_f^i=C\va$ such that $C$ is strictly positive and, for all $j$,
$T_{k_j}=C\vu$ with $C\in\CFB$ and $\vu|_C=\va|_C$.

\item [\bf Assignment compatibility.] A status assignment $stat$ is
{\em compatible} with a precedence $\ge_\cF$ if $f =_\cF g$ implies
$stat_f= stat_g$, $SP(f)=SP(g)$ and, for all $i\in SP(f)$, $T_f^i=T_g^i$.

\item [\bf Status ordering.] Let $>$ be an ordering on terms and
$stat= lex~\vm$ be a status of arity $n$. The {\em extension} of $>$
to the sequences of terms of length $n$ is the ordering $>_{stat}$
defined as follows:
\begin{lst}{--}
\item $\vu >_{stat} \vv$ ~if~ $\vm\vxu ~(>^m)\lex~ \vm\vxv$,
\item $mul~\vu >^m mul~\vv$ ~if~ $\{\vu\} >\mul \{\vv\}$.
\end{lst}
\end{lst}
\end{dfn}

For instance, if $stat= lex(mul~x_2)(mul~x_1~x_3)$ then $(u_1,u_2,u_3)
>_{stat} (v_1,v_2,v_3)$ iff $(\{u_2\},\{u_1$, $u_3\})$ $(>\mul)\lex$
$(\{v_2\},\{v_1,v_3\})$. An important property of $>_{stat}$ is that
it is well-founded whenever $>$ is.

We now define the computability closure of a rule $R=(l\a r,\G,\r)$
with $l=f\vl$, $f:{(\vx:\vT)U}$ and $\g=\vxl$.


\begin{dfn}[Ordering on symbol arguments]
\label{def-arg-ord}

The ordering $>_R$ on arguments of $f$ is an adaptation of $>\sf$
where the ordering $>$ depends on the type (basic or strictly
positive) of the argument. Assume that $stat_f= lex~m_1\ldots
m_k$. Then:

\begin{lst}{\bu}
\item $\vt:\vT >_R \vu:\vU$ ~if~ $\vm\{\vx\to(\vt:\vT)\}
~(>^1,\ldots,>^k)\lex~ \vm\{\vx\to(\vu:\vU)\}$.

\item $mul(\vt:\vT) >^i mul(\vu:\vU)$ ~if~ $i\in SP(f)$ and
$\{\vt:\vT\} ~(>^i_R)\mul~ \{\vu:\vU\}$,
\item $mul(\vt:\vT) >^i mul(\vu:\vU)$ ~if~ $i\notin SP(f)$ and
$\{\vt:\vT\} ~(\tgt_\r^+)\mul~ \{\vu:\vU\}$,

\item $t:T ~>^i_R~ u:U$ ~if:
\begin{lst}{--}
\item $t=f\vt$, $f:(\vx:\vT)C\vv$, $\g=\vxt$ and no $D=_\cC C$ occurs
in $\vv\g\r$,
\item $u=x\vu$ with $x\in\dom(\G)$,
\item $t:T ~\tgt_\r^+~ x:V$,
\item $V\r= x\G= (\vy:\vU)C\vw$, $\d=\vyu$, $U\r= C\vw\d$ and no
$D=_\cC C$ occurs in $\vU\d$,
\item $\vv\g\r|_C=\vw\d|_C$.
\end{lst}
\end{lst}
\end{dfn}


One can easily check that, for the addition on ordinals, $lim~f:ord
>_R^1 fn:ord$. Indeed, for this rule, one can take $\G= x:ord, f:nat\A
ord$ and the identity for $\r$. Then, $f\in\dom(\G)$, $f\G= nat\A ord$
and $lim~f:ord \tgt_\r f:nat\A ord$.

\newcommand{\thc}{\th_\mr{\!\!c}}


\begin{figure}[ht]
\centering
\caption{Computability closure of $R=(f\vl\a r,\G,\r)$ with
$f:(\vx:\vT)U$ and $\g=\vxl$\label{fig-thc}}

\begin{tabular}{rcc}
\\ (ax) & $\cfrac{}{\thc\st:\B}$\\

\\ (symb$^<$) & $\cfrac{\thc\tg:s_g}{\thc g:\tg}$
& $(g <_\cF f)$\\

\\ (symb$^=$) & $\cfrac{
\thc\tg:s_g \quad \d:\G_g\leadsto_c\D}
{\D\thc g\vy\d:V\d}$ &
$\begin{array}{c}
(g =_\cF f,\, g:(\vy:\vU)V,\\
\vy\d:\vU\d <_R \vx\g:\vT\g)\\
\end{array}$\\

\\ (var) & $\cfrac{\D\thc T:s_x}{\D,x:T\thc x:T}$
& $(x\notin\dom(\D))$\\

\\ (weak) & $\cfrac{\D\thc t:T \quad \D\thc U:s_x}{\D,x:U\thc t:T}$
& $(x\notin\dom(\D))$\\

\\ (prod) & $\cfrac{\D,x:U\thc V:s}{\D\thc (x:U)V:s}$\\

\\ (abs) & $\cfrac{\D,x:U\thc v:V \quad \D\thc (x:U)V:s}
{\D\thc [x:U]v:(x:U)V}$\\

\\ (app) & $\cfrac{\D\thc t:(x:U)V \quad \D\thc u:U}{\D\thc tu:V\xu}$\\

\\ (conv) & $\cfrac{\D\thc t:T \quad \D\thc T:s \quad \D\thc T':s}
{\D\thc t:T'}$ & $(T \ad T')$\\
\end{tabular}
\end{figure}


\begin{dfn}[Computability closure]
\label{def-thc}

Let $\cF'=\cF\,\cup\,\dom(\G)$, $\cX'=\cX\moins\FV(l)$,
$\cT=\cT(\cF',\cX')$ and $\cE'=\cE(\cF',\cX')$. The {\em computability
closure} of $R$ w.r.t. a precedence $\ge_\cF$ and a status assignment
$stat$ compatible with $\ge_\cF$ is the smallest relation $\thc\,\sle
\cE'\times\cT'\times\cT'$ defined by the inference rules of Figure
\ref{fig-thc} where, for all $x\in\dom(\G)$, $\tau_x=x\G$ and $x<_\cF
f$, and where $\d:\G_g\leadsto_c\D$ means that, for all
$y\in\dom(\G_g)$, $\D\thc x\d:x\G_g\d$.
\end{dfn}


Note that the computability closure can easily be extended by adding
new inference rules. Then, for preserving strong normalization, it
suffices to complete the proof of Theorem \ref{thm-cor-thc} where we
prove that the rules of the computability closure indeed preserve
computability.


\begin{dfn}[Well-formed rule]
\label{def-wf-rule}

$R$ is {\em well-formed} if:
\begin{lst}{--}
\item $\G\th l\r:U\g\r$,
\item $\all x\in\dom(\G)$, $\ex i$, $l_i:T_i\g ~\tgt_\r^*~ x:x\G$,
\item $\dom(\r)\sle \FV(l)\moins\dom(\G)$.
\end{lst}
\end{dfn}


For instance, consider the rule:

\begin{rewc}
app~p~(cons~x~n~\ell)~n'~\ell' & cons~x~(n+n')~(app~n~\ell~n'~\ell')\\
\end{rewc}

\noindent
with $\G= x:nat, n:nat, \ell:listn, n':nat, \ell':listn'$ and
$\r=\{p\to sn\}$. We have $\G\th l\r:list(p+n')\r$. For $x$, we have
$cons~x~n~\ell:listp \tgt_\r x:nat$. One can easily check that the
conditions are also satisfied for the other variables.


\begin{dfn}[Computable system]
\label{def-rec-sys}

  $R$ satisfies the {\em General Schema} w.r.t. a precedence $\ge_\cF$
  and a status assignment $stat$ compatible with $\ge_\cF$ if it is
  well-formed and if $\thc r:U\g\r$. A set of rules $\cR$ is {\em
  computable} if there exists a precedence $\ge_\cF$ and a status
  assignment $stat$ compatible with $\ge_\cF$ for which every rule of
  $\cR$ satisfies the General Schema w.r.t. $\ge_\cF$ and $stat$.
\end{dfn}


To summarize, the rule $(l\a r,\G,\r)$ is well-typed and satisfies the
General Schema if:

\begin{lst}{--}
\item $\G\th l\r:U\g\r$,
\item $\all \D,\s,T$, if $\D\th l\s:T$ then $\s:\G\leadsto\D$ and $\s\ad\r\s$,
\item $\all x\in\dom(\G)$, $\ex i$, $l_i:T_i\g ~\tgt_\r^*~ x:x\G$,
\item $\dom(\r)\sle \FV(l)\moins\dom(\G)$,
\item $\thc r:U\g\r$.
\end{lst}


Because of the (conv) rule, the relation $\thc$ may be undecidable. On
the other hand, if we restrict the (conv) rule to a confluent and
strongly normalizing fragment of $\a$, then $\thc$ becomes decidable
(with an algorithm similar to the one for $\th$). This is quite
reasonable since, in practice, the symbols and the rules are often
added one after the other (or by groups, but the argument can be
generalized), thus confluence and strong normalization can be shown
incrementally.

For instance, let $(\cF,\cR)$ be a confluent and strongly normalizing
system, $f\notin\cF$ and $\cR_f$ be a set of rules defining $f$ and
whose symbols belong to $\cF'=\cF\cup\{f\}$. Then, $(\cF',\cR)$ is
also confluent and strongly normalizing. Thus, we can check that the
rules of $\cR_f$ satisfy the General Schema with the rule (conv)
restricted to the case where $T \ad_{\b\cR} T'$. This does not seem a
big restriction: it would be surprising that the typing of a rule
requires the use of the rule itself !


We now detail the case of $app~p~(cons~x~n~\ell)~n'~\ell' \a
cons~x~(n+n')~(app~n~\ell~n'~\ell')$. We take $stat_{app}=
lex(mul~x_2)$; $app >_\cF cons, +$; $cons >_\cF nat$ and $+>_\cF
s,0>_\cF nat$. We have already seen that this rule is well-formed. Let
us show that $\thc r:list(sn)$.

For applying (symb$^<$), we must show that $\thc\tau_{cons}:\st$,
$\thc x:nat$, $\thc n+n':nat$ and $\thc
app~n~\ell~n'~\ell':list(n+n')$. The first assertions follow from the
fact that the same judgements holds in $\th$ without using
$app$. Hence, we are left to prove the last assertion.

For applying (symb$^=$), we must show that $\thc\tau_{app}:\st$, $\thc
n:nat$, $\thc \ell:listn$, $\thc n':nat$, $\thc \ell':listn'$ and
$cons~x~n~\ell:list(sn) \tgt_\r \ell:listn$. The first assertions
follow from the fact that the same judgements hold in $\th$ without
using $app$. The last assertion has already been shown when proving
that the rule is well-formed.




\subsection{Strong normalization conditions}


\newcommand{\domB}{\dom^\B}

\begin{dfn}
\label{def-rew-sys}

  Let $\cG\sle\cF$. The {\em rewrite system} $(\cG,\cR_\cG)$ is:

\begin{lst}{\bu}
\item {\em first-order} if every rule of $\cR_\cG$ has an algebraic
right hand-side and, for all $g\in\cG$, either $g\in\FB$ or
$g:(\vx:\vT)C\vv$ with $C\in\CFB$ primitive.

\item {\em primitive} if all the rules of $\cR_\cG$ have a right
hand-side of the form $[\vx:\vT]g\vu$ with $g$ a symbol of $\cG$ or a
primitive constant predicate symbol.

\item {\em simple} if there is no critical pair between $\cR_\cG$
and $\cR$.

\item {\em small} if, for every rule $g\vl\a r\in \cR_\cG$, $\all
x\in\FVB(r)$, $\ex \ka_x$, $l_{\ka_x}=x$.

\item {\em positive} if, for every $g\in\cG$, for every rule $l\a r
\in\cR_\cG$, $\pos(g,r) \sle \pp(r)$.

\item {\em safe} if for every rule $(g\vl\a r,\G,\r)\in \cR_\cG$ with
  $g:(\vx:\vT)U$ and $\g=\vxl$:
\begin{lst}{--}
\item $\all x\in \FVB(\vT U)$, $x\g\r\in\domB(\G)$,
\item $\all x,x'\in\FVB(\vT U)$, $x\g\r=x'\g\r \A x=x'$.\footnote{All
  this means that $\g\r$ is an injection from $\FVB(\vT U)$ to
  $\domB(\G)$.}
\end{lst}
\end{lst}
\end{dfn}


\begin{dfn}[Strong normalization conditions]\hfill
\label{def-cond}

\begin{bfenumi}{A}
\setcounter{enumi}{-1}

\item All the rules are well-typed.

\item The relation $\a\,=\,\ar\cup\ab$ is confluent on $\cT$.

\item There exists an admissible inductive structure.

\item There exists a precedence $\cge$ on $\DFB$ which is compatible
with $\cR_\DFB$ and whose equivalence classes form a system which is
either:
\begin{lst}{}
\item [\bf(p)] primitive,
\item [\bf(q)] positive, small and simple,
\item [\bf(r)] computable, small and simple.
\end{lst}

\item There exists a partition $\cF_1 \uplus \cF_\w$ of $\DF$ ({\em
first-order} and {\em higher-order} symbols) such that:
\begin{bfenumalphai}
\item $(\cF_\w,\cR_\w)$ is computable,
\item $(\cF_\w,\cR_\w)$ is safe,
\item no symbol of $\cF_\w$ occurs in the rules of $\cR_1$,
\item $(\cF_1,\cR_1)$ is first-order,
\item if $\cR_\w \neq \vide$ then $(\cF_1,\cR_1)$ is non-duplicating,
\item $\a_{\cR_1}$ is strongly normalizing on first-order algebraic
terms.
\end{bfenumalphai}

\end{bfenumi}
\end{dfn}

The condition (A1) ensures, among other things, that $\b$ preserves
typing. This condition may seem difficult to fulfill since confluence
is often proved by using strong normalization and local confluence of
critical pairs \cite{nipkow91lics}.

We know that $\ab$ is confluent and that there is no critical pair
between $\cR$ and $\b$ since the left hand-sides of rules are
algebraic. M\"uller \cite{muller92ipl} showed that, in this case, if
$\ar$ is confluent and all the rules of $\cR$ are left-linear, then
$\ar\cup\ab$ is confluent. Thus, the possibility we have introduced of
linearizing some rules (substitution $\r$) appears to be very useful
(see Definition \ref{def-wt-rule}).

In the case of left-linear rules, and assuming that $\a_{\cR_1}$ is
strongly normalizing as it is required in (f), how can we prove that
$\a$ is confluent? In the case where $\a_{\cR_1}$ is non-duplicating
if $\cR_\w\neq\vide$, we show in Theorem \ref{thm-sn-rew} that
$\a_{\cR_1} \cup \a_{\cR_\w}$ is strongly normalizing. Therefore, it
suffices to check that the critical pairs of $\cR$ are confluent
(without using any $\b$-reduction).\\

In (A4), in the case where $\cR_\w\neq\vide$, we require that the
rules of $\cR_1$ are non-duplicating. Indeed, strong normalization is
not a modular property \cite{toyama87ipl}, even with confluent systems
\cite{drosten89thesis}. On the other hand, strong normalization is
modular for disjoint and non duplicating systems
\cite{rusinowitch87ipl}. Here, $\cR_1$ and $\cR_\w$ are not disjoint
but hierarchically defined: by (c), no symbol of $\cF_\w$ occurs in
the rules of $\cR_1$. In \cite{dershowitz94ctrs}, Dershowitz gathers
some results on the modularity of strong normalization for first-order
rewrite systems. It would be very interesting to study the modularity
of strong normalization in the case of higher-order rewriting and, in
particular, other conditions than non-duplication which, for example,
does not allow us to accept the following definition:

\begin{rewc}
0/y & 0\\
(s~x)/y & s((x-y)/y)\\[2mm]

0-y & 0\\
(s~x)-0 & s~x\\
(s~x)-(s~y) & x-y\\
\end{rewc}

\noindent
This system is a duplicating first-order system not satisfying the
General Schema: it can be put neither in $\cR_1$ nor in $\cR_\w$. Note
that Gim\'enez \cite{gimenez98icalp} has developped a termination
criterion for the Calculus of Inductive Constructions that accepts
this example.\\

In (A3), the smallness condition for computable and positive systems
is equivalent in the Calculus of Inductive Constructions to the
restriction of strong elimination to ``small'' inductive types, that
is, to the types whose constructors have no other predicate parameters
than the ones of the type. For example, the type $list$ of polymorphic
list is small since, in the type $(A:\st)A\A listA\A listA$ of its
constructor $cons$, $A$ is a parameter of $list$. On the other hand, a
type $T$ having a constructor $c$ of type $\st\A T$ is not small. So,
we cannot define a function $f$ of type $T\A\st$ with the rule
$f(c~A)\a A$. Such a rule is not small and does not form a primitive
system either. In some sense, primitive systems can always be
considered as small systems since they contain no projection and
primitive predicate symbols have no predicate argument. This
restriction is not only technical: elimination on big inductive types
may lead to logical inconsistencies \cite{coquand86lics}.\\

Finally, in (A4), the safeness condition for higher-order symbols
means that one cannot do matching or equality tests on predicate
arguments that are necessary for typing other arguments. In her
extension of HORPO \cite{jouannaud99lics} to the Calculus of
Constructions, Walukiewicz \cite{walukiewicz02jfp} requires a similar
condition. This has to be related to the fact that the polymorphism of
CC is essentially parametric, that is, a polymorphic function uses the
same algorithm at all types \cite{reynolds83ifip}. Girard already
demonstrated in \cite{girard71} that normalization fails if a
non-parametric operator $J:(A:\st)(B:\st)A\A B$ defined by $J~A~A~x\a
x$ is added to system F. See \cite{harper99ipl} for an analysis of
Girard's $J$ operator. On the other hand, the rule
$map~A~A~[x:A]x~\ell \a \ell$, which does not seem problematic, does
not satisfy the safeness condition either (note however that the left
hand-side if not algebraic).\\


We can now state our main result whose proof is the subject of Section
\ref{sec-correctness}:\\

\begin{center}
\fbox{\begin{minipage}{117mm}\u{\bf THEOREM:} If a CAC satisfies the
conditions of Definition \ref{def-cond} then its reduction relation
$\a= \ar\cup\ab$ preserves typing and is strongly normalizing.
\end{minipage}}\\[5mm]
\end{center}


In \cite{blanqui01thesis}, we prove that most of CIC can be encoded
into a CAC satisfying our conditions, and that our conditions can also
be applied to prove the cut-elimination property in Natural Deduction
Modulo \cite{dowek98types}. But let us give a more concrete example:

\begin{center}
\begin{tabular}{ccc}
\begin{rew}
\non\top & \bot\\
\non\bot & \top\\
\end{rew}
&
\begin{rew}
P\ou\top & \top\\
P\ou\bot & P\\
\end{rew}
&
\begin{rew}
P\et\top & P\\
P\et\bot & \bot\\
\end{rew}\\
\end{tabular}\\[2mm]
\end{center}

\begin{center}
\begin{tabular}{cc}
\begin{rew}
x+0 & x\\
0+x & x\\
x+(s~y) & s(x+y)\\
(s~x)+y & s(x+y)\\
(x+y)+z & x+(y+z)\\
\end{rew}
&
\begin{rew}
x\times 0 & 0\\
0\times x & 0\\
x\times (s~y) & (x\times y) + x\\
(s~0)\times x & x\\
x\times (s~0) & x\\
\end{rew}\\
\end{tabular}\\[2mm]

\begin{rew}
eq~A~0~0 & \top\\
eq~A~0~(s~x) & \bot\\
eq~A~(s~x)~0 & \bot\\
eq~A~(s~x)~(s~y) & eq~nat~x~y\\
\end{rew}

\begin{rul}{}
app~A~(nil~A')~\ell &\a& \ell\\
app~A~(cons~A'~x~\ell)~\ell' &\a& cons~A~x~(app~A~\ell~\ell')\\
app~A~(app~A'~\ell~\ell')~\ell'' &\a& app~A~\ell~(app~A~\ell'~\ell'')\\[2mm]

len~A~(nil~A') &\a& 0\\
len~A~(cons~A'~x~\ell) &\a& s~(len~A~\ell)\\
len~A~(app~A'~\ell~\ell') &\a& (len~A~\ell)+(len~A~\ell')\\[2mm]

in~A~x~(nil~A') &\a& \bot\\
in~A~x~(cons~A'~y~l) &\a& (eq~A~x~y) \ou (in~A~x~l)\\[2mm]

incl~A~(nil~A')~l &\a& \top\\
incl~A~(cons~A'~x~l)~l' &\a& (in~A~x~l') \et (incl~A~l~l')\\[2mm]

sub~A~(nil~A')~l &\a& \top\\
sub~A~(cons~A'~x~l)~(nil~A'') &\a& \bot\\
sub~A~(cons~A'~x~l)~(cons~A''~x'~l') &\a& ((eq~A~x~x') \et (sub~A~l~l'))\\
&& \ou (sub~A~(cons~A~x~l)~l')\\[2mm]

eq~L~(nil~A)~(nil~A') &\a& \top\\
eq~L~(nil~A)~(cons~A'~x~l) &\a& \bot\\
eq~L~(cons~A'~x~l)~(nil~A) &\a& \bot\\
eq~L~(cons~A~x~l)~(cons~A'~x'~l')
&\a& (eq~A~x~x') \et (eq~(list~A)~l~l')\\[2mm]
\end{rul}
\end{center}

This rewriting system is computable, simple, small, safe and confluent
(this can be automatically proved by CiME \cite{cime}). Since the
rules are left-linear, the combination with $\ab$ is also
confluent. Therefore, the conditions of strong normalization are
satisfied. For example, for the last rule, take $\G= A:\st, x:A, x':A,
\ell:listA, \ell':listA$ and $\r= \{A'\to A, L\to listA\}$. The rule
is well-formed ($cons(A',x',\ell'):L ~\tgt_\r~ x':A'$, \ldots) and
satisfies the General Schema ($\{cons(A,x,\ell):L,
cons(A',x',\ell'):L\} ~(\tgt_\r)\mul~ \{x:A, x':A'\}$ and
$\{\ell:listA, \ell':listA\}$).

However, the system lacks several important rules to get a complete
decision procedure for classical propositional tautologies (Figure
\ref{fig-hsiang} in Section \ref{sec-intro}) or other simplification
rules on the equality (Figure \ref{fig-int-ring} in Section
\ref{sec-intro}). To accept these rules, we must consider rewriting
modulo associativity and commutativity and get rid of the simplicity
conditions. Moreover, the distributivity rule $P\et(Q\oplus R)\a (P\et
Q)\oplus(P\et R)$ is not small. Rewriting modulo AC does not seem to
be a difficult extension, except perhaps in the case of
predicate-level rewriting. On the other hand, confluence, simplicity
and smallness seem difficult problems.\\


From strong normalization, we can deduce the decidability of the
typing relation, which is the essential property on which proof
assistants like Coq \cite{coq02} or LEGO \cite{lego92} are based.

\begin{thm}[Decidability of type-checking]
\label{thm-th-dec}

Let $\G$ be a valid environment and $T$ be $\B$ or a term typable in
$\G$. In a CAC satisfying the conditions of Definition \ref{def-cond},
checking whether a term $t$ is of type $T$ in $\G$ is decidable.
\end{thm}

\begin{prf}
  Since $\G$ is valid, it is possible to say whether $t$ is typable
  and, if so, it is possible to infer a type $T'$ for $t$. Since types
  are convertible, it suffices to check that $T$ and $T'$ have the
  same normal form. The reader is invited to look at
  \cite{coquand91book,barras99thesis} for more details.\cqfd
\end{prf}




\section{Correctness of the conditions}
\label{sec-correctness}

\newcommand{\SN}{\cS\cN}
\newcommand{\CR}{\cC\cR}
\newcommand{\WN}{\cW\cN}

Our strong normalization proof is based on Tait and Girard's method of
computability predicates and reducibility candidates
\cite{girard88book}. The idea is to interpret each type $T$ as a set
$\I{T}$ of strongly normalizable terms and to prove that every term of
type $T$ belongs to $\I{T}$. The reader not familiar with these
notions is invited to read the Chapter 3 of the Ph.D. thesis of Werner
\cite{werner94thesis} for an introduction, and the paper of Gallier
for a more detailed presentation \cite{gallier90book}.

It is worth noting several differences with previous strong
normalization proofs:

\begin{lst}{--}
\item The present proof is an important simplification of the proof
given in \cite{blanqui01thesis}, which uses candidates {\em \`a la}
Coquand and Gallier \cite{coquand90lf} where only well-typed terms are
considered. Here, candidates are made of well-typed and not well-typed
terms. This leads to simpler notations and less properties to be care
of.

\item In \cite{geuvers94types}, Geuvers uses candidates with possibly
not well-typed terms too. However, the way dependent types are
interpreted does not allow this proof to be extended to type-level
rewriting. Indeed, in this proof, dependencies are simply ignored but,
if one has a predicate symbol $F:nat\A\st$ defined by $F0\a nat$ and
$F(sn)\a nat\A nat$, then one expects $F0$ to be interpreted as $nat$,
and $F(sn)$ as $nat\A nat$.

\item In \cite{werner94thesis}, Werner uses candidates with (not
well-typed) pure $\la$-terms, that is, terms without type annotation
in abstractions, in order to deal with $\eta$-conversion, whose
combination with $\b$ is not confluent on annotated terms. As a
consequence, he has to define a translation from annotated terms to
pure terms that implies the strong normalization of annotated
terms. Here, we give a direct proof.
\end{lst}




\subsection{Reducibility candidates}

We denote by:

\begin{lst}{--}
\item $\SN$ the set of strongly normalizable terms,
\item $\WN$ the set of weakly normalizable terms,
\item $\CR$ the set of terms from which reductions are confluent.
\end{lst}


\begin{dfn}[Neutral terms]
A term $t$ is {\em neutral} if it is not of the following form:

\begin{lst}{--}
\item abstraction: $[x:T]u$,
\item partial application: $f\vt$ with $f\in\DF$ and $|\vt|<|\vl|$
for some rule $f\vl\a r\in\cR$,
\item constructor: $f\vt$ with $f:(\vx:\vT)C\vv$ and $C\in\CFB$.
\end{lst}

\noindent
Let $\cN$ be the set of neutral terms.
\end{dfn}

Note that, if $t$ is neutral, then $tu$ is neutral and not
head-reducible.


\begin{dfn}[Reducibility candidates]
We inductively define the set $\cR_t$ of the interpretations for the
terms of type $t$, the ordering $\le_t$ on $\cR_t$, the element
$\top_t\in\cR_t$, and the operation $\biget_t$ from the powerset of
$\cR_t$ to $\cR_t$ as follows. If $t\neq\B$ and $\G\not\th t:\B$ then:

\begin{lst}{--}
\item $\cR_t=\{\vide\}$, $\le_t=\sle$, $\top_t=\vide$ and
$\biget_t(\Re)=\top_t$.
\end{lst}

\noindent
Otherwise:

\begin{lst}{--}
\item $\cR_s$ is the set of all the subsets $R$ of $\cT$ such that:
\begin{bfenumii}{R}
\item $R\sle\SN$ (strong normalization).
\item If $t\in R$ then $\a\!\!(t)\sle R$ (stability by reduction).
\item If $t\in\cN$ and $\a\!\!(t)\sle R$ then $t\in R$ (neutral terms).
\end{bfenumii}
Furthermore, $\le_s=\sle$, $\top_s=\SN$, $\biget_s(\Re)=\bigcap\Re$ if
$\Re\neq\vide$, and $\biget_s(\vide)= \top_s$.

\item $\cR_{(x:U)K}$ is the set of functions $R$ from $\cT\times\cR_U$
to $\cR_K$ such that $R(u,S)=R(u',S)$ whenever $u\a u'$,
$\top_{(x:U)K}(u,S)=\top_K$, $\biget_{(x:U)K}(\Re)(u,S)=
\biget_K(\{R(u,S)~|~R\in\Re\})$, and $R\le_{(x:U)K} R'$ iff, for all
$(u,S)$, $R(u,S)\le_K R'(u,S)$.
\end{lst}
\end{dfn}


\begin{lem}
\label{lem-var-cand}

$\cV=\{x\vt\in\cT~|~x\in\cX,\vt\in\SN\}\neq\vide$ and, for all
$R\in\cR_s$, $\cV\sle R$.
\end{lem}

\begin{prf}
$\cV\neq\vide$ since $\cX\neq\vide$. Let $R\in\cR_s$. We prove that
$x\vt\in R$ by induction on $\vt$ with $\a\lex$ as well-founded
ordering ($\vt\in\SN$). Since $x\vt\in\cN$, it suffices to prove that
$\a\!\!(x\vt)\sle R$, which is the induction hypothesis.\cqfd
\end{prf}


\begin{lem}
\label{lem-cand-conv}

\begin{enumalphai}
\item If $T\CVG T'$ then $\cR_T=\cR_{T'}$.
\item If $\G\th T:s$ and $\t:\G\leadsto\D$ then $\cR_T=\cR_{T\t}$.
\end{enumalphai}
\end{lem}

\begin{prf}
\begin{enumalphai}
\item By induction on the size of $T$. If $\G\th T:\st$ then $\G\th
T':\st$ and $\cR_T=\{\vide\}=\cR_{T'}$. Assume now that $\G\th
T:\B$. If $T=\st$ then $T'=\st$ and $\cR_T=\cR_{T'}$. If $T=(x:U)K$
then $T'=(x:U')K'$ with $U\CVG U'$ and $K\CV[\G,x:U] K'$. By induction
hypothesis, $\cR_U=\cR_{U'}$ and $\cR_K=\cR_{K'}$. Therefore,
$\cR_T=\cR_{T'}$.
\item By induction on the size of $T$. If $\G\th T:\st$ then $\D\th
T\t:\st$ and $\cR_T=\{\vide\}=\cR_{T\t}$. Assume now that $\G\th
T:\B$. If $T=\st$, this is immediate. If $T=(x:U)K$ then
$T\t=(x:U\t)K\t$. By induction hypothesis, $\cR_U=\cR_{U\t}$ and
$\cR_K=\cR_{K\t}$. Therefore, $\cR_T=\cR_{T\t}$.\cqfd
\end{enumalphai}
\end{prf}


\begin{lem}[Completeness of the candidates lattice]
\label{lem-cand-lat-compl}

  $(\cR_t,\le_t)$ is a complete lattice with greatest element $\top_t$
  and the lower bound of $\Re\sle\cR_t$ given by $\biget_t(\Re)$.
\end{lem}

\begin{prf}
  It suffices to prove that $(\cR_t,\le_t)$ is a complete
  inf-semi-lattice and that $\top_t$ is its greatest element. One can
  easily check by induction on $t$ that $\le_t$ is an ordering ({\em
  i.e.} is reflexive, transitive and anti-symmetric), $\top_t$ is the
  greatest element of $\cR_t$, and $\biget_t(\Re)$ is the lower bound
  of $\Re\sle\cR_t$.\cqfd
\end{prf}


\begin{lem}[Smallest element]
\label{lem-bottom}

Let $\bot_0\!=\!\vide$ and
$\bot_{i+1}\!=\!\bot_i\cup\{t\in\cN~|\a\!\!(t)\sle\bot_i\}$. The set
$\bot_s\!=\!\bigcup\{\bot_i~|~i<\w\}$ is the smallest element of
$\cR_s$: $\bot_s=\bigcap\cR_s$.
\end{lem}

\begin{prf}
Let $R\in\cR_s$. We prove by induction on $i$ that $\bot_i\sle R$. For
$i=0$, this is immediate. Assume that $\bot_i\sle R$ and let
$t\in\bot_{i+1}\moins\bot_i$. We have $t\in\cN$ and $\a\!\!(t)\sle R$
by induction hypothesis. Therefore, by (R3), $t\in R$ and $\bot_s\sle
R$ for all $R\in\cR_s$. Thus, $\bot_s\sle\bigcap\cR_s$.

We now prove that $\bot_s\in\cR_s$, hence that $\bot_s=\bigcap\cR_s$.

\begin{bfenumi}{R}
\item We prove that $\bot_i\sle\SN$ by induction on $i$. For $i=0$,
this is immediate. Assume that $\bot_i\sle\SN$ and let
$t\in\bot_{i+1}\moins\bot_i$. We have $\a\!\!(t)\sle\SN$ by induction
hypothesis. Therefore, $t\in\SN$.

\item Let $t\in\bot_s$. Since $\bot_0=\vide$,
$t\in\bot_{i+1}\moins\bot_i$ for some $i$. So,
$\a\!\!(t)\sle\bot_i\sle\bot_s$.

\item Let $t\in\cN$ with $\a\!\!(t)\sle\bot_s$. Since $\a$ is assumed
to be finitely branching, ${\a\!\!(t)}=\{t_1,\ldots,t_n\}$. For all
$i$, there exists $k_i$ such that $t_i\in\bot_{k_i}$. Let $k$ be the
max of $\{k_1,\ldots,k_n\}$. We have $\a\!\!(t)\sle\bot_k$ and thus
$t\in\bot_{k+1}\sle\bot_s$.\cqfd
\end{bfenumi}
\end{prf}





\subsection{Interpretation schema}
\label{subsec-schema-int}

The interpretation $\I{t}$ of a term $t$ is defined by using a {\em
candidate assignment} $\xi$ for the free variables and an
interpretation $I$ for the predicate symbols. The interpretation of
constant predicate symbols will de defined in Section
\ref{subsec-int-const}, and the interpretation of defined predicate
symbols in Section \ref{subsec-int-def}.


\begin{dfn}[Interpretation schema]
\label{def-schema-int}

A {\em candidate assignment} is a function $\xi$ from $\cX$ to
$\bigcup \,\{\cR_t ~|~ t\in\cT\}$. A candidate assignment $\xi$ {\em
validates} an environment $\G$ or is a {\em $\G$-assignment}, written
$\xi\models\G$, if, for all $x\in\dom(\G)$, $x\xi\in \cR_{x\G}$. An
{\em interpretation} of a symbol $f$ is an element of $\cR_\tf$. An
{\em interpretation} of a set $\cG$ of symbols is a function which, to
each symbol $g\in\cG$, associates an interpretation of $g$.

The {\em interpretation} of $t$ w.r.t. a candidate assignment $\xi$,
an interpretation $I$ and a substitution $\t$, is defined by induction
on $t$ as follows:

\begin{lst}{\bu}
\item $\I{t}^I_{\xi,\t}= \top_t$ if $t$ is an object or a sort,
\item $\I{f}^I_{\xi,\t}= I_f$,
\item $\I{x}^I_{\xi,\t}= x\xi$,
\item $\I{(x:U)V}^I_{\xi,\t}= \{t\in\cT~|~ \all u\in\I{U}^I_{\xi,\t},
\all S\in\cR_U, tu\in\I{V}^I_{\xi_x^S,\t_x^u}\}$,
\item $\I{[x:U]v}^I_{\xi,\t}(u,S)= \I{v}^I_{\xi_x^S,\t_x^u}$,
\item $\I{tu}^I_{\xi,\t}= \I{t}^I_{\xi,\t}(u\t,\I{u}^I_{\xi,\t})$,
\end{lst}

\noindent
where $\t_x^u=\t\cup\xu$ and $\xi_x^S=\xi\cup\xS$. In the case where
$\G\th t:s$, the elements of $\I{t}^I_{\xi,\t}$ are called {\em
computable}. A substitution $\t$ is {\em adapted} to a $\G$-assignment
$\xi$ if $\dom(\t)\sle\dom(\G)$ and, for all $x\in\dom(\t)$, $x\t\in
\I{x\G}^I_{\xi,\t}$. A pair $(\xi,\t)$ is {\em $\G$-valid}, written
$\xi,\t\models\G$, if $\xi\models\G$ and $\t$ is adapted to $\xi$.
\end{dfn}

After Lemma \ref{lem-var-cand}, the identity substitution is adapted
to any $\G$-candidate assignment.


\begin{lem}[Correctness of the interpretation schema]
\label{lem-cor-schema-int}

If $\G\th t:T$ and $\xi\models\G$ then
$\I{t}^I_{\xi,\t}\in\cR_T$. Moreover, if $\t\a\t'$ then
$\I{t}^I_{\xi,\t}=\I{t}^I_{\xi,\t'}$.
\end{lem}

\begin{prf}
By induction on $\G\th t:T$.

\begin{lst}{}
\item [\bf(ax)] $\I{\st}^I_{\xi,\t}=\top_\st=\SN\in\cR_\B$ and
$\I{\st}^I_{\xi,\t}$ does not depend on $\t$.

\item [\bf(symb)] $\I{f}^I_{\xi,\t}=I_f\in\cR_\tf$ by assumption on
$I$ and $\I{f}^I_{\xi,\t}$ does not depend on $\t$.

\item [\bf(var)] $\I{x}^I_{\xi,\t}$ does not depend on $\t$. Now,
if $x\in\Xs$ then $\I{x}^I_{\xi,\t}= \vide\in\cR_T=
\{\vide\}$. Otherwise, $\I{x}^I_{\xi,\t}=x\xi\in\cR_{T}$ since
$\xi\models\G,x:T$.

\item [\bf(weak)] By induction hypothesis.

\item [\bf(prod)] $R=\I{(x:U)V}^I_{\xi,\t}= \{t\in\cT~|~ \all
u\in\I{U}^I_{\xi,\t}, \all S\in\cR_U, tu\in
\I{V}^I_{\xi_x^S,\t_x^u}\}\in\cR_s$ if it satisfies the properties
(R1) to (R3):

\begin{bfenumii}{R}
\item Strong normalization. Let $t\in R$. By induction hypothesis,
$\I{U}^I_{\xi,\t}\in\cR_{s'}$ for some $s'$, and
$\I{V}^I_{\xi_x^S,\t_x^u}\in\cR_s$. Therefore,
$\cX\sle\I{U}^I_{\xi,\t}$ and $\I{V}^I_{\xi_x^S,\t_x^u}\sle\SN$. Take
$u=x\in\cX$. Then, $tx\in\I{V}^I_{\xi_x^S,\t}$ and $t\in\SN$.

\item Stability by reduction. Let $t\in R$ and $t'\in{\a\!\!(t)}$. Let
$u\in\I{U}^I_{\xi,\t}$ and $S\in\cR_U$. Then,
$tu\in\I{V}^I_{\xi_x^S,\t_x^u}$ which, by induction hypothesis, is
stable by reduction. Therefore, since $t'u\in {\a\!\!(tu)}$,
$t'u\in\I{V}^I_{\xi_x^S,\t_x^u}$ and $t'\in R$.

\item Neutral terms. Let $t$ be a neutral term such that
$\a\!\!(t)\sle R$. Let $u\in\I{U}^I_{\xi,\t}$ and $S\in\cR_U$. Since
$t$ is neutral, $tu$ is neutral and, by induction hypothesis, $tu\in
R'=\I{V}^I_{\xi_x^S,\t_x^u}$ if $\a\!\!(tu)\sle R'$. We prove it by
induction on $u$ with $\a$ as well-founded ordering ($u\in\SN$ by
induction hypothesis). Since $t$ is neutral, $tu$ is not
head-reducible and a reduct of $tu$ is either of the form $t'u$ with
$t'\in{\a\!\!(t)}$, or of the form $tu'$ with $u'\in{\a\!\!(u)}$. In
the former case, $t'u\in R'$ by assumption. In the latter case, we
conclude by induction hypothesis.
\end{bfenumii}

Assume now that $\t\a\t'$. Let $R'=\I{(x:U)V}^I_{\xi,\t'}=
\{t\in\cT|\all u\in\I{U}^I_{\xi,\t'}, \all S\in\cR_U,
tu\in\I{V}^I_{\xi_x^S,{\t'}_x^u}\}$. By induction hypothesis,
$\I{U}^I_{\xi,\t'}= \I{U}^I_{\xi,\t}$ and
$\I{V}^I_{\xi_x^S,{\t'}_x^u}= \I{V}^I_{\xi_x^S,\t_x^u}$. Therefore,
$R'=R$.

\item [\bf(abs)] Let $R=\I{[x:U]v}^I_{\xi,\t}$. $R(u,S)=
\I{v}^I_{\xi_x^S,\t_x^u}$. By induction hypothesis,
$R(u,S)\in\cR_V$. Assume now that $u\a u'$. Then, $R(u',S)=
\I{v}^I_{\xi_x^S,\t_x^{u'}}$. By induction hypothesis,
$\I{v}^I_{\xi_x^S,\t_x^{u'}}= \I{v}^I_{\xi_x^S,\t_x^u}$. Therefore,
$R\in \cR_{(x:U)V}$. Assume now that $\t\a\t'$. Let
$R'=\I{[x:U]v}^I_{\xi,\t'}$. $R'(u,S)=
\I{v}^I_{\xi_x^S,{\t'}_x^u}$. By induction hypothesis,
$R'(u,S)=R(u,S)$. Therefore, $R=R'$.

\item [\bf(app)] Let $R=\I{tu}^I_{\xi,\t}=
\I{t}^I_{\xi,\t}(u\t,\I{u}^I_{\xi,\t})$. By induction hypothesis,
$\I{t}^I_{\xi,\t}\in\cR_{(x:U)V}$ and
$\I{u}^I_{\xi,\t}\in\cR_U$. Therefore, $\I{tu}^I_{\xi,\t}\in
\cR_V=\cR_{V\xu}$ by Lemma \ref{lem-cand-conv}. Assume now that
$\t\a\t'$. Then, $R'=\I{tu}^I_{\xi,\t'}=
\I{t}^I_{\xi,\t'}(u\t',\I{u}^I_{\xi,\t'})$. By induction hypothesis,
$\I{t}^I_{\xi,\t'}= \I{t}^I_{\xi,\t}$ and $\I{u}^I_{\xi,\t'}=
\I{u}^I_{\xi,\t}$. Finally, since $\I{t}^I_{\xi,\t}$ is stable by
reduction and $u\t\a^* u\t'$, we have $R=R'$.

\item [\bf(conv)] By induction hypothesis since, by Lemma
\ref{lem-cand-conv}, $\cR_T=\cR_{T'}$.\cqfd
\end{lst}
\end{prf}


\begin{lem}
\label{lem-int-dep}

  Let $I$ and $I'$ be two interpretations equal on the predicate
  symbols occurring in $t$, $\xi$ and $\xi'$ be two candidate
  assignments equal on the predicate variables free in $t$, and $\t$
  and $\t'$ be two substitutions equal on the variables free in
  $t$. If $\G\th t:T$ and $\xi\models\G$ then $\I{t}^{I'}_{\xi',\t'}=
  \I{t}^I_{\xi,\t}$.
\end{lem}

\begin{prf}
By induction on $t$.
\end{prf}


\begin{lem}[Candidate substitution]
\label{lem-cand-subs}

If $\G\th t:T$, $\s:\G\leadsto\D$ and $\xi\models\D$ then, for all
$\t$, $\I{t\s}^I_{\xi,\t}= \I{t}^I_{\xi',\s\t}$ with $x\xi'=
\I{x\s}^I_{\xi,\t}$ and $\xi'\models\G$.
\end{lem}

\begin{prf}
We first check that $\xi'\models\G$. Let $x\in\dom(\G)$. $x\xi'=
\I{x\s}_{\xi,\t}$. By Lemma \ref{lem-cor-schema-int},
$x\xi'\in\cR_{x\G\s}$ since $\D\th x\s:x\G\s$ and $\xi\models\D$. By
Lemma \ref{lem-cand-conv}, $\cR_{x\G\s}=\cR_{x\G}$ since $\G\th
x\G:s_x$ and $\s:\G\leadsto\D$. We now prove the lemma by induction on
$t$. If $t$ is an object then $t\s$ is an object too and
$\I{t\s}^I_{\xi,\t}= \vide= \I{t}^I_{\xi',\s\t}$. If $t$ is not an
object then $t\s$ is not an object either. We proceed by case on $t$:

\begin{lst}{\bu}
\item $\I{s\s}^I_{\xi,\t}= \top_s= \I{s}^I_{\xi',\s\t}$.

\item $\I{f\s}^I_{\xi,\t}= I_f= \I{f}^I_{\xi',\s\t}$.

\item $\I{x\s}^I_{\xi,\t}= x\xi'= \I{x}^I_{\xi',\s\t}$.

\item Let $R= \I{(x:U\s)V\s}^I_{\xi,\t}= \{t\in\cT~|~ \all
u\in\I{U\s}^I_{\xi,\t}, \all S\in\cR_{U\s}=\cR_U$, $tu\in
\I{V\s}^I_{\xi_x^S,\t_x^u}\}$ and $R'= \I{(x:U)V}^I_{\xi',\s\t}=
\{t\in\cT~|~ \all u\in\I{U}^I_{\xi',\s\t}, \all S\in\cR_U,
tu\in\I{V}^I_{{\xi'}_x^S,(\s\t)_x^u}\}$. By induction hypothesis,
$\I{U\s}^I_{\xi,\t}= \I{U}^I_{\xi',\s\t}$ and
$\I{V\s}^I_{\xi_x^S,\t_x^u}= \I{V}^I_{\xi'',\s(\t_x^u)}$ with $y\xi''=
\I{y\s}_{\xi_x^S,\t_x^u}$. Since $\s(\t_x^u)=(\s\t)_x^u$ ($x\notin
\dom(\s)\cup \dom(\t)\cup \FV(\s)$) and $\xi''={\xi'}_x^S$
($x\notin\dom(\s)\cup \FV(\s)$), we have $R=R'$.

\item Let $R=\I{[x:U\s]v\s}^I_{\xi,\t}$ and
$R'=\I{[x:U]v}^I_{\xi',\s\t}$. By Lemma \ref{lem-cand-conv}, $R$ and
$R'$ have the same domain $\cT\times\cR_U$ and the same codomain
$\cR_V$. Moreover, $R(u,S)= \I{v\s}^I_{\xi_x^S,\t_x^u}$ and $R'(u,S)=
\I{v}^I_{{\xi'}_x^S,(\s\t)_x^u}$. By induction hypothesis, $R(u,S)=
\I{v}^I_{\xi'',\s(\t_x^u)}$ with $y\xi''=
\I{y\s}_{\xi_x^S,\t_x^u}$. Since $\s(\t_x^u)=(\s\t)_x^u$ and
$\xi''={\xi'}_x^S$, we have $R=R'$.

\item Let $R=\I{t\s u\s}^I_{\xi,\t}=
\I{t\s}^I_{\xi,\t}(u\s\t,\I{u\s}^I_{\xi,\t})$ and
$R'=\I{tu}^I_{\xi',\s\t}=
\I{t}^I_{\xi',\s\t}(u\s\t,\I{u}^I_{\xi',\s\t})$. By induction
hypothesis, $\I{t\s}^I_{\xi,\t}= \I{t}^I_{\xi',\s\t}$ and
$\I{u\s}^I_{\xi,\t}= \I{u}^I_{\xi',\s\t}$. Therefore, $R=R'$.\cqfd
\end{lst}
\end{prf}




\subsection{Interpretation of constant predicate symbols}
\label{subsec-int-const}

Like Mendler \cite{mendler87thesis} or Werner \cite{werner94thesis},
we define the interpretation of constant predicate symbols as the
fixpoint of some monotonic function on a complete lattice. The
monotonicity is ensured by the positivity conditions of admissible
inductive structures (Definition \ref{def-adm-ind-str}). The main
difference with these works is that we have a more general notion of
constructor since it includes any function symbol whose output type is
a constant predicate symbol. This allows us to define functions and
predicates by matching not only on constant constructors but also on
defined symbols.


\begin{dfn}[Monotonic interpretation]
\label{def-mon-int}
Let $I$ be an interpretation of $C:(\vx:\vT)\st$,
$\va=(\vt,\vS)$\footnote{For simplicity, we write $(\vt,\vS)$ instead
of $(t_1,S_1),\ldots,(t_n,S_n)$.} and $\va'=(\vt',\vS')$ be arguments
of $I$. Let $\va\le_i\va'$ iff $\vt=\vt'$, $S_i\le S_i'$ and, for all
$j\neq i$, $S_j=S_j'$. We say that $I$ is {\em monotonic} if, for all
$i\in\mon(C)$, $\va\le_i\va'\A I(\va)\le I(\va')$.
\end{dfn}

We define the monotonic interpretation $I$ of $\CFB$ by induction on
$>_\cC$ {\bf(A2)}. Let $C\in\CFB$ and assume that we already defined a
monotonic interpretation $K$ for every symbol smaller than $C$. Let
$\cI$ (resp. $\cI^m$) be the set of (resp. monotonic) interpretations
of $\{D\in\CFB~|~D=_\cC C\}$, and $\le$ be the relation on $\cI$
defined by $I \le I'$ iff, for all $D=_\cC C$, $I_D \le_\tD I'_D$. For
simplicity, we denote $\I{t}^{K\cup I}$ by $\I{t}^I$.


\begin{lem}
\label{lem-mon-int-lat-comp}
$(\cI^m,\le)$ is a complete lattice.
\end{lem}

\begin{prf}
  First of all, $\le$ is an ordering since, for all $D =_\cC C$,
  $\le_\tD$ is an ordering.
  
  The function $I^\top$ defined by $I^\top_D= \top_\tD$ is the
  greatest element of $\cI$. We show that it belongs to $\cI^m$. Let
  $D=_\cC C$ with $D:(\vx:\vT)U$, $i\in\mon(D)$ and
  $\va\le_i\va'$. Then, $I^\top_D(\va)= \top_U= I^\top_D(\va')$.
  
  We now show that every part of $\cI^m$ has an inf. Let $\Im\sle
  \cI^m$ and $I^\et$ be the function defined by $I^\et_D=
  \biget_\tD(\Re_D)$ where $\Re_D= \{I_D~|~I\in\Im\}$. We show that
  $I^\et\in\cI^m$. Let $D=_\cC C$ with $D:(\vx:\vT)U$, $i\in\mon(D)$
  and $\va\le_i\va'$. Then, $I^\et_D(\va)= \biget_U \{I_D(\va) ~|~
  I\in\Im\}$ and $I^\et_D(\va')= \biget_U \{I_D(\va') ~|~
  I\in\Im\}$. Since each $I_D$ is monotonic, $I_D(\va)\le_U
  I_D(\va')$. Therefore, $I^\et_D\le_\tD I^\et_D$.
  
  We are left to show that $I^\et$ is the inf of $\Im$. For all
  $I\in\Im$, $I^\et\le I$ since, for all $D=_\cC C$, $I^\et_D$ is the
  inf of $\Re_D$. Assume now that there exists $I'\in \cI^m$ such
  that, for all $I\in\Im$, $I'\le I$. Then $I'\le I^\et$ since
  $I^\et_D$ is the inf of $\Re_D$.\cqfd
\end{prf}


\begin{dfn}[Interpretation of constant predicate symbols]
\label{def-int-const}

Let $\vphi$ be the function which, to $I\in\cI^m$, associates the
interpretation $\vphi^I\in\cI^m$ such that $\vphi^I_D(\vt,\vS)$ is the
set of terms $u\in\SN$ such that if $u$ reduces to $f\vu$ with
$f:(\vy:\vU)D\vv$ and $|\vu|=|\vy|$ then, for all $j\in\acc(f)$,
$u_j\in \I{U_j}^I_{\xi,\t}$ with $\t=\vyu$ and $y\xi= S_{\io_y}$. We
show hereafter that $\vphi$ is monotonic. Therefore, we can take
$I=\lfp(\vphi)$, the least fixpoint of $\vphi$.
\end{dfn}

Since $\vphi^I_D(\vt,\vS)$ does not depend on $\vt$, we may sometimes
write $I_D(\vS)$ instead of $I_D(\vt,\vS)$. The aim of this definition
is to ensure the correctness of the accessibility relations (Lemma
\ref{lem-cor-acc}): if $f\vu$ is computable then each accessible $u_j$
is computable. This will allow us to ensure the computability of the
variables of the left hand-side of a rule if the arguments of the left
hand-side are computable, and thus the computability of the right
hand-sides that belong to the computability closure.


\begin{lem}
\label{lem-cor-int-const}

  $\vphi^I$ is a well defined interpretation.
\end{lem}

\begin{prf}
  We first prove that $\vphi^I$ is well defined. The existence of
  $\io_y$ is the hypothesis {\bf(I6)}. The interpretations necessary
  for computing $\I{U_j}_{\xi,\t}$ are all well defined. The
  interpretation of constant predicate symbols smaller than $D$ is
  $K$. The interpretation of constant predicate symbols equivalent to
  $D$ is $I$. By {\bf(I4)} and {\bf(I5)}, constant predicate symbols
  greater than $D$ and defined predicate symbols do not occur in
  $U_j$. Finally, we must make sure that $\xi\models\G$ where $\G$ is
  the environment made of the declarations $y_i:U_i$ such that
  $y_i\in\FVB(U_j)$ for some $j$. Let $y\in\dom(\G)$. We must prove
  that $y\xi\in\cR_{y\G}$. Assume that $D:(\vx:\vT)U$. Then,
  $y\xi=S_{\io_y}\in \cR_{T_{\io_y}}$. Let $\g=\{\vx\to\vv\}$. Since
  $\g:\G_D\leadsto\G_f$, by Lemma \ref{lem-cand-conv},
  $\cR_{T_{\io_y}}=\cR_{T_{\io_y}\g}$. By {\bf(I6)},
  $v_{\io_y}=y$. So, $\G_f\th y:T_{\io_y}\g$ and $T_{\io_y}\g\CV[\G_f]
  y\G$. Therefore, by Lemma \ref{lem-cand-conv}, $\cR_{T_{\io_y}\g}=
  \cR_{y\G}$ and $y\xi\in \cR_{y\G}$.
  
  We now prove that $\vphi^I_D \in \cR_\tD$. It is clearly stable by
  reduction since it does not depend on $\vt$. Furthermore, $R=
  \vphi^I_D(\vt,\vS)$ satisfies the properties (R1) to (R3):

\begin{bfenumi}{R}
\item Strong normalization. By definition.
  
\item Stability by reduction. Let $u\in R$ and
$u'\in{\a\!\!(u)}$. Since $u\in\SN$, $u'\in \SN$. Assume furthermore
that $u'\a^* f\vu$ with $f:(\vy:\vU)D\vv$. Then, $u\a^*
f\vu$. Therefore, for all $j\in\acc(f)$, $u_j\in \I{U_j}_{\xi,\t}$ and
$u'\in R$.
  
\item Neutral terms. Let $u$ be a neutral term such that
$\a\!\!(u)\sle R$. Then, $u\in\SN$. Assume now that $u\a^* f\vu$ with
$f:(\vy:\vU)D\vv$. Since $u$ is neutral, $u\neq f\vu$ and there exists
$u'\in{\a\!\!(u)}$ such that $u'\a^* f\vu$. Therefore, for all
$j\in\acc(f)$, $u_j\in\I{U_j}_{\xi,\t}$ and $u\in R$.\cqfd
\end{bfenumi}
\end{prf}


\begin{lem}
\label{lem-xi-mon}

Let $\le^+=\le$, $\le^-=\ge$ and $\xi\le_x\xi'$ iff $x\xi\le x\xi'$
and, for all $y\neq x$, $y\xi=y\xi'$. If $I$ is monotonic,
$\xi\le_x\xi'$, $\pos(x,t)\sle\pd(t)$, $\G\th t:T$ and
$\xi,\xi'\models\G$ then $\I{t}^I_{\xi,\t}\le^\d \I{t}^I_{\xi',\t}$.
\end{lem}

\begin{prf}
By induction on $t$.

\begin{lst}{\bu}
\item $\I{s}^I_{\xi,\t}= \top_s= \I{s}^I_{\xi',\t}$.

\item $\I{x}^I_{\xi,\t}= x\xi\le x\xi'= \I{x}^I_{\xi',\t}$ and $\d=+$
necessarily.

\item $\I{y}^I_{\xi,\t}= y\xi= y\xi'= \I{y}^I_{\xi',\t}$ ($y\neq x$).

\item Let $R=\I{F\vt}^I_{\xi,\t}$ and
$R'=\I{F\vt}^I_{\xi',\t}$. $R=I_F(\va)$ with
$a_i=(t_i\t,\I{t_i}^I_{\xi,\t})$ and $R'=I_F(\va')$ with
$a_i'=(t_i\t,\I{t_i}^I_{\xi',\t})$. $\pd(F\vt)=
\{1^{|\vt|}~|~\d=+\}\cup\, \bigcup \{1^{|\vt|-i}2.\pd(t_i)~|~
i\in\mon(F)\}$. If $i\in\mon(F)$ then $\pos(x,t_i)\sle\pd(t_i)$ and,
by induction hypothesis, $\I{t_i}^I_{\xi,\t}\le^\d
\I{t_i}^I_{\xi',\t}$. Otherwise, $\pos(x,t_i)=\vide$ and
$\I{t_i}^I_{\xi,\t}= \I{t_i}^I_{\xi',\t}$. Therefore, in both cases,
$R\le^\d R'$ since $I_F$ is monotonic.

\item Let $R=\I{(x:U)V}^I_{\xi,\t}$ and
$R'=\I{(x:U)V}^I_{\xi',\t}$. $R=\{t\in\cT~|~ \all u\in
\I{U}^I_{\xi,\t}, \all S\in\cR_U, tu\in
\I{V}^I_{\xi_x^S,\t_x^u}\}$. $R'= \{t\in\cT~|~ \all u\in
\I{U}^I_{\xi',\t}, \all S\in\cR_U, tu\in
\I{V}^I_{{\xi'}_x^S,\t_x^u}\}$. Since $\pd((x:U)V)= 1.\pmd(U)\cup
2.\pd(V)$, $\pos(x,U)$ $\sle\pmd(U)$ and
$\pos(x,V)\sle\pd(V)$. Therefore, by induction hypothesis,
$\I{U}^I_{\xi,\t}\le^{-\d} \I{U}^I_{\xi',\t}$ and
$\I{V}^I_{\xi_x^S,\t_x^u}\le^\d \I{V}^I_{{\xi'}_x^S,\t_x^u}$. So,
$R\le^\d R'$. Indeed, if $\d=+$, $t\in R$ and $u\in
\I{U}^I_{\xi',\t}\sle \I{U}^I_{\xi,\t}$ then $tu\in
\I{V}^I_{\xi_x^S,\t_x^u}\sle \I{V}^I_{{\xi'}_x^S,\t_x^u}$ and $t\in
R'$. If $\d=-$, $t\in R'$ and $u\in \I{U}^I_{\xi,\t}\sle
\I{U}^I_{\xi',\t}$ then $tu\in \I{V}^I_{{\xi'}_x^S,\t_x^u}\sle
\I{V}^I_{\xi_x^S,\t_x^u}$ and $t\in R$.

\item Let $R=\I{[x:U]v}^I_{\xi,\t}$ and
$R'=\I{[x:U]v}^I_{\xi',\t}$. $R$ and $R'$ have the same domain
$\cT\times\cR_U$ and the same codomain
$\cR_V$. $R(u,S)=\I{v}^I_{\xi_x^S,\t_x^u}$ and
$R'(u,S)=\I{v}^I_{{\xi'}_x^S,\t_x^u}$. Since $\pd([x:U]v)= 2.\pd(v)$,
$\pos(x,v)\sle\pd(v)$. Therefore, by induction hypothesis,
$R(u,S)\le^\d R'(u,S)$ and $R\le^\d R'$.

\item Let $R=\I{tu}^I_{\xi,\t}$ and $R'=\I{tu}^I_{\xi',\t}$ ($t\neq
  f\vt$). $R=\I{t}^I_{\xi,\t}(u\t,S)$ with
  $S=\I{u}^I_{\xi,\t}$. $R'=\I{t}^I_{\xi',\t}(u\t,S')$ with
  $S'=\I{u}^I_{\xi',\t}$. Since $\pd(tu)= 1.\pd(t)$,
  $\pos(x,t)\sle\pd(t)$ and $\pos(x,u)=\vide$. Therefore, $S=S'$ and,
  by induction hypothesis, $\I{t}^I_{\xi,\t}\le^\d
  \I{t}^I_{\xi',\t}$. So, $R\le^\d R'$.\cqfd
\end{lst}
\end{prf}


\begin{lem}
\label{lem-int-const-mon-arg-ind}

$\vphi^I$ is monotonic.
\end{lem}

\begin{prf}
  Let $D=_\cC C$ with $D:(\vx:\vT)U$, $i\in\mon(D)$ and $\va\le_i\va'$
  with $\va=(\vt,\vS)$ and $\va'=(\vt,\vS')$. We have to show that
  $\vphi^I_D(\va)\sle \vphi^I_D(\va')$. Let $u\in \vphi^I_D(\va)$. We
  prove that $u\in \vphi^I_D(\va')$. First, we have $u\in\SN$. Assume
  now that $u$ reduces to $f\vu$ with $f:(\vy:\vU)D\vv$. Let
  $j\in\acc(f)$. We have to prove that $u_j\in \I{U_j}_{\xi',\t}$ with
  $\t=\vyu$ and, for all $y\in \FVB(U_j)$, $y\xi'= S'_{\io_y}$. Since
  $u\in \vphi^I_D(\va)$, we have $u_j\in \I{U_j}_{\xi,\t}$ with, for
  all $y\in\FVB(U_j)$, $y\xi= S_{\io_y}$. If, for all $y\in\FVB(U_j)$,
  $\io_y\neq i$, then $\xi$ and $\xi'$ are equal on
  $\FVB(U_j)$. Therefore, $\I{U_j}_{\xi,\t}= \I{U_j}_{\xi',\t}$ and
  $u_j\in \I{U_j}_{\xi',\t}$. If there exists $y\in\FVB(U_j)$ such
  that $\io_y=i$ then $\xi\le_y \xi'$. By {\bf(I2)}, $\pos(y,U_j)\sle
  \pp(U_j)$. Therefore, by Lemma \ref{lem-xi-mon}, $\vphi^I_D(\va)\sle
  \vphi^I_D(\va')$ and $u_j\in \I{U_j}_{\xi',\t}$.\cqfd
\end{prf}


\begin{lem}
\label{lem-I-mon}

Let $I\le_F I'$ iff $I_F\le I'_F$ and, for all $G\neq F$,
$I_G=I'_G$. If $I$ is monotonic, $I\le_F I'$, $\pos(F,t)\sle \pd(t)$,
$\G\th t:T$ and $\xi\models\G$ then $\I{t}^I_{\xi,\t}\le^\d
\I{t}^{I'}_{\xi,\t}$.
\end{lem}

\begin{prf}
By induction on $t$.

\begin{lst}{\bu}
\item $\I{s}^I_{\xi,\t}= \top_s= \I{s}^{I'}_{\xi,\t}$.

\item $\I{x}^I_{\xi,\t}= x\xi= \I{x}^{I'}_{\xi,\t}$.

\item Let $R=\I{G\vt}^I_{\xi,\t}$ and
$R'=\I{G\vt}^{I'}_{\xi,\t}$. $R=I_G(\va)$ with
$a_i=(t_i\t,\I{t_i}^I_{\xi,\t})$. $R'=I'_G(\va')$ with
$a_i'=(t_i\t,\I{t_i}^{I'}_{\xi,\t})$. $\pd(G\vt)=
\{1^{|\vt|}~|~\d=+\}\cup \,\bigcup \{1^{|\vt|-i}2.\pd(t_i)~|~
i\in\mon(G)\}$. If $i\in\mon(G)$ then $\pos(F,t_i)\sle\pd(t_i)$ and,
by induction hypothesis, $\I{t_i}^I_{\xi,\t}\le^\d
\I{t_i}^{I'}_{\xi,\t}$. Otherwise, $\pos(F,t_i)=\vide$ and
$\I{t_i}^I_{\xi,\t}= \I{t_i}^{I'}_{\xi,\t}$. Therefore,
$I_G(\va)\le^\d I_G(\va')$ since $I_G$ is monotonic. Now, if $G=F$
then $\d=+$ and $I_G(\va)\le I_G(\va')= I_F(\va')\le I'_F(\va')=
I'_G(\va')$. Otherwise, $I_G(\va)\le^\d I_G(\va')= I'_G(\va')$.

\item Let $R=\I{(x:U)V}^I_{\xi,\t}$ and
$R'=\I{(x:U)V}^{I'}_{\xi,\t}$. $R=\{t\in\cT~|~ \all u\in
\I{U}^I_{\xi,\t}, \all S\in\cR_U, tu\in
\I{V}^I_{\xi_x^S,\t_x^u}\}$ and $R'=\{t\in\cT~|~ \all u\in
\I{U}^{I'}_{\xi,\t}, \all S\in\cR_U, tu\in
\I{V}^{I'}_{\xi_x^S,\t_x^u}\}$. Since $\pd((x:U)V)=
1.\pmd(U)\cup 2.\pd(V)$, $\pos(F,U)\sle\pmd(U)$ and
$\pos(F,V)\sle\pd(V)$. Therefore, by induction hypothesis,
$\I{U}^I_{\xi,\t}\le^{-\d} \I{U}^{I'}_{\xi,\t}$ and
$\I{V}^I_{\xi_x^S,\t_x^u}\le^\d
\I{V}^{I'}_{\xi_x^S,\t_x^u}$. So, $\I{t}^I_{\xi,\t}\le^\d
\I{t}^{I'}_{\xi,\t}$. Indeed, if $\d=+$, $t\in R$ and $u\in
\I{U}^{I'}_{\xi,\t}\sle \I{U}^I_{\xi,\t}$ then $tu\in
\I{V}^I_{\xi_x^S,\t_x^u}\sle \I{V}^{I'}_{\xi_x^S,\t_x^u}$
and $t\in R'$. If $\d=-$, $t\in R'$ and $u\in \I{U}^I_{\xi,\t}\sle
\I{U}^{I'}_{\xi,\t}$ then $tu\in \I{V}^{I'}_{\xi_x^S,\t_x^u}\sle
\I{V}^I_{\xi_x^S,\t_x^u}$ and $t\in R$.

\item Let $R=\I{[x:U]v}^I_{\xi,\t}$ and
$R'=\I{[x:U]v}^{I'}_{\xi,\t}$. $R$ and $R'$ have the same domain
$\cT\times\cR_U$ and same codomain $\cR_V$. $R(u,S)=
\I{v}^I_{\xi_x^S,\t_x^u}$ and $R'(u,S)=
\I{v}^{I'}_{\xi_x^S,\t_x^u}$. Since $\pd([x:U]v)= 2.\pd(v)$,
$\pos(F,v)\sle\pd(v)$. Therefore, by induction hypothesis,
$R(u,S)\le^\d R'(u,S)$ and $R\le^\d R'$.

\item Let $R=\I{tu}^I_{\xi,\t}$ and $R'=\I{tu}^{I'}_{\xi,\t}$ ($t\neq
f\vt$). $R=\I{t}^I_{\xi,\t}(u\t,S)$ with $S=\I{u}^I_{\xi,\t}$. $R'=
\I{t}^{I'}_{\xi,\t}(u\t,S')$ with $S'=\I{u}^{I'}_{\xi,\t}$. Since
$\pd(tu)= 1.\pd(t)$, $\pos(F,t)\sle\pd(t)$ and
$\pos(F,u)=\vide$. Therefore, $S=S'$ and, by induction hypothesis,
$\I{t}^I_{\xi,\t}\le^\d \I{t}^{I'}_{\xi,\t}$. So, $R\le^\d R'$.\cqfd
\end{lst}
\end{prf}


\begin{lem}
\label{lem-int-const-mon}
$\vphi$ is monotonic.
\end{lem}

\begin{prf}
  Let $I,I'\in\cI^m$ such that $I\le I'$. We have to prove that, for
  all $D=_\cC C$, $\vphi^I_D\le \vphi^{I'}_D$, that is,
  $\vphi^I_D(\va)\sle \vphi^{I'}_D(\va)$ for all $\va$. Let $u\in
  \vphi^I_D(\va)$. We prove that $u\in \vphi^{I'}_D(\va)$. First, we
  have $u\in\SN$. Assume now that $u$ reduces to $f\vu$ with
  $f:(\vy:\vU)D\vv$. Let $j\in \acc(f)$. We have to prove that $u_j\in
  \I{U_j}^{I'}_{\xi,\t}$ with $\t=\vyu$ and, for all $y\in \FVB(U_j)$,
  $y\xi= S_{\io_y}$. Since $u\in \vphi^I_D(\va)$, we have $u_j\in
  \I{U_j}^I_{\xi,\t}$. Since $j\in\acc(f)$, by {\bf(I3)}, for all
  $E=_\cC D$, $\pos(E,U_j)\sle \pp(U_j)$. Now, only a finite number of
  symbols $E=_\cC D$ can occur in $U_j$, say $E_0,\ldots,E_{n-1}$. Let
  $I^0=I$ and, for all $i<n$, $I^{i+1}_D=I^i_D$ if $D\neq E_i$, and
  $I^{i+1}_D=I'_{E_i}$ otherwise. We have $I=I^0\le_{E_0} I^1\le_{E_1}
  \ldots I^{n-1}\le_{E_{n-1}} I^n=I'$. Hence, by Lemma
  \ref{lem-I-mon}, $\I{U_j}^I_{\xi,\t}\le \I{U_j}^{I'}_{\xi,\t}$ and
  $u\in \vphi^{I'}_D(\va)$.\cqfd
\end{prf}

Since $(\cI^m,\le)$ is a complete lattice, $\vphi$ has a least fixpoint
$I$ which is an interpretation for all the constant predicate symbols
equivalent to $C$. Hence, by induction on $>_\cC$, we obtain an
interpretation $I$ for all the constant predicate symbols.


In the case of a primitive constant predicate symbol, the
interpretation is simply the set of strongly normalizable terms of
this type:

\begin{lem}[Interpretation of primitive constant predicate symbols]
\label{lem-int-const-prim}

If $C$ is a primitive constant predicate symbol then $I_C=\top_\tC$.
\end{lem}

\begin{prf}
  Since $I_C\le\top_\tC$, it suffices to prove that $\top_\tC\le
  I_C$. Since, by assumption, $\th\tC:\B$, $\tC$ is of the form
  $(\vx:\vT)\st$. If $\va$ are arguments of $\top_\tC$ then
  $\top_\tC(\va)=\top_\st=\SN$ and it suffices to prove that, for all
  $u\in\SN$, $C$ primitive and $\va$ arguments of $I_C$, $u\in
  I_C(\va)$, by induction on $u$ with $\a\cup\,\tgt$ as well-founded
  ordering. Assume that $u\a^* f\vu$ with $f:(\vy:\vU)C\vv$. If $u\a^+
  f\vu$, we can conclude by induction hypothesis. So, assume that
  $u=f\vu$. In this case, we have to prove that, for all
  $j\in\acc(f)$, $u_j\in \I{U_j}_{\xi,\t}$ with $\t=\vyu$ and, for all
  $y\in\FVB(U_j)$, $y\xi= S_{\iota_y}$. By definition of primitive
  constant predicate symbols, for all $j\in\acc(f)$, $U_j$ is of the
  form $D\vw$ with $D$ primitive too. Hence, $\I{U_j}_{\xi,\t}=
  I_D(\va')$ with $a_i'= (w_i\t, \I{w_i}_{\xi,\t})$. Since
  $u_j\in\SN$, by induction hypothesis, $u_j\in I_D(\va')$. Therefore,
  $u\in I_C(\va)$.\cqfd
\end{prf}




\subsection{Computability ordering}
\label{sec-red-ord}

In this section, we assume given an interpretation $J$ for defined
predicate symbols and denote $\I{T}^{I\cup J}$ by $\I{T}$. The
fixpoint of the function $\vphi$ defined in the previous section can
be reached by transfinite iteration from the smallest element of
$\cI^m$, $\bot_C(\vt,\vS)= \bot_\st$. Let $I^\fa$ be the
interpretation reached after $\fa$ iterations of $\vphi$.


\begin{dfn}[Order of a computable term]
\label{def-ord}

The {\em order} of a term $t\in I_C(\vS)$, written $o_{C(\vS)}(t)$, is
the smallest ordinal $\fa$ such that $t\in I^\fa_C(\vS)$.
\end{dfn}


This notion of order will enable us to define a well-founded ordering
in which recursive definitions on strictly positive predicates
strictly decrease. Indeed, in this case, the subterm ordering is not
sufficient. In the example of the addition on ordinals, we have the
rule:

\begin{rewc}
x + (lim~f) & lim~([n:nat]x+fn)\\
\end{rewc}

We have a recursive call with $(fn)$ as argument, which is not a
subterm of $(lim~f)$. However, thanks to the definition of the
interpretation for constant predicate symbols and products, we can say
that, if $(lim~f)$ is computable then $f$ is computable and thus that,
for all computable $n$, $(fn)$ is computable. So, the order of
$(lim~f)$ is greater than the one of $(fn)$: $o(lim~f) > o(fn)$.


\begin{dfn}[Computability ordering]
\label{def-red-ord}

Let $f\in\cF$ with $stat_f= lex~m_1$ \ldots $m_k$. Let $\Theta_f$ be
the set of tuples $(g,\xi,\t)$ such that $g =_\cF f$ and
$\xi,\t\models\G_g$. We equip $\Theta_f$ with the ordering $\qgt_f$
defined by:

\begin{lst}{\bu}
\item $(g,\xi,\t) \,\qgt_f (g',\xi',\t')$ ~if~ $\vm\t
~(\qgt_f^{1,m},\ldots,\qgt_f^{k,m})\lex~ \vm\t'$,
\item $mul~\vt \,\qgt_f^{i,m} mul~\vt'$ ~if~ $\{\vt\} ~(\qgt_f^i)\mul~
  \{\vt'\}$,
\item $t\qgt_f^i t'$ ~if~ $i\in SP(f)$, $T_f^i=C\va$,
  $\I{\va}_{\xi,\t}= \I{\va}_{\xi',\t'}= \vS$ and
  $o_{C(\vS)}(t)>o_{C(\vS)}(t')$,
\item $t\qgt_f^i t'$ ~if~ $i\notin SP(f)$ and $t \,(\a\cup\,\tgt)\, t'$.
\end{lst}

\noindent
We equip $\Theta= \bigcup\, \{\Theta_f ~|~f\in\cF\}$ with the {\em
computability ordering} $\qgt$ defined by $(f,\xi,\t) \,\qgt
(f',\xi',\t')$ if $f>_\cF f'$ or, $f=_\cF f'$ and $(f,\xi,\t) \,\qgt_f
(f',\xi',\t')$.
\end{dfn}


\begin{lem}
\label{lem-red-ord-wf}

The computability ordering is well-founded and compatible with $\a$,
that is, if $\t\a\t'$ then $(g,\xi,\t) \qge (g,\xi,\t')$.
\end{lem}

\begin{prf}
  The computability ordering is well-founded since ordinals are
  well-founded and lexicographic and multiset orderings preserve
  well-foundedness. It is compatible with $\a$ by definition of the
  interpretation of constant predicate symbols.\cqfd
\end{prf}

We check hereafter that the accessibility relation is correct, that
is, an accessible subterm of a computable term is computable. Then, we
check that the ordering on arguments is correct too, that is, if
$t>_R^i u$ and $t$ is computable then $u$ is computable and $o(t)>
o(u)$.


\begin{lem}[Correctness of accessibility]
\label{lem-cor-acc}

If $t:T \tgt_\r u:U$ and $t\s\in \I{T\r}^{I^\fa}_{\xi,\s}$ with $\fa$
as small as posssible then $\fa=\fb+1$ and $u\s\in
\I{U\r}^{I^\fb}_{\xi,\s}$.
\end{lem}

\begin{prf}
  By definition of $\tgt_\r$, we have $t=f\vu$, $f:(\vy:\vU)C\vv$,
  $C\in\CFB$, $u=u_j$, $j\in \acc(f)$, $T\r= C\vv\g\r$, $U\r=U_j\g\r$,
  $\g=\vyu$ and no $D=_\cC C$ occurs in $\vu\r$. Hence, $t\s\in
  \I{C\vv\g\r}^{I^\fa}_{\xi,\s}= I^\fa_C(\vS)$ with $\vS=
  \I{\vv\g\r}^{I^\fa}_{\xi,\s}$. Assume that $\fa=0$. Then,
  $I^\fa_C(\vS)= \bot_\st$. But $f\vu\notin \bot_\st$ since $f\vu$ is
  not neutral (see Lemma \ref{lem-bottom}). So, $\fa\neq 0$. Assume
  now that $\fa$ is a limit ordinal. Then, $I^\fa_C(\vS)= \bigcup
  \{I^\fb_C(\vS)~|~ \fb<\fa\}$ and $t\s\in I^\fb_C(\vS)$ for some
  $\fb<\fa$, which is not possible since $\fa$ is as small as
  possible. Therefore, $\fa=\fb+1$ and, by definition of $I_C$,
  $u_j\s\in \I{U_j}^{I^\fb}_{\xi',\g\r\s}$ with $y\xi'= S_{\io_y}$. By
  {\bf(I6)}, $v_{\io_y}=y$. Thus, $y\xi'=
  \I{y\g\r}^{I^\fa}_{\xi,\s}$. Now, since no $D=_\cC C$ occurs in
  $\vu\r$, $y\xi'= \I{y\g\r}^{I^\fb}_{\xi,\s}$. Hence, by candidate
  substitution, $\I{U_j}^{I^\fb}_{\xi',\g\r\s}=
  \I{U_j\g\r}^{I^\fb}_{\xi,\s}$ and $u\s\in \I{U\r}^{I^\fb}_{\xi,\s}$
  since $U\r=U_j\g\r$.\cqfd
\end{prf}


\begin{lem}[Correctness of the ordering on arguments]
\label{lem-cor-arg-ord}

Assume that $t:T >_R^i u:U$ as in Definition \ref{def-arg-ord},
$t\s\in \I{T\r}_{\xi,\s}$ and $\vu\s\in \I{\vU\d}_{\xi,\s}$. Then,
$u\s\in \I{U\r}_{\xi,\s}$ and $o_{C(\vS)}(t\s)> o_{C(\vS)}(u\s)$ with
$\vS= \I{\vv\g\r}_{\xi,\s}$.
\end{lem}

\begin{prf}
Since $t:T \tgt_\r^+ x:V$, $T\r= C\vv\g\r$. Hence, $t\s\in
I^\fa_C(\vS)$ with $\fa= o_{C(\vS)}(t\s)$. By Lemma \ref{lem-cor-acc},
$\fa=\fb+1$ and $x\s\in \I{V\r}^{I^\fb}_{\xi,\s}$. Since no $D=_\cC C$
occurs in $\vU\d$, $\I{\vU\d}_{\xi,\s}=
\I{\vU\d}^{I^\fb}_{\xi,\s}$. Since $V\r= (\vy:\vU)C\vw$ and $\vu\s\in
\I{\vU\d}^{I^\fb}_{\xi,\s}$, $u\s\in
\I{C\vw}^{I^\fb}_{\xi_\vy^\vR,\s_\vy^{\vu\s}}$ with $\vR=
\I{\vu}^{I^\fb}_{\xi,\s}$. By candidate substitution,
$\I{C\vw}^{I^\fb}_{\xi_\vy^\vR,\s_\vy^{\vu\s}}=
\I{C\vw\d}^{I^\fb}_{\xi,\s}= I_C^\fb(\vS')$ with $\vS'=
\I{\vw\d}^{I^\fb}_{\xi,\s}$. Since $\vw\d|_C=\vv\g\r|_C$, $\vS'=
\I{\vv\g\r}^{I^\fb}_{\xi,\s}$. Since no $D=_\cC C$ occurs in
$\vv\g\r$, $\vS'=\vS$. Therefore, $u\s\in I_C^\fb(\vS)$ and
$o_{C(\vS)}(t\s)> o_{C(\vS)}(u\s)$.
\end{prf}




\subsection{Interpretation of defined predicate symbols}
\label{subsec-int-def}

We define the interpretation $J$ for defined predicate symbols by
induction on $\cgt$ ({\bf A3}). Let $F$ be a defined predicate symbol
and assume that we already defined an interpretation $K$ for every
symbol smaller than $F$. There are three cases depending on the fact
that the equivalence class of $F$ is primitive, positive or
computable. For simplicity, we denote $\I{T}^{I\cup K\cup J}$ by
$\I{T}^J$.




\subsubsection{Primitive systems}

\begin{dfn}
\label{def-int-def-prim}

  For every $G\simeq F$, we take $J_G= \top_\tG$.
\end{dfn}




\subsubsection{Positive, small and simple systems}
\label{sec-sys-pos}

Let $\cJ$ be the set of interpretations of the symbols equivalent to
$F$ and $\le$ be the relation on $\cJ$ defined by $J\le J'$ if, for
all $G\simeq F$, $J_G\le_\tG J'_G$. Since $(\cR_\tG,\le_\tG)$ is a
complete lattice, it is easy to see that $(\cJ,\le)$ is a complete
lattice too.


\begin{dfn}
\label{def-int-pos}

  Let $\psi$ be the function which, to $J\in\cJ$ and $G\simeq F$ with
  $G:(\vx:\vT)U$, associates the interpretation $\psi^J_G$ defined by:

$\psi^J_G(\vt,\vS) = \left\{
\begin{array}{@{}l@{}}
\I{r}^J_{\xi,\s} ~~\mbox{if}~ \vt\in\WN\!\cap\CR,
\,\vt\!\ad=\vl\s ~\mbox{and}~ (G\vl\a r,\G,\r)\in\cR\\[2mm]
\top_U ~~\mbox{otherwise}
\end{array}
\right.$

\noindent
where $x\xi= S_{\ka_x}$. We show hereafter that $\psi$ is
monotonic. So, we can take $J=\lfp(\psi)$.
\end{dfn}


\begin{lem}
\label{lem-cor-int-pos}

$\psi^J$ is a well defined interpretation.
\end{lem}

\begin{prf}
  By simplicity, at most one rule can be applied at the top of
  $G(\vt\!\ad)$. The existence of $\ka_x$ is the smallness condition
  {\bf(q)}. We now prove that $\psi^J_G\in \cR_\tG$. By {\bf(S3)},
  $\G\th r:U\g\r$ with $\g=\vxl$. Now, we prove that
  $\xi\models\G$. Let $x\in\FVB(r)$, $x\xi= S_{\ka_x}\in\cR_{x\s}$
  since $S_{\ka_x}\in \cR_{t_{\ka_x}}$ and, by smallness, $t_{\ka_x}=
  l_{\ka_x}\s= x\s$. Therefore, by Lemma \ref{lem-cor-schema-int},
  $\I{r}_{\xi,\s}\in \cR_{U\g\r}=\cR_U$. We are left to check that
  $\psi^J_G$ is stable by reduction. Assume that $\vt\a\vt'$. By
  {\bf(A1)}, $\a$ is confluent. Therefore, $\{\vt\}\sle\WN$ iff
  $\{\vt'\}\sle\WN$. Furthermore, if $\{\vt\}\sle\WN$, then $\vt\!\ad=
  \vt'\!\ad$ and $\psi^J_G(\vt,\vS)= \psi^J_G(\vt',\vS)$.\cqfd
\end{prf}


\begin{lem}
\label{lem-mon-int-pos}

  $\psi$ is monotonic.
\end{lem}

\begin{prf}
  As in Lemma \ref{lem-int-const-mon}.\cqfd
\end{prf}




\subsubsection{Computable, small and simple systems}


Let $\cD$ be the set of tuples $(G,\vt,\vS)$ such that $G\simeq F$,
and $\vxS,\vxt\models\G_G$. We equip $\cD$ with the well-founded
ordering $(G,\vt,\vS) \qgt_{_\cD} (G',\vt',\vS')$ iff $(G,\vxS,\vxt)
\qgt (G',\{\vx\to\vS'\},\{\vx\to\vt'\})$ (see Definition
\ref{def-red-ord}).

\begin{dfn}
\label{def-int-rec}

We first define $J'$ on $\cD$ by induction on $\qgt_{_\cD}$. Let
$G\simeq F$ with $G:(\vx:\vT)U$.

$J'_G(\vt,\vS) = \left\{
\begin{array}{@{}l@{}}
\I{r}^{J'}_{\xi,\s} ~~\mbox{if}~ \vt\in\WN\!\cap\CR,
\,\vt\!\ad=\vl\s ~\mbox{and}~ (G\vl\a r,\G,\r)\in\cR\\[2mm]
\top_U ~~\mbox{otherwise}\\
\end{array}
\right.$

\noindent
where $x\xi= S_{\ka_x}$. Then, $J_G(\vt,\vS)= J'_G(\vt\!\ad,\vS)$ if
$\vt\in\WN\!\cap\CR$, and $J_G(\vt,\vS)= \top_U$ otherwise.
\end{dfn}


\begin{lem}
\label{lem-cor-int-rec}

  $J$ is a well defined interpretation.
\end{lem}

\begin{prf}
  As in Lemma \ref{lem-cor-int-pos}. The well-foundedness of the
  definition comes from Lemma \ref{lem-red-ho} and Theorem
  \ref{thm-cor-thc}. In Lemma \ref{lem-red-ho}, we show that, starting
  from a sequence in $\cD$, we can apply Theorem \ref{thm-cor-thc}
  where we show that, in a recursive call $G'\vt'$, $(G,\vt,\vS)\qgt
  (G',\vt',\vS')$ for some $\vS'$.\cqfd
\end{prf}




\subsection{Correctness of the conditions}


\begin{dfn}[Cap and aliens]
\label{def-cap}

  Let $\z$ be an injection from classes of terms modulo $\aa^*$ to
  $\cX$. The {\em cap} of a term $t$ w.r.t. a set $\cG$ of symbols is
  the term $cap_\cG(t)= t[x_1]_{p_1} \ldots [x_n]_{p_n}$ such that,
  for all $i$, $x_i= \z(t|_{p_i})$ and $t|_{p_i}$ is not of the form
  $g\vt$ with $g\in\cG$. The $t|_{p_i}$'s are the {\em aliens} of
  $t$. We denote by $aliens_\cG(t)$ their multiset.
\end{dfn}


\begin{lem}[Pre-computability of first-order symbols]
\label{lem-pre-red-fo}

  If $f\in\cF_1$ and $\vt\in\SN$ then $f\vt\in\SN$.
\end{lem}

\begin{prf}
  We prove that every reduct $t'$ of $t=f\vt$ is in $\SN$. Hereafter,
  $cap=cap_{\cF_1}$.
  
  \u{\bf Case $\cR_\w \neq \vide$}. By induction on
  $(aliens(t),cap(t))\lex$ with $((\a\cup\,\tgt)\mul,\a_{\cR_1})\lex$
  as well-founded ordering (the aliens are strongly normalizable and,
  by {\bf(f)}, $\a_{\cR_1}$ is strongly normalizing on first-order
  algebraic terms).
  
  If the reduction takes place in $cap(t)$ then this is a
  $\cR_1$-reduction. By {\bf(c)}, no symbol of $\cF_\w$ occurs in the
  rules of $\cR_1$. And, by {\bf(d)}, the right hand-sides of the
  rules of $\cR_1$ are algebraic. Therefore, $cap(t) \a_{\cR_1}
  cap(t')$. By {\bf(e)}, the rules of $\cR_1$ are non
  duplicating. Therefore, $aliens(t) \tgt\mul aliens(t')$ and we can
  conclude by induction hypothesis.
  
  If the reduction takes place in an alien then $aliens(t)
  ~(\a\cup\,\tgt)\mul~ aliens(t')$ and we can conclude by induction
  hypothesis.
  
  \u{\bf Case $\cR_\w = \vide$}. Since the $t_i$'s are strongly
  normalizable and no $\b$-reduction can take place at the top of $t$,
  $t$ has a $\b$-normal form. Let $cap\b(t)$ be the cap of its
  $\b$-normal form. We prove that every immediate reduct $t'$ of $t$
  is strongly normalizable, by induction on $(\b
  cap(t),aliens(t))\lex$ with $(\a_{\cR_1},(\a\cup\,\tgt)\mul)\lex$ as
  well-founded ordering (the aliens are strongly normalizable and, by
  {\bf(f)}, $\a_{\cR_1}$ is strongly normalizing on first-order
  algebraic terms).
  
  If the reduction takes place in $cap(t)$ then this is a
  $\cR_1$-reduction. By {\bf(d)}, the right hand-sides of the rules of
  $\cR_1$ are algebraic. Therefore, $t'$ has a $\b$-normal form and
  $cap\b(t) \a_{\cR_1} cap\b(t')$. Hence, we can conclude by
  induction hypothesis. If the reduction is a $\b$-reduction in an
  alien then $cap\b(t)= cap\b(t')$ and $aliens(t)$
  $(\a\cup\,\tgt)\mul~$ $aliens(t')$. Hence, we can conclude by
  induction hypothesis.
 
  We are left with the case where the reduction is a $\cR_1$-reduction
  taking place in an alien $u$. Then, $aliens(t) \a\mul aliens(t')$,
  $cap\b(t) \a_{\cR_1}^* cap\b(t')$ and we can conclude by induction
  hypothesis. To see that $cap\b(t) \a_{\cR_1}^* cap\b(t')$, it
  suffices to remark that, if we $\b$-normalize $u$, then all the
  residuals of the $\cR_1$-redex are still reducible (left and right
  hand-sides of first-order rules are algebraic).\cqfd
\end{prf}


\begin{lem}[Computability of first-order symbols]
\label{lem-red-fo}

For all $f\in\cF_1$, $f\in\I{\tf}$.
\end{lem}

\begin{prf}
  Assume that $f:(\vx:\vT)U$. $f\in\I{\tf}$ iff, for all $\G_f$-valid
  pair $(\xi,\t)$, $f\vx\t\in R=\I{U}_{\xi,\t}$. For first-order
  symbols, $U=\st$ or $U=C\vv$ with $C$ primitive. If $U=\st$ then
  $R=\top_\st=\SN$. If $U=C\vv$ with $C:(\vy:\vU)V$ then $R=I_C(\va)$
  with $a_i=(v_i\t,\I{v_i}_{\xi,\t})$. Since $C$ is primitive, by
  Lemma \ref{lem-int-const-prim}, $I_C=\top_\tC$ and $R=\top_V$. By
  assumption, $\th\tC:\B$ and $\th\tf:s_f$. After Lemma
  \ref{lem-kind}, $s_f=\st$ and $V=\st$. Therefore,
  $R=\top_\st=\SN$. Now, since $\xi,\t\models\G_f$, we have $x_i\t\in
  \I{T_i}_{\xi,\t}\sle\SN$ by {\bf(R1)}. Hence, by pre-computability
  of first-order symbols, $f\vx\t\in \I{U}_{\xi,\t}$.\cqfd
\end{prf}


\begin{thm}[Strong normalization of $\ar$]
\label{thm-sn-rew}

\hfill The relation $\ar= \a_{\cR_1}\cup \a_{\cR_\w}$ is strongly
normalizing.
\end{thm}

\begin{prf}
  By induction on the structure of terms. The only difficult case is
  $f\vt$. If $f$ is first-order, we use the Lemma of pre-computability
  of first-order symbols. If $f$ is higher-order, we have to show
  that, if $\vt\in\SN_\cR$, then $t=f\vt\in\SN_\cR$, where $\SN_\cR$
  is the set of terms that are strong normalizable w.r.t. $\ar$.
  
  Let $\varpi(t)=0$ if $t$ is not of the form $g\vu$ and $\varpi(t)=1$
  otherwise. We prove that every reduct $t'$ of $t$ is strongly
  normalizable by induction on $(f,\varpi(\vt),\vt,\vt)$ with
  $(>_\cF,(>_\mb{N})\sf,(\tgt\,\cup\ar)\sf,(\ar)\lex)\lex$ as
  well-founded ordering. Assume that $t'= f\vt'$ with $t_i\ar t_i'$
  and, for all $j\neq i$, $t_j= t_j'$. Then, $\vt ~(\ar)\lex~ \vt'$
  and $\varpi(t_i) \ge \varpi(t_i')$ since if $t_i$ is not of the form
  $g\vu$ then $t_i'$ is not of the form $g\vu$ either.

  Assume now that there exists $f\vl\a r \in \cR_\w$ such that $\vt=
  \vl\s$ and $t'= r\s$. By {\bf(a)}, $r$ belongs to the computability
  closure of $l$. It is then easy to prove that $r\s$ is strongly
  normalizable by induction on the structure of $r$. Again, the only
  difficult case is $g\vu$. But then, either $g$ is smaller than $f$,
  or $g$ is equivalent to $f$ and its arguments are smaller than
  $\vl$. If $l_i >_1 u_j$ then $l_i \tgt u_j$ and $\FV(u_j) \sle
  \FV(l_i)$. Therefore $l_i\s \tgt u_j\s$ and $\varpi(l_i\s)=1 \ge
  \varpi(u_j\s)$. If now $l_i >_2 u_j$ then $u_j$ is of the form
  $x\vv$ and $\varpi(l_i\s)=1 > \varpi(u_j\s)=0$.\cqfd
\end{prf}


\begin{lem}[Invariance by reduction]
\label{lem-inv-by-red}

If $\G\th t:T$, $t\a t'$, $\xi\models\G$ and $t\t\in\WN$ then
$\I{t}_{\xi,\t}= \I{t'}_{\xi,\t}$.
\end{lem}

\begin{prf}
By induction on $t$. If $t$ is an object then $t'$ is an object too
and $\I{t}_{\xi,\t}= \vide= \I{t'}_{\xi,\t}$. Otherwise, we proceed by
case on $t$ and $t'$:

\begin{lst}{\bu}
\item Let $R= \I{F\vl\s}_{\xi,\t}$ and $R'= \I{r\s}_{\xi,\t}$ with
$(F\vl\a r,\G_0,\r)\in\cR$. $R= I_F(\va)$ with
$a_i=(l_i\s\t,\I{l_i\s}_{\xi,\t})$. By {\bf(A3)}, there are two
sub-cases:

\begin{lst}{--}
\item {\bf $F$ belongs to a primitive system}. Then, $I_F=\top_\tF$
and $r$ is of the form $[\vx:\vT]$ $G\vu$ with $G\simeq F$ or $G$ a
primitive constant predicate symbol. In both cases,
$I_G=\top_\tG$. Therefore, $R=R'$.
  
\item {\bf $F$ belongs to a positive or computable, small and simple
system}. Since $l_i\s\t\in\WN$, by {\bf(A1)}, $l_i\s\t$ has a unique
normal form $t_i$. By simplicity, the symbols in $\vl$ are
constant. Therefore, $t_i$ is of the form $l_i\t'$ with $\s\t\a^*
\t'$, and $R= \I{r}_{\xi',\t'}$ with $x\xi'=
\I{l_{\ka_x}\s}_{\xi,\t}$. By smallness, $l_{\ka_x}=x$ and $x\xi'=
\I{x\s}_{\xi,\t}$. By Lemma \ref{lem-cor-schema-int},
$\I{r}_{\xi',\t'}= \I{r}_{\xi',\s\t}$. By {\bf(S4)},
$\s:\G_0\leadsto\G$. Therefore, by candidate substitution, $R=R'$.
\end{lst}

\item Let $R= \I{[x:U]v~u}_{\xi,\t}$ and $R'= \I{v\xu}_{\xi,\t}$. Let
$S=\I{u}_{\xi,\t}$. $R= \I{[x:U]v}(u\t,S)= \I{v}_{\xi_x^S,\t'}$ with
$\t'=\t_x^{u\t}=\xu\t$. Since $\xu:(\G,x:U)\a\G$, by candidate
substitution, $R'= \I{v}_{\xi_x^S,\t'}= R$.

\item Let $R= \I{tu}_{\xi,\t}$ and $R'= \I{t'u'}_{\xi,\t}$ with $t\a
t'$ and $u\a u'$. $R= \I{t}_{\xi,\t}(u\t,\I{u}_{\xi,\t})$ and $R'=
\I{t'}_{\xi,\t}(u'\t,\I{u'}_{\xi,\t})$. By induction hypothesis,
$\I{t}_{\xi,\t}= \I{t'}_{\xi,\t}$ and $\I{u}_{\xi,\t}=
\I{u'}_{\xi,\t}$. Finally, since candidates are stable by reduction,
$R=R'$.

\item Let $R= \I{[x:U]v}_{\xi,\t}$ and $R'= \I{[x:U']v'}_{\xi,\t}$
with $U\a U'$ and $v\a v'$. Since $\cR_U=\cR_{U'}$, $R$ and $R'$ have
the same domain $\cT\times\cR_U$ and codomain $\cR_V$, where $V$ is
the type of $v$. $R(u,S)= \I{v}_{\xi_x^S,\t_x^u}$ and $R'(u,S)=
\I{v'}_{\xi_x^S,\t_x^u}$. By induction hypothesis, $R(u,S)=
R'(u,S)$. Therefore, $R=R'$.

\item Let $R= \I{(x:U)V}_{\xi,\t}$ and $R'=
\I{(x:U')V'}_{\xi,\t}$. $R= \{t\in\cT~|~ \all u\in\I{U}_{\xi,\t}, \all
S\in\cR_U, tu\in \I{V}_{\xi_x^S,\t_x^u}\}$ and $R'= \{t\in\cT~|~
\all u\in\I{U'}_{\xi,\t}, \all S\in\cR_U, tu\in
\I{V'}_{\xi_x^S,\t_x^u}\}$. By induction hypothesis,
$\I{U}_{\xi,\t}= \I{U'}_{\xi,\t}$ and $\I{V}_{\xi_x^S,\t_x^u}=
\I{V'}_{\xi_x^S,\t_x^u}$. Therefore, $R=R'$.\cqfd
\end{lst}
\end{prf}


\begin{lem}[Pre-computability of well-typed terms]
\label{lem-red}

Assume that, for all $f$, $f\in\I{\tf}$. If $\G\th t:T$ and
$\xi,\t\models\G$ then $t\t\in \I{T}_{\xi,\t}$.
\end{lem}

\begin{prf}
By induction on $\G\th t:T$.

\begin{lst}{}
\item [\bf(ax)] $\st\t=\st\in \I{\B}_{\xi,\t}= \top_\B= \SN$.

\item [\bf(symb)] By assumption.

\item [\bf(var)] $x\t\in \I{T}_{\xi,\t}$ since $\t$ is adapted to $\xi$.

\item [\bf(weak)] By induction hypothesis.

\item [\bf(prod)] We have to prove that $(x:U\t)V\t\in
\I{s'}_{\xi,\t}= \top_{s'}= \SN$. By induction hypothesis, $U\t\in
\I{s}_{\xi,\t}= \SN$. Now, let $\xi'= \xi_x^{\top_U}$. Since
$\xi',\t\models\G,x:U$, by induction hypothesis, $V\t\in
\I{s'}_{\xi',\t}= \SN$.

\item [\bf(abs)] Let $t=[x:U]v$. We have to prove that $t\t\in
\I{(x:U)V}_{\xi,\t}$. First note that $U\t,v\t\in\SN$. Indeed, let
$\xi'= \xi_x^{\top_U}$. Since $\xi',\t\models\G,x:U$, by induction
hypothesis, $v\t\in \I{V}_{\xi',\t}$. Furthermore, by inversion,
$\G\th U:s$ for some $s$. So, by induction hypothesis, $U\t\in
\I{s}_{\xi,\t}= \SN$. Now, let $u\in\I{U}_{\xi,\t}\sle\SN$ and
$S\in\cR_U$. We must prove that $t\t u\in S'=
\I{V}_{\xi_x^S,\t_x^u}$. Since $t\t u$ is neutral, it suffices to
prove that $\a\!\!(t\t u)\sle S'$. We prove it by induction on
$(U\t,v\t,u)$ with $\a\lex$ as well-founded ordering. We have $t\t u\a
v\t\xu=v\t'$. Since $\xi_x^S,\t_x^u\models\G,x:U$, by induction
hypothesis, $v\t'\in S'$. For the other cases, we can conclude by
induction hypothesis on $(U\t,v\t,u)$.

\item [\bf(app)] We have to prove that $t\t u\t\in
\I{V\xu}_{\xi,\t}$. By induction hypothesis, $t\t\in
\I{(x:U)V}_{\xi,\t}$ and $u\t\in \I{U}_{\xi,\t}$. Since $S=
\I{u\t}_{\xi,\t}\in \cR_{U\t}=\cR_U$, by definition of
$\I{(x:U)V}_{\xi,\t}$, $t\t u\t\in \I{V}_{\xi_x^S,\t'}$ with
$\t'=\t_x^{u\t}$. By candidate substitution, $\I{V\xu}_{\xi,\t}=
\I{V}_{\xi',\xu\t}$ with $y\xi'= \I{y\xu}_{\xi,\t}$. Since
$\xi'=\xi_x^S$ and $\xu\t=\t'$, $t\t u\t\in \I{V\xu}_{\xi,\t}$.

\item [\bf(conv)] In \cite{blanqui01thesis}, we show that adding the
hypothesis $\G\th T:s$ does not change the typing relation. Therefore,
by induction hypothesis, $t\t\in \I{T}_{\xi,\t}$, $T\t\in
\I{s}_{\xi,\t}= \top_s= \SN$ and $T'\t\in \I{s}_{\xi,\t}= \SN$. Hence,
by invariance by reduction, $\I{T}_{\xi,\t}= \I{T'}_{\xi,\t}$ and
$t\t\in \I{T'}_{\xi,\t}$.\cqfd
\end{lst}
\end{prf}


\begin{thm}[Computability closure correctness]
\label{thm-cor-thc}

Let $(f\vl\a r,\G,\r)$ be a well-formed rule with $f\in\cF_\w$,
$f:(\vx:\vT)U$ and $\g=\vxl$. Assume that $\eta,\g\s\models\G_f$,
$\xi,\s\models\G$, $x\eta= \I{x\g\r}_{\xi,\s}$, and $\vl\s\in
\I{\vT\g\r}_{\xi,\s}$. Assume also that:

\begin{lst}{\bu}
\item $\all g<_\cF f$, $g\in\I{\tg}$,
\item $\all g=_\cF f$, if $g:(\vy:\vU)V$ and $(f,\eta,\g\s)\qgt
(g,\xi'',\t)$ then $g\vy\t\in \I{V}_{\xi'',\t}$.
\end{lst}

\noindent
If $\D\thc t:T$ and $\xi\xi',\s\s'\models\G,\D$ then $t\s\s'\in
\I{T}_{\xi\xi',\s\s'}$.
\end{thm}

\begin{prf}
  By induction on $\D\thc t:T$, we prove that $t\s\s'\in
  \I{T}_{\xi\xi',\s\s'}$ as in the previous lemma. We only detail the
  case (symb$^=$). Let $\vu=\vy\d$. By induction hypothesis,
  $\vu\s\s'\in \I{\vU\d}_{\xi\xi',\s\s'}$. By candidate substitution,
  there exists $\xi''$ such that $\I{\vU\d}_{\xi\xi',\s\s'}=
  \I{\vU}_{\xi'',\d\s\s'}$, $\I{V\d}_{\xi\xi',\s\s'}$
  $=\I{V}_{\xi'',\d\s\s'}$ and $\xi''\models\G_g$. Therefore,
  $\xi'',\d\s\s'\models\G_g$.

  We now prove that $(f,\eta,\g\s)\qgt (g,\xi'',\d\s\s')$. If
  $l_i:T_i\g \tgt_\r^+ u_j:U_j\d$. Then, $l_i\tgt u_j$ and
  $\FV(u_j)\sle \FV(l_i)$. Therefore, $l_i\s= l_i\s\s'\tgt
  u_j\s\s'$. Assume now that $l_i:T_i\g >^k_R u_j:U_j\d$, $k\in SP(f)$
  and $T_f^k=C\va$. By definition of $>^k_R$, $l_i=h\vt'$,
  $h:(\vx':\vT')C\vv$, $u_j=x\vu'$, $x\in\dom(\G)$, $l_i:T_i\g
  \tgt_\r^+ x:V$ and $V\r= x\G= (\vy':\vU')C\vw$, where
  $\g'=\{\vx'\mapsto\vt'\}$ and $\d'=\{\vy'\mapsto\vu'\}$. We must
  prove that $\I{\va}_{\eta,\g\s}= \I{\va}_{\xi'',\d\s\s'}= \vS$ and
  $o_{C(\vS)}(l_i\s)> o_{C(\vS)}(u_j\s\s')$.

  Assume that $T_i=C\vt$ and $U_j=C\vu$. Since $k\in SP(f)$,
  $\vt|_C=\vu|_C=\va|_C$. By definition of $\tgt_\r$, $T_i\g\r=
  C\vv\g'\r$. Hence, $\va\g\r|_C= \vv\g'\r|_C$. By definition of
  $>^k_R$, $\vv\g'\r|_C= \vw\d'|_C$ and $U_j\d\r=C\vw\d'$. Therefore,
  $\va\g\r|_C= \vw\d'|_C= \vu\d\r|_C= \va\d\r|_C= \va\d|_C$ since
  $\dom(\r)\sle\FV(l)$, $\FV(\d)\sle\dom(\D)$ and $\dom(\D)\cap\FV(l)=
  \vide$. By {\bf(S5)}, $\I{\va}_{\eta,\g\s}=
  \I{\va}_{\eta,\g\r\s}$. Since $x\eta= \I{x\g\r}_{\xi,\s}$, by
  candidate substitution, $\I{\va}_{\eta,\g\r\s}=
  \I{\va\g\r}_{\xi,\s}$. So, $\I{\va}_{\eta,\g\s}= \I{\va\d}_{\xi,\s}=
  \I{\va\d}_{\xi\xi',\s\s'}= \I{\va}_{\xi'',\d\s\s'}$. Now, by
  induction hypothesis, $\vu'\s\s'\in
  \I{\vU'\d'}_{\xi\xi',\s\s'}$. Therefore, since $l_i\s=l_i\s\s'\in
  \I{T_i\g\r}_{\xi,\s}= \I{T_i\g\r}_{\xi\xi',\s\s'}$, by Lemma
  \ref{lem-cor-arg-ord}, $u_j\s\s'\in \I{U_j\d\r}_{\xi\xi',\s\s'}$ and
  $o_{C(\vR)}(l_i\s)> o_{C(\vR)}(u_j\s\s')$ where $\vR=
  \I{\vv\g'\r}_{\xi\xi',\s\s'}= \vS$.
\end{prf}


\begin{lem}[Computability of higher-order symbols]
\label{lem-red-ho}

For all $f\in\cF_\w$, $f\in\I{\tf}$.
\end{lem}

\begin{prf}
  Assume that $f:(\vx:\vT)U$. $f\in\I{\tf}$ iff, for all $\G_f$-valid
  pair $(\eta,\t)$, $f\vx\t\in\I{U}_{\eta,\t}$. We prove it by
  induction on $((f,\eta,\t),\t)$ with $(\qgt,\a)\lex$ as well-founded
  ordering. Let $t_i= x_i\t$ and $t=f\vt$. By assumption (see
  Definition \ref{def-typing}), for all rule $f\vl\a r\in\cR$,
  $|\vl|\le|\vt|$. So, if $U\neq C\vv$ with $C\in\CFB$ then $t$ is
  neutral and it suffices to prove that $\a\!\!(t)\sle
  \I{U}_{\eta,\t}$. Otherwise, $\I{U}_{\eta,\t}= I_C(\va)$ with
  $a_i=(v_i\t,\I{v_i}_{\eta,\t})$. Since $\eta,\t\models\G_f$, $t_j\in
  \I{T_j}_{\eta,\t}$. Therefore, in this case too, it suffices to
  prove that $\a\!\!(t)\sle \I{U}_{\eta,\t}$.
  
  If the reduction takes place in one $t_i$ then we can conclude by
  induction hypothesis since reducibility candidates are stable by
  reduction and $\qgt$ is compatible with reduction. Assume now that
  there exist $(l\a r,\G,\r)\in\cR$ and $\s$ such that $l=f\vl$ and
  $t=l\s$. Then, $\t=\g\s$ with $\g=\vxl$. Furthermore, by {\bf(S5)},
  $\s\ad\r\s$. Hence, by Lemma \ref{lem-cor-schema-int},
  $\I{U}_{\eta,\t}= \I{U}_{\eta,\g\r\s}$ and $\I{\vT}_{\eta,\t}=
  \I{\vT}_{\eta,\g\r\s}$. Now, since rules are well-formed, $\G\th
  l\r:U\g\r$. Therefore, by inversion, $\G\th l_i\r:T_i\g\r$ and
  $\g\r:\G_f\leadsto\G$.

  We now define $\xi$ such that $\I{U}_{\eta,\g\r\s}=
  \I{U\g\r}_{\xi,\s}$ and $\I{\vT}_{\eta,\g\r\s}=
  \I{\vT\g\r}_{\xi,\s}$. By safeness {\bf(b)}, for all $x\in\FVB(\vT
  U)$, $x\g\r\in\dom(\G)$ and, for all $x,x'\in \FVB(\vT U)$,
  $x\g\r=x'\g\r \A x=x'$. Let $y\in\domB(\G)$. If there exists
  $x\in\dom(\G_f)$ (necessarily unique) such that $y=x\g\r$, we take
  $y\xi=x\eta$. Otherwise, we take $y\xi= \top_{y\G}$. We check that
  $\xi\models\G$. If $y\neq x\g\r$, $y\xi= \top_{y\G}\in
  \cR_{y\G}$. If $y=x\g\r$ then $y\xi=x\eta$. Since $\eta\models\G_f$,
  $x\eta\in \cR_{x\G_f}$. Since $\g\r:\G_f\leadsto\G$, $\G\th
  y:x\G_f\g\r$. Therefore, $y\G\CVG x\G_f\g\r$ and, by Lemma
  \ref{lem-cand-conv}, $y\xi=x\eta\in \cR_{x\G_f}= \cR_{x\G_f\g\r}=
  \cR_{y\G}$. So, $\xi\models\G$. Now, by candidate substitution,
  $\I{U\g\r}_{\xi,\s}= \I{U}_{\xi',\g\r\s}$ with $x\xi'=
  \I{x\g\r}_{\xi,\s}$. Let $x\in\FV(\vT U)$. By {\bf(b)}, $x\g\r= y\in
  \domB(\G)$ and $x\xi'= y\xi= x\eta$. Since $\xi'$ and $\eta$ are
  equal on $\FVB(\vT U)$, $\I{U}_{\xi',\g\r\s}= \I{U}_{\eta,\g\r\s}=
  \I{U\g\r}_{\xi,\s}$ and $\I{\vT}_{\xi',\g\r\s}=
  \I{\vT}_{\eta,\g\r\s}= \I{\vT\g\r}_{\xi,\s}$.
  
  We now prove that $\s$ is adapted to $\xi$. Let
  $x\in\dom(\G)$. Since rules are well-formed, there exists $i$ such
  that $l_i:T_i\g ~\tgt_\r^*~ x:x\G$ and
  $\dom(\r)\sle\FV(l)\moins\dom(\G)$. Since $l_i\s\in
  \I{T_i\g\r}_{\xi,\s}$, by correctness of accessibility, $x\s\in
  \I{x\G\r}_{\xi,\s}$. Since $\dom(\r)\cap\dom(\G)= \vide$, $x\G\r=
  x\G$ and $x\s\in \I{x\G}_{\xi,\s}$. Therefore, $\s$ is adapted to
  $\xi$ and, by correctness of the computability closure, $r\s\in
  \I{U\g\r}_{\xi,\s}= \I{U}_{\eta,\t}$.\cqfd
\end{prf}


\begin{lem}[Computability of well-typed terms]
\label{lem-red-wt-terms}

  If $\G\th t:T$ and $\xi,\t\models\G$ then $t\t\in \I{T}_{\xi,\t}$.
\end{lem}

\begin{prf}
  After Lemmas \ref{lem-red-fo}, \ref{lem-red} and \ref{lem-red-ho}.\cqfd
\end{prf}


\begin{thm}[Strong normalization]
\label{thm-sn}

  Every typable term is strongly normalizable.
\end{thm}

\begin{prf}
  Assume that $\G\th t:T$. Let $x\xi=\top_{x\G}$ for all
  $x\in\dom(\G)$. Since $\xi\models\G$ and the identity substitution
  $\io$ is adapted to $\xi$, $t\in S=\I{T}_{\xi,\io}$. Now, either
  $T=\B$ or $\G\th T:s$ for some $s$. If $T=\B$ then $S= \top_\B=
  \SN$. If $\G\th T:s$ then $S\in\cR_s$ and $S\sle\SN$ by
  {\bf(R1)}. So, in both cases, $t\in\SN$.\cqfd
\end{prf}




\section{Future directions of research}
\label{sec-future}

We conclude by giving some directions of research for improving our
conditions of strong normalization.\\

\noindent{\bf Rewriting modulo.} We did not consider rewriting modulo
some equational theories like associativity and commutativity. While
this does not create too much difficulties at the object level
\cite{blanqui03rta}, it is less clear for rewriting at the type
level.\\

\noindent{\bf Quotient types.} We have seen that rewrite rules on
constructors allows us to formalize some quotient types. However, to
prove properties by induction on such types requires to know what the
normal forms are \cite{jouannaud86lics} and may also require a
particular reduction strategy \cite{courtieu01csl} or conditional
rewriting.\\

\noindent{\bf Confluence.} Among our strong normalization conditions,
we not only require rewriting to be confluent but also its combination
with $\b$-reduction. This is a strong condition since we cannot rely
on strong normalization for proving confluence
\cite{nipkow91lics,blanqui00rta}. Except for first-order rewriting
systems without dependent types \cite{breazu94ic} or left-linear
higher-order rewrite systems \cite{muller92ipl,oostrom94thesis}, few
results are known on modularity of confluence for the combination of
higher-order rewriting and $\b$-reduction. Therefore, it would be
interesting to study this problem more deeply.\\

\noindent{\bf Local confluence.} We believe that local confluence is
sufficient for establishing strong normalization since local
confluence and strong normalization together imply confluence. But,
then, it seems necessary to prove many properties simultaneously
(subject reduction, strong normalization and confluence), which seems
difficult.\\

\noindent{\bf Simplicity.} For non-primitive predicate symbols, we
require that their defining rules have no critical pairs between them
or with the other rules. These strong conditions allow us to define a
valid interpretation in a simple way. It is important to be able to
weaken these conditions in order to capture more decision procedures.\\

\noindent{\bf Local definitions.} In our work, we considered only
globally defined symbols, that is, symbols whose type is typable in
the empty environment. However, in practice, during a formal proof in
a system like Coq \cite{coq02}, it may be very useful to introduce
symbols and rules using some hypothesis. We should study the problems
arising from local definitions and how our results can be used to
solve them. Local abbreviations are studied by Poll and Severi
\cite{poll94lfcs} and local definitions by rewriting are considered by
Chrzaszcz \cite{chrzaszcz00types}.\\

\noindent{\bf HORPO.} For higher-order definitions, we have chosen to
extend the General Schema of Jouannaud and Okada
\cite{jouannaud97tcs}. But the Higher-Order Recursive Path Ordering
(HORPO) of Jouannaud and Rubio \cite{jouannaud99lics}, which is an
extension of RPO to the simply typed $\la$-calculus, is naturally more
powerful. Walukiewicz recently extended this ordering to the Calculus
of Constructions with symbols at the object level only
\cite{walukiewicz00lfm,walukiewicz02jfp}. The combination of the two
works should allow us to extend RPO to the Calculus of Constructions
with type-level rewriting too.\\

\noindent{\bf $\eta$-Reduction.} Among our conditions, we require the
confluence of $\ar\cup\ab$. Hence, our results cannot be directly
extended to $\eta$-reduction, which is well known to create important
difficulties \cite{geuvers93thesis} since $\ab\cup\ae$ is not
confluent on not well-typed terms.\\

\noindent{\bf Non-strictly positive predicates.} The ordering used in
the General Sche\-ma for comparing the arguments of function symbols
can capture recursive definitions on basic and strictly-positive
types, but cannot capture recursive definitions on non-strictly
positive types \cite{matthes00}. However, Mendler
\cite{mendler87thesis} showed that such definitions are strongly
normalizing. In \cite{blanqui03tlca}, we recently showed how to deal
with such definitions in the Calculus of Algebraic Constructions.\\

\noindent{\bf Acknowledgments:} I would like to thank very much Daria
Walukiewicz who pointed to me several errors or imprecisions in
previous versions of this work. I also thank Jean-Pierre Jouannaud,
Gilles Dowek, Christine Paulin, Herman Geuvers, Thierry Coquand and
the anonymous referees for their useful comments on previous versions
of this work.


\end{document}